%% file: main.tex
\documentclass[%
     aip,
     amsmath,amssymb,
     reprint,%
]{revtex4-1}

\usepackage{graphicx}
\usepackage{dcolumn}
\usepackage{bm}

\usepackage[utf8]{inputenc}
\usepackage[T1]{fontenc}
\usepackage{mathptmx}
\usepackage{etoolbox}

\makeatletter
\def\@email#1#2{%
 \endgroup
 \patchcmd{\titleblock@produce}
  {\frontmatter@RRAPformat}
  {\frontmatter@RRAPformat{\produce@RRAP{*#1\href{mailto:#2}{#2}}}\frontmatter@RRAPformat}
  {}{}
}%
\makeatother

\usepackage[margin=2cm]{geometry}
\allowdisplaybreaks

\include{./commands}

\usepgfplotslibrary{fillbetween}

\begin{document}
\preprint{AIP/123-QED}

\title[]{Quantum Signal Processing (QSP) for simulating cold plasma waves}
\author{I. Novikau}
\email{inovikau@pppl.gov}
\author{E. A. Startsev}
\affiliation{Princeton Plasma Physics Laboratory, Princeton, New Jersey 08543, USA}

\author{I. Y. Dodin}
\affiliation{Princeton Plasma Physics Laboratory, Princeton, New Jersey 08543, USA}
\affiliation{Department of Astrophysical Sciences, Princeton University, Princeton, New Jersey 08544, USA}
\date{\today}

\begin{abstract}
Numerical modeling of radio-frequency waves in plasma with sufficiently high spatial and temporal resolution remains challenging even with modern computers.
However, such simulations can be sped up using quantum computers in the future.
Here, we propose how to do such modeling for cold plasma waves, in particular, for an X wave propagating in an inhomogeneous one-dimensional plasma.
The wave system is represented in the form of a vector Schr\"odinger equation with a Hermitian Hamiltonian.
Block-encoding is used to represent the Hamiltonian through unitary operations that can be implemented on a quantum computer.
To perform the modeling, we apply the so-called Quantum Signal Processing algorithm and construct the corresponding circuit.
Quantum simulations with this circuit are emulated on a classical computer, and the results show agreement with traditional classical calculations.
We also discuss how our quantum circuit scales with the resolution.
\end{abstract}
\maketitle

\input{intr}

\input{waves}

\input{discr}

\input{qsp}

\input{BE}

\input{results}

\input{conclusions}

\begin{acknowledgments}
The work was supported by the U.S. DOE through Contract No. DE-AC02-09CH11466.
The authors also thank Alexander Engel for valuable discussions.
\end{acknowledgments}

\input{appendix}

\bibliography{main}

\end{document}

%% file: commands.tex
\usepackage{subfig}

\usepackage{hyperref}
\usepackage[hyphenbreaks]{breakurl}
\usepackage[dvipsnames]{xcolor}
\usepackage{amsmath}

\usepackage{tikz}
\usetikzlibrary{quantikz}

\usepackage{nicematrix}
\usepackage{arydshln}
\NiceMatrixOptions{letter-for-dotted-lines=V}

\usepackage{pgfplots}
\pgfplotsset{compat=1.17}

\usepackage{mathbbol}

\usepackage[titletoc,title]{appendix}

\usepackage{hhline}
\usepackage{multirow}

\colorlet{ygray}{gray!20}

\newcommand{\aeq}{\begin{equation}}
\newcommand{\eeq}{\end{equation}}
\newcommand{\aeqn}{\begin{eqnarray}}
\newcommand{\eeqn}{\end{eqnarray}}
\newcommand{\aeqns}{\begin{eqnarray*}}
\newcommand{\eeqns}{\end{eqnarray*}}

\newcommand{\ybs}[1]{\boldsymbol{#1}}

\newcommand{\yi}{\mathrm{i}}

\newcommand{\oO}{\mathcal{O}}

\newcommand{\ydiff}{\mathrm{d}}

\newcommand{\ylb}{\left(}
\newcommand{\yrb}{\right)}

\newcommand{\yatop}[2]{ \genfrac{}{}{0pt}{3}{#1}{#2} }

\newcommand{\ygfirstlast}{\gate{\yatop{0}{\text{-}1}}}
\newcommand{\ygsecondprev}{\gate{\yatop{1}{\text{-}2}}}
\newcommand{\ygsecond}{\gate{1}}
\newcommand{\ygprev}{\gate{\text{-}2}}

\newcommand{\ygfl}{\gate{\yatop{0}{\text{-}1}}\vqw{-1}}
\newcommand{\ygsp}{\gate{\yatop{1}{\text{-}2}}\vqw{-1}}
\newcommand{\ygp}{\gate{\text{-}2}\vqw{-1}}
\newcommand{\ygq}{\gate{j_Q}\vqw{-1}}
\newcommand{\ytarg}[1]{\targ{}\vqw{#1}}

\newcommand{\yqsp}{\text{qsp}}
\newcommand{\yout}{\text{out}}



%% file: intr.tex
\section{Introduction}

In recent years, there has been a significant development of quantum-computing (QC) applications to simulating classical physical systems. Various methods have been proposed to solve the standard wave equation\cite{Suau21}, Poisson's equation\cite{Cao13, Wang20}, Maxwell's equations\cite{Sinha10, Scherer17}, first-order linear hyperbolic systems\cite{Gourdeau19}, the Navier--Stokes equation\cite{Gaitan20, Gaitan21}, the Boltzmann equation\cite{Todorova20}, and to simulate advection--diffusion processes\cite{Budinski21}. Recently, it has also been noted\cite{Dodin20} that particularly fitting the QC framework may be simulations of the highly dispersive radiofrequency (RF) waves in inhomogeneous classical plasmas, which are of interest due to their rich physics and importance for practical applications. This paper proposes a quantum algorithm for modeling such waves for the first time.

\subsection{Introduction to plasma waves}

Classical plasmas support a broad variety of waves, which are present there naturally and are also launched with external antennas for plasma diagnostics and control.\cite{Stix92} 
In the simplest case of cold nonmagnetized plasma, these waves consist of two types. 
One is the electrostatic Langmuir oscillations of the electron fluid relative to the ion fluid, which occur at the frequency $\omega = \omega_p$, where  $\omega_p$ is the so-called plasma frequency determined by the local background density (Sec.~\ref{ssec:basic-equations}). 
The other type of waves are electromagnetic light waves, which are similar to the light waves in vacuum but have a different dispersion relation, $\omega = \smash{(\omega_p^2 + k^2 c^2)^{1/2}}$, 
where $\ybs{k}$ is the wavevector and $c$ is the speed of light. The Langmuir waves have their electric field along $\ybs{k}$, and the light waves have their electric field in the plane perpendicular to $\ybs{k}$; i.e., there are two electromagnetic modes with the same frequency but different polarization. When a background magnetic field $\ybs{B}_0$ is added, the Langmuir and electromagnetic modes hybridize, and also additional branches appear due to the ion motion relative to $\ybs{B}_0$, for example, Alfv\'en waves at low frequencies. All these branches are classified as~O and~X waves depending on their polarization, which in the general case is neither strictly parallel nor strictly perpendicular to $\ybs{k}$. Once thermal and kinetic effects are taken into account, even more modes appear (sound waves, electron and ion Bernstein waves, etc.). Even in homogeneous plasma, the number of dispersion branches becomes infinite in this case, and additional modes caused by plasma inhomogeneity are also possible.

In laboratory plasmas such as those in magnetic fusion experiments, the typical frequencies of interest include the ion cyclotron frequencies $\Omega_i$ (tens of MHz), lower-hybrid frequency $\omega_{\rm LH} \sim |\Omega_i\Omega_e|^{1/2}$ (few GHz), and the electron cyclotron frequency $\Omega_e$ ($\sim 10^2$ GHz). 
The latter is usually comparable to $\omega_p$ and to the so-called upper-hybrid (UH) frequency $\omega_{\rm UH} = \smash{(\Omega_e^2 + \omega_p^2)^{1/2}}$, at which a collective resonance occurs for $\ybs{k} \perp \ybs{B}_0$. The corresponding waves are widely used to control fusion plasmas through RF heating~\cite{Pinsker01, Prater04} and also current drive~\cite{Fisch87}, which is mainly done with lower-hybrid~\cite{Fisch78} and electron-cyclotron waves~\cite{Fisch80} and can also help suppress plasma instabilities~\cite{Reiman83, Reiman18}. Many of these RF techniques are practiced on existing fusion devices~\cite{Ding18, Tsujimura20}, and current drive in particular is now envisioned to play a large role for achieving steady-state operation and for suppressing instabilities in future devices~\cite{Bonoli87, Cesario10, Wan19, Wallace21}. Hence, precise modeling of RF waves in plasma is of significant practical interest.

Various approximations are used for such modeling, depending on the physics of interest, but for many purposes, the waves can be considered linear and collisions can be neglected. (Quantum effects are entirely negligible for most purposes as well.) Also, the propagation, albeit not absorption, can often be described within the cold-fluid approximation. Furthermore, in the ``electron frequency range'' ($\omega \sim \Omega_e \sim \omega_p$), ions can be considered stationary; then left of interest are only nondissipative linear oscillations of the electron fluid. Although simplified, this model retains rich wave physics that can be difficult to simulate on a classical computer. The difficulty is due to the fact that the corresponding wavelengths are in the mm range, while the device size is in the meter range.
This makes multi-dimensional simulations computationally expensive\cite{Fasoli16, Svidzinski18} or even entirely unrealistic. 
Modeling becomes even more difficult when electromagnetic waves linearly transform into each other or into electrostatic oscillations, which have even smaller wavelengths. (This process is called mode conversion. \cite{book:tracy, Stix92}) For example, the latter can happen at $\omega \sim \omega_{\rm LH}$ and $\omega \sim \omega_{\rm UH}$. Various reduced schemes have been used to speed up simulations of waves in the electron frequency range
\footnote{For example, see the recent series of Refs.~\onlinecite{my:quasiop1, my:quasiop2, my:quasiop3, my:quasiop4, Yanagihara21} and the references cited therein.}, 
but they are fundamentally limited. First-principle simulations could be beneficial, and this is where QC could help.

\subsection{Quantum simulations}

In this work, we report a quantum algorithm that simulates the propagation of a linear X~wave in the electron frequency range in a plasma with an inhomogeneous static magnetic field and inhomogeneous density.
The considered problem is one-dimensional, with $\ybs{k} \perp \ybs{B}_0$, in which case the X wave is the branch whose polarization lies in the plane perpendicular to $\ybs{B}_0$. (The remaining O wave has polarization parallel to $\ybs{B}_0$.)
We also report test simulations using this algorithm on a classical emulator of a quantum computer. To describe these results, let us introduce the necessary vocabulary first.


We assume the circuit model of quantum computation, where information is stored in a set of $n_q$ quantum bits called \textit{qubits}.\cite{Nielsen10, Rieffel11} When entangled, the qubits create a configuration space described by a $2^{n_q}$-dimensional complex vector, and such an exponential scaling with $n_q$ can be beneficial in large-dimensional problems.
A quantum circuit consists of a sequence of so-called \textit{gates}. Some of the gates operate in parallel, and the longest path between the input and output points of the circuit is called \textit{circuit depth}, which is roughly proportional to the runtime\cite{CircuitDepth}. A gate that acts on $m_q$ qubits can be represented by a $2^{m_q}\times 2^{m_q}$ unitary matrix, while the whole quantum circuit is described by a $2^{n_q}\times2^{n_q}$ unitary matrix $U$. 
This matrix is applied to an initial state $\psi(0)$, generating an output state $\psi(t) = U\psi(0)$, from which classical information is then extracted via a classical measurement. 
In particular, in this paper we focus on \textit{Quantum Hamiltonian Simulations} (QHS) of systems with Hermitian Hamiltonians $\mathcal{H}$ that are independent of time $t$; then, $U = \exp(-\yi \mathcal{H} t)$.

Here, we bring the X-wave equations to the form suitable for QHS using the analytic model from Ref.~\onlinecite{Dodin20}. Then, we implement the QHS using the \textit{Quantum Signal Processing} (QSP) paradigm, which was originally developed for efficient QHS.\cite{Low17, Low19} The essence of the QSP is in encoding polynomials of given matrices into sequences of rotations. In the QHS in particular, the QSP searches for a polynomial to approximate the exponential function of $\mathcal{H}$, which is \textit{block-encoded} into an auxiliary unitary. 
This state-of-the-art quantum method is now extensively studied in the QC community\cite{Gilyen19, Haah20, Chao20, Dong21, Burg21} and holds promise as a universal numerical framework applicable to any linear Hamiltonian problem. For instance, it was recently applied to simulations of one-dimensional kinetic plasma waves in the spectral representation\cite{Engel19}. However, an application of the QSP to modeling plasma waves in inhomogeneous plasma, which is a more practical problem, is reported here for the first time.

We show that for the problem of our choice, a linear cold-X-wave simulation, the QSP implementing QHS can be constructed efficiently. Using the QuEST computing toolkit\cite{Jones19}, we also implement the corresponding circuit explicitly. Then we run this algorithm on a classical emulator of a quantum computer and show that our results agree with those of conventional classical simulations. We also discuss how the resulting QSP quantum circuit scales with the grid resolution, the precision of the QSP approximation, and the simulated time interval. 

Our paper is organized as follows. In Sec.~\ref{sec:cold-plasma}, we outline our analytical model of a linear cold X~wave in inhomogeneous plasma and its Schr\"odinger representation. In Sec.~\ref{sec:xwave}, we construct the corresponding one-dimensional model and derive the Hamiltonian that is used later in our QHS. It is also shown in Sec.~\ref{sec:xwave} how to encode the resulting plasma system into a quantum circuit and how to initialize the circuit. The QSP algorithm is explained in Sec.~\ref{sec:qsp}, and the block-encoding of the wave Hamiltonian is constructed in Sec.~\ref{sec:oracle}. The comparison of the QSP with classical simulations and the scaling of the QSP circuit are presented in Sec.~\ref{sec:results}. Finally, the advantages and challenges of applying the QSP to classical plasma problems are discussed in Sec.~\ref{sec:conclusions}. 
A reader not familiar with quantum computing is encouraged to read a brief introduction into the field presented in Appendix~\ref{app:qc-basics}.

%% file: waves.tex
\section{Cold-plasma waves}
\label{sec:cold-plasma}


\begin{figure*}
\centering
\subfloat{\includegraphics[]{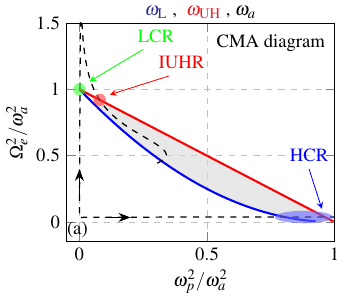}\label{fig:cma}}
\subfloat{\includegraphics[]{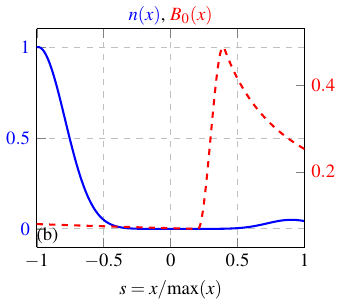}\label{fig:nB-profiles}}
\subfloat{\includegraphics[]{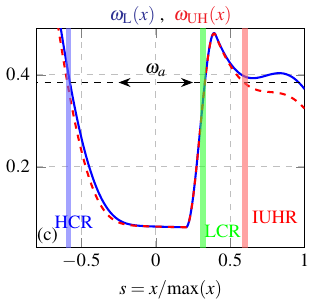}\label{fig:w-profiles}}
\caption{
(a): CMA diagram showing $\omega_{\rm L}(x)$ (blue bottom solid), $\omega_{\rm UH}(x)$ (red upper solid), and the X-wave frequency $\omega_a$ (black dashed) in the parameter space $(\Omega_e^2/\omega_a^2, \omega_p^2/\omega_a^2)$.
Here, HCR is the high-density cutoff--resonance pair; LCR is the low-density cutoff--resonance pair; IUHR is an isolated UH resonance. The gray area corresponds to the forbidden zone between the low-density cutoff and the UH resonance. The black arrows indicate the wave propagation from $s = 0$ to $s < 0$ (horizontal arrow) and from $s = 0$ to $s > 0$ (vertical arrow). (b): the corresponding background density $n(x)$ (solid left blue) and magnetic field $B_0(x)$ (red right dashed), Eqs.~\ref{sys:nB-profiles}. The peaks in $B_0(x)$ and $n(x)$ at $s > 0$ are added ad hoc to reduce wave reflection from the right boundary. They guarantee the presence of the LCR and the IUHR in~(a). 
The parameters of the background profiles are specified in Sec.~\ref{sec:results}, Eqs.~\ref{eq:n-profile}-\ref{eq:b-profile}. 
(c): $\omega_{\rm L}$ (solid blue), $\omega_{\rm UH}$ (dashed red), and $\omega_a$ (dashed black). 
}
\end{figure*}

\subsection{Basic equations}\label{ssec:basic-equations}

We assume a cold fluid electron plasma with density $n(\ybs{x})$ immersed into a background magnetic field $\ybs{B}_0(\ybs{x})$. Linear waves in such plasma can be described by the following equations:
\begin{subequations}\label{sys:plasma}
\begin{eqnarray}
    &&\partial_{t}\tilde{\ybs{v}} = - \tilde{\ybs{v}} \times \ybs{B}_0 - \tilde{\ybs{E}},\label{eq:3d-v}\\
    &&\partial_{t}\tilde{\ybs{E}} = n\tilde{\ybs{v}} + \ybs{\nabla}\times\tilde{\ybs{B}},\\
    &&\partial_{t}\tilde{\ybs{B}} = - \ybs{\nabla}\times\tilde{\ybs{E}},\label{eq:3d-b}
\end{eqnarray}
\end{subequations}
where $\tilde{\ybs{v}}$ is the electron fluid velocity, and $\tilde{\ybs{E}}$ and $\tilde{\ybs{B}}$ are the  wave electric and magnetic fields, respectively. Time is measured in units of the maximum plasma frequency $\omega_{p, 0} = \omega_p(n_0)$, where $\omega_p(n) = \smash{(4\pi n e^2/m)^{1/2}}$ is the local plasma frequency (in CGS units), $e$ is the absolute value of the electron charge, $m$ is the electron mass, and $n_0$ is the maximum value of $n(\ybs{x})$. The velocity is normalized to $c$, while the fields are normalized to $c\sqrt{4\pi n_0 m}$, and the space coordinate is normalized to $\kappa_x = c/\omega_{p, 0}$. Equations~\ref{sys:plasma} satisfy Poynting's theorem
\begin{eqnarray}\label{eq:pt}
\partial_t \int W\ydiff V + \int\ybs{N}\cdot \ydiff\ybs{S} = 0,
\end{eqnarray}
where $W = n|\tilde{\ybs{v}}|^2 + |\tilde{\ybs{E}}|^2 + |\tilde{\ybs{B}}|^2$ is the system energy density, and $\ybs{N} = \tilde{\ybs{E}}\times\tilde{\ybs{B}}$ is the Poynting vector. With appropriate boundary conditions (such as the Dirichlet boundary conditions), the integral over the surface $\ybs{S}$ disappears so the total energy $\int W\ydiff V$ is conserved.

One can also rewrite Eqs.~\ref{sys:plasma} as a vector Schr\"odinger equation\cite{Dodin20}
\begin{equation}
    \partial_t \ket{\psi} = - \yi \mathcal{H} \ket{\psi},\label{eq:schrodinger}
\end{equation}
where $\ket{\psi} = (\sqrt{n}\tilde{\ybs{v}}, \tilde{\ybs{E}}, \tilde{\ybs{B}})$, and $\mathcal{H}$ serves as a time-independent Hamiltonian, which is Hermitian if the system has suitable boundary conditions (periodic or Dirichlet). The corresponding dynamics is described by
\begin{eqnarray}
    &&\ket{\psi(t)} = e^{-\yi\mathcal{H}t}\ket{\psi(0)}.
\end{eqnarray}
Because $\exp(-\yi\mathcal{H}t)$ is unitary, $\langle \psi|\psi \rangle$ is conserved, which is an alternative representation of Eq.~\ref{eq:pt}.\cite{Dodin20}

\subsection{CMA diagram, resonances, and cutoffs}

The parameter space of a linear wave with a fixed frequency $\omega$ in cold stationary magnetized plasma is fully determined by $n$ and $B_0$, or equivalently, by $\smash{\omega_p^2/\omega^2}$ and $\smash{\Omega_e^2/\omega^2}$, where $\Omega_e = e B_0/(mc)$ in CGS units. 
Thus, it is convenient to explore the wave propagation on the plane $(\smash{\omega_p^2/\omega^2}, \smash{\Omega_e^2/\omega^2})$, The corresponding plot is called a Clemmow--Mullaly--Allis (CMA) diagram.\cite{Stix92, book:swanson} Notable in this diagram are the curves that correspond to \textit{resonances} and \textit{cutoffs}. Those are defined in the geometrical-optics limit, when the inverse inhomogeneity scale is much smaller than (loosely speaking) the local wavenumber, which satisfies the local dispersion relation
\footnote{For an introduction on the geometrical-optics approximation, see, for example, Refs.~\onlinecite{book:tracy, Stix92} or Secs.~7.1-7.3 in I.~Y. Dodin, arXiv:2201.08562.}
\begin{equation}\label{eq:dr}
\omega(\ybs{x}, \ybs{k}) = \text{const}.
\end{equation}
A resonance is a point where $k \rightarrow \infty$. (Collisionless dissipation is never negligible at $k \rightarrow \infty$, so in reality, a wave always experiences damping near a resonance.) A cutoff is a point where $k$ turns to zero, meaning that a wave experiences reflection. (Definitions of cutoffs can be subtle in multi-dimensional systems, but that is unimportant for the purposes of this paper.) The area beyond the cutoff, where Eq.~\ref{eq:dr} has no real solutions for $k$, is called a forbidden zone. If the forbidden zone is sufficiently narrow, a wave can tunnel through it much like a quantum particle tunnels through a potential barrier.

For X waves, the CMA diagram for the regime that we consider is shown in Fig.~\ref{fig:cma}. (There, $\omega_a$ is a proxy for $\omega$; the reason why a separate notation $\omega_a$ is introduced is explained in Sec.~\ref{sec:xwave}.) It exhibits the ``low-density'' cutoff (blue curve) where $\omega_{\rm L}(x) = \omega$,\cite{book:swanson}
\begin{equation}
    \omega_{\rm L} = \frac{1}{2}\left(|\Omega_e| + \sqrt{\Omega_e^2 + 4\omega_p^2} \right),\label{eq:wlr} 
\end{equation}
and the UH resonance (red curve)  where $\omega_{\rm UH}(x) = \omega$,
\begin{equation}
    \omega_{\rm UH} = \sqrt{\Omega_e^2 + \omega_p^2}\label{eq:wuh}.
\end{equation}
The gray area between the cutoff and the resonance is a forbidden zone. This zone narrows at $\smash{\omega_p^2/\omega^2} \to 0$ and at $\smash{\omega_p^2/\omega^2} \to 1$. Thus, if the wave trajectory enters and exits the forbidden zone in these regions, the entrance and the exit are located close to each other (so a significant amount of the wave energy can tunnel through this zone), forming a pair. There are two such pairs in our case: the low-density cutoff--resonance pair (LCR) and the high-density cutoff--resonance pair (HCR). They will be further discussed in Sec.~\ref{sec:xwave}, after we explain our choice of the density and magnetic-field profiles, which determine the wave trajectory on the CMA diagram.

%% file: discr.tex
\section{One-dimensional model}
\label{sec:xwave}


\begin{figure}[!b]
\centering
\subfloat{\includegraphics[]{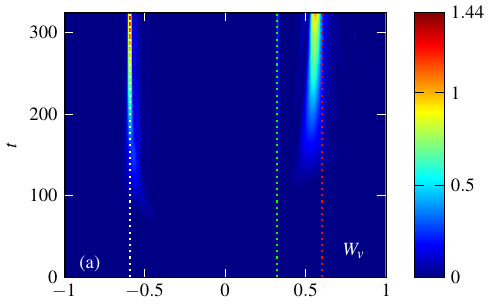}\label{fig:W-kin-xt}}\\
\subfloat{\includegraphics[]{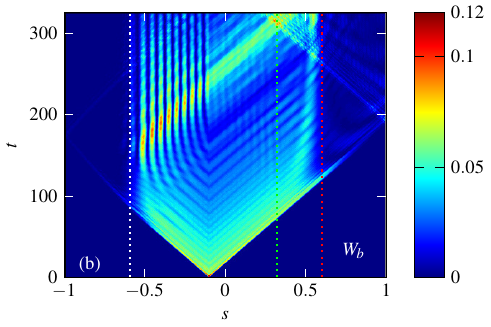}\label{fig:W-magn-xt}}
\caption{Classical simulation: dynamics of kinetic energy $W_v$ (a) and magnetic energy $W_b$ (b) in space and time. 
The input parameters are described in Sec.~\ref{sec:results}.
The white line indicates the HCR; 
the red line indicates the IUHR; 
the green line indicates the LCR (Fig.~\ref{fig:cma}).
The left-propagating wave reaches the HCR, partially reflects from it, and partially tunnels to the UH resonance.
The right-propagating wave is trapped within the IUHR, and only a small part of its energy reaches the right boundary.
}
\end{figure}

\subsection{Field configuration}
We reduce Eqs.~\ref{sys:plasma} to a one-dimensional system of size $2r_0$. The corresponding space grid $x \in [-r_0/\kappa_x,r_0/\kappa_x]$ has $N_x$ points.
For convenience, we define also a space grid $s$ via
\begin{equation}
    s = x/\max(x)\label{eq:s}
\end{equation}
to have $s\in[-1,1]$. 
Because we focus on X-wave propagation, it is enough for us to keep only 
$\tilde{v}_x, \tilde{v}_y, \tilde{E}_x, \tilde{E}_y$, and $\tilde{B}_z$, and we impose the Dirichlet boundary conditions
\begin{eqnarray}
    \tilde E_y(-r_0) = \tilde E_y(r_0) = 0,\\
    \tilde B_z(-r_0) = \tilde B_z(r_0) = 0.
\end{eqnarray}
To mimic the effect of the antenna that launches the wave, we introduce an auxiliary linear oscillator $Q$ with frequency $\omega_a$ (the index $a$ stands for ``antenna'') and initial amplitude $Q_0$. Coupled to the magnetic field $\tilde{B_z}$, the oscillator gradually transfers its energy to the X wave, while the energy of the whole system remains constant. Provided that this energy transfer is slow, the wave being launched with an approximately constant frequency, $\omega \approx \omega_a$.

The source is placed at the center ($x = 0$), where the background density (represented by $\smash{\omega_p^2}$) and magnetic field (represented by $\smash{\Omega_e^2}$) are small. Thus, the wave trajectory starts at the bottom left corner of the CMA diagram (Fig.~\ref{fig:cma}). Then the wave propagates both to the left and to the right in the $x$ space. The horizontal black arrow corresponds to the wave propagation toward the HCR. The X wave is partly reflected at the cutoff. But it also partly tunnels through the narrow forbidden zone (gray area) beyond the cutoff (blue line) and accumulates at the UH resonance (red curve), where the wavelength ends up decreasing indefinitely (strictly speaking, until dissipation ceases to be negligible). 
This is the effect of practical interest that we seek to model, and naturally, doing so requires high resolution near the resonance. 
However, modeling of this effect can be obscured by wave reflection from the right boundary. 
To suppress this reflection, auxiliary background profiles are introduced at $x > 0$ (Fig.~\ref{fig:nB-profiles}) such that they ensure X-wave trapping inside an additional resonance. Specifically, variations of the density and magnetic field give rise to a LCR and also an additional, \textit{isolated}, UH resonance (IUHR), as shown in Fig.~\ref{fig:w-profiles}. Because the LCR is located in the region of extremely small density, the forbidden zone there is practically transparent. Then the wave energy propagating from the source to the right ($x > 0$) goes through the LCR without noticeable reflection and accumulates at the IUHR, so no reflection from the right boundary ever occurs. Classical simulations of this system (Figs.~\ref{fig:W-kin-xt}-\ref{fig:W-magn-xt}) show that the wave behaves as described indeed.

\subsection{Rescaled variables}
In terms of the rescaled velocity $\xi = \tilde{v} \sqrt{n}$, our one-dimensional model can be written explicitly as
\begin{subequations}\label{sys:1d-model}
\begin{eqnarray}
    &&\yi\partial_t\xi_x(x,t)       = -\yi B_0(x)\xi_y(x,t) - \yi\sqrt{n(x)}\tilde{E}_x(x,t),\label{eq:vx}\\
    &&\yi\partial_t\xi_y(x,t)       =  \yi B_0(x)\xi_x(x,t) - \yi\sqrt{n(x)}\tilde{E}_y(x,t),\\
    &&\yi\partial_t\tilde{E}_x(x,t) =  \yi\sqrt{n(x)}\xi_x(x,t),\\
    &&\yi\partial_t\tilde{E}_y(x,t) =  \yi\sqrt{n(x)}\xi_y(x,t) -\yi\partial_x\tilde{B}_z(x,t),\\
    &&\yi\partial_t\tilde{B}_z(x\neq x_q,t) = -\yi\partial_x \tilde{E}_y(x\neq x_q,t),\label{eq:bz-pure}\\
    &&\yi\partial_t\tilde{B}_z(x_q,t) = -\yi\partial_x \tilde{E}_y(x_q,t)  - \beta Q(x_q,t),\label{eq:bz}\\
    &&\yi\partial_t Q(x_q,t)          = -\omega_a Q(x_q,t)              - \beta \tilde B_z(x_q,t),\label{eq:q}\\
    &&Q(x_{q_1},0) = Q(x_{q_2},0) = Q_0,\label{eq:q0}
\end{eqnarray}
\end{subequations}
where the source $Q$ with a constant frequency $\omega_a$ is coupled to the magnetic field $\tilde B_z$ using an ad hoc coupling coefficient $\beta$. 
The 1-D model can be rewritten in the Hamiltonian form (Eq.~\ref{eq:schrodinger}) with
\begin{equation}
    \psi = (\xi_x, \xi_y, \tilde E_x, \tilde E_y, \tilde B_z, Q)^\intercal,\label{eq:psi}
\end{equation}
where $^\intercal$ denotes transposition. 
The energy density $W_\text{tot} = \psi\psi^\dagger$ is then represented as $W_\text{tot} = W_v + W_{eb} + W_q$, where
\begin{subequations}
\begin{eqnarray}
&&W_v = |\xi_x|^2 + |\xi_y|^2,\label{eq:wv}\\
&&W_{eb} = |\tilde E_x|^2 + |\tilde E_y|^2 + |\tilde B_z|^2,\label{eq:web}\\
&&W_q = |Q|^2.\label{eq:wq}
\end{eqnarray}
\end{subequations}

\subsection{Discretization}

        
        

\begin{figure}[!b]
\centering
\includegraphics[]{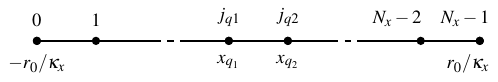}
\caption{\label{fig:x-grid} One-dimensional space grid $x\in[-r_0/\kappa_x,r_0/\kappa_x]$ has $N_x$ points numerated by index $j$, starting with $j = 0$. 
The source $Q$ is placed at points $x_{q_1}$ and $x_{q_2}$ that correspond to the indices $j_{q_1}$ and $j_{q_2}$.}
\end{figure}

The $x$-axis is represented by $N_x = 2^{n_x}$ points with a step $h$ (Fig.~\ref{fig:x-grid}).
Equations \ref{sys:1d-model} are discretized in space using the central differencing scheme.
The boundary conditions for the velocity are 
\begin{subequations}
\begin{eqnarray}
&&i\partial \xi_{y,0} = \yi B_{0,0}\xi_{x,0},\\ 
&&i\partial \xi_{y,N_x - 1} = \yi B_{0,N_x - 1}\xi_{x,N_x - 1},
\end{eqnarray}
\end{subequations}
for the wave electric field they are
\begin{subequations}
\begin{eqnarray}
&&i\partial_t\tilde E_{y,0} = 0,\\     
&&i\partial_t\tilde E_{y,1} = \yi \sqrt{n_1}\xi_{y,1} - \frac{\yi}{2h}\tilde B_{z,2},\\
&&i\partial_t\tilde E_{y,N_x-2} = \yi \sqrt{n_{N_x-2}}\xi_{y,N_x-2} + \frac{\yi}{2h}\tilde B_{z,N_x-3},\\ 
&&i\partial_t\tilde E_{y,N_x-1} = 0,
\end{eqnarray}
\end{subequations}
and for the magnetic field they are
\begin{subequations}
\begin{eqnarray}
&&i\partial_t\tilde B_{z,0} = 0,\\
&&i\partial_t\tilde B_{z,1} = - \frac{\yi}{2h}\tilde E_{y,2},\\
&&i\partial_t\tilde B_{z,N_x-2} = \frac{\yi}{2h}\tilde E_{y,N_x-3},\\
&&i\partial_t\tilde B_{z,N_x-1} = 0.
\end{eqnarray}
\end{subequations}
Since the source $Q$ interacts with the wave magnetic field only at the center of the system (Eqs.~\ref{eq:bz}-\ref{eq:q}, Fig.~\ref{fig:x-grid}), it does not enter the boundary conditions.
The corresponding Hamiltonian can be expressed as
\begin{eqnarray}
&&\mathcal{H} =  
\begin{pmatrix}
    0   & -\yi B_{0}   & -\yi\sqrt{n}   & 0   & 0   & 0 \\
    \yi B_{0}   &  0   & 0   & -\yi\sqrt{n}\epsilon  & 0   & 0\\
    \yi\sqrt{n}  & 0   & 0   & 0   & 0   & 0 \\
    0   & \yi\sqrt{n}\epsilon  & 0   & 0   & M_h   & 0 \\
    0   & 0   & 0   & M_h   & 0   & M_\beta \\
    0   & 0   & 0   & 0   & M_\beta   & M_{\omega_a}
\end{pmatrix}.\label{eq:H}
\end{eqnarray}
Here, $\epsilon = {\rm diag}(0, 1, ..., 1, 0)$, $M_h$ is the matrix representation of the operator $\partial_x$:
\begin{eqnarray}
&&M_h = 
\begin{pmatrix}
    0 & 0 & 0 & 0 & ... \\
    0 & 0 & - \frac{\yi}{2h} & 0 & ... \\
    0 & \frac{\yi}{2h} & 0 & - \frac{\yi}{2h} & ... \\
    ...& ...& ...& ...& ...\\
    ... & \frac{\yi}{2h} & 0 & - \frac{\yi}{2h} & 0 \\
    ... & 0 & \frac{\yi}{2h} & 0 & 0 \\
    ... & 0 & 0 & 0 & 0
\end{pmatrix}.
\end{eqnarray}
The matrix $M_\beta$ describes the source-wave coupling
\begin{eqnarray}
&&M_\beta = 
\begin{pmatrix}
    ...& ...& ... & ... & ...    & ...\\
    ...&  0 & 0      &      0 & 0 & ...\\
    ...&  0 & -\beta &      0 & 0 & ...\\
    ...&  0 &      0 & -\beta & 0 & ...\\
    ...&  0 &      0 &      0 & 0 & ...\\
    ...& ...& ... & ... & ...    & ...\\
\end{pmatrix},
\end{eqnarray}
while $M_{\omega_a}$ encodes the source frequency:
\begin{eqnarray}
&&M_{\omega_a} = 
\begin{pmatrix}
    ...& ...& ... & ... & ...       & ...\\
    ...&  0 &         0 &          0 & 0 & ...\\
    ...&  0 & -\omega_a &         0 & 0 & ...\\
    ...&  0 &         0 & -\omega_a & 0 & ...\\
    ...&  0 &         0 &         0 & 0 & ...\\
    ...& ...& ... & ... & ...       & ...\\
\end{pmatrix}.
\end{eqnarray}
Here, $B_0$ is a diagonal $N_x\times N_x$ matrix with values of the background magnetic field on the diagonal. 
The matrix that encodes $\sqrt{n(x)}$ has the same form.
The coefficients $\beta$ and $\omega_a$ are placed at the diagonal elements $(j_{q_1}, j_{q_1})$ and $(j_{q_2}, j_{q_2})$ of the matrices $M_\beta$ and $M_{\omega_a}$.
Finally, the values $\yi/2h$ are shifted by $+1$ and $-1$ with respect to the diagonal of the matrix $M_h$.

\subsection{Quantum encoding of plasma signals}\label{ssec:main-registers}
To encode our discretized system into a quantum circuit, we map $\psi$ on two registers: $\ket d$ and $\ket j$.
The register $\ket j$ has $n_x = \log_2 N_x$ qubits and stores the space dependence of every variable.
That is, $\ket j$ contains the binary representation of the spatial-point indices in the $x$-grid.
The register $\ket d$ encodes the variable index: 
\begin{subequations}\label{sys:label-d}
\begin{eqnarray}
&&d = 0 \longleftrightarrow \xi_x,\\
&&d = 1 \longleftrightarrow \xi_y,\\
&&d = 2 \longleftrightarrow \tilde E_x
\end{eqnarray}
\end{subequations}
etc.
Since we have six independent fields in $\psi$, the register $\ket d$ must have at least three qubits.
Then,
\begin{equation}
\psi = A_{d,j}\ket{d}\ket{j} \equiv A_{d,j}\ket{d_2d_1d_0}_d\ket{j_{n_x-1}...j_2j_1j_0}_j,\label{eq:qc-enc}
\end{equation}
where $d_k$ and $j_k$ take values of $0$ or $1$; $A_{d,j}$ stores the amplitude of the variable with the index $d$ at $x_j$.
For instance, the value of $\tilde B_z(x=x_5)$ is stored in $A_{d=4,j=5}$, which is the amplitude of the quantum state $\ket{100}_d\ket{0...0101}_j$.
We assume that the rightmost qubit is the least-significant one (which is responsible for the parity), which is the bottom qubit in the quantum circuit.
At the same time, once qubits are represented by a classical column vector, the amplitude of the least-significant qubit is stored in the first two top elements of the column vector as shown in Eq.~\ref{eq:sup-one-qubit}.

\subsection{Initialization}


\begin{figure}[t!]
\centering
\includegraphics[]{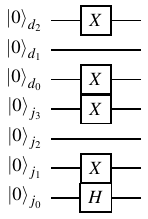}
\caption{\label{fig:source-init} An example of the quantum-state initialization in a system with $n_x = 4$, where the source $Q$, which is stored in $\ket{5}_d\equiv\ket{101}_d$, is placed at $x$ points with $j_q = 10$ and $11$.}
\end{figure}

The source $Q$ is initialized at two spatial points with indices $j_q$ and $(j_q+1)$ such that 
$j_q$ is even and close to $N_x/2$, i.e., at the center of the system. There, the plasma density is low, so the X-wave wavenumber is $k_x = \omega_a/c$.
To guarantee the state normalization $\braket{\psi}{\psi} = 1.0$, we set $Q_0 = 1/\sqrt{2}$ (Eq.~\ref{eq:q0}). 
To prepare this initial state, one sets $\ket{0}_j$ and $\ket{5}_d$.
Then, a Hadamard gate (Eq.~\ref{eq:plus-state}) is placed at the least-significant qubit of the register $\ket j$. 
Apart from that, every qubit that must have a value $1$ according to the bit representation of the integer $j_q$, is inverted by a Pauli $X$ gate (Eq.~\ref{eq:x-gate}).
An example of the initialization circuit for $n_x = 4$ is shown in Fig.~\ref{fig:source-init}.
The proper initialization is confirmed by the comparison of the source time evolution from classical and QC modeling shown in Figs.~\ref{fig:comp-Q-real-t} and~\ref{fig:comp-Q-imag-t}.
If one wants to initialize a source with various amplitudes at different spatial points, one can use one or several rotation gates $R_y$ (Eq.~\ref{eq:ry-matrix}) instead of the Hadamard gate.
The number of gates in the initialization circuit is proportional to $n_x$.

%% file: qsp.tex
\begin{figure}[b!]
\centering
\includegraphics[]{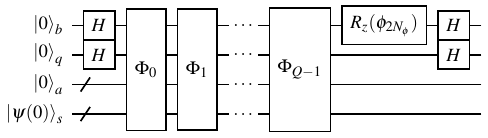}
\caption{\label{fig:qsp-gen} 
General QSP circuit (Eq.~\ref{eq:qsp}). 
The subcircuit for each $\Phi_j(\phi_{2j+1}, \phi_{2j})$ is shown in Fig.~\ref{fig:qsp-element}. 
The rotation gate $R_z$ is defined in Eq.~\ref{eq:rz}.
The ancillae $\ket{a}, \ket{q}, \ket{b}$ are initialized in the zero state, while the register $\ket{s}$ stores the initial conditions.
The Hadamard gates $H$ prepare the necessary superposition state $\ket+$ (Eq.~\ref{eq:plus-state}) for the ancillae $\ket{q}$ and $\ket{b}$.
If all ancillae are output in the zero state, then the register $\ket{s}$ contains the QSP approximation of  $\exp(-\yi \mathcal{H}t)\ket{\psi(0)}_s$ for the given $t$ and error $\epsilon_\yqsp$.
}
\end{figure}


\begin{figure*}[t!]
\centering
\includegraphics[]{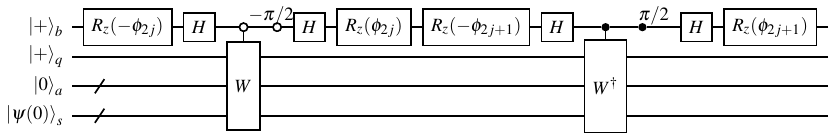}
\caption{\label{fig:qsp-element} Quantum circuit of each operator 
$\Phi_j(\phi_{2j+1}, \phi_{2j})$ that enters Eq.~\ref{eq:qsp}.
}
\end{figure*}


\begin{figure}
\centering
\includegraphics[]{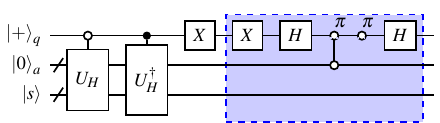}
\caption{\label{fig:w} 
Quantum circuit of the operator $W$ (Eq.~\ref{eq:w}).
The controlled unitary $U_H$ (Eq.~\ref{eq:u}) and its complex-conjugate $U_H^\dagger$ form the extended operator $U_H^\prime$ (Eq.~\ref{eq:u-prime}).
The left Pauli $X$ gate implements the operator $S$ (Eq.~\ref{eq:s-operator}).
The subcircuit within the blue box implements the reflector $U_R$ (Eq.~\ref{eq:reflector}).
}
\end{figure}


\section{Quantum Signal Processing framework}
\label{sec:qsp}

\subsection{QSP basics}
Quantum Signal Processing (QSP)~\cite{Low16, Low17, Low19} is an algorithm for constructing a polynomial $f(A)$ of a given matrix $A$.
In the case of Hamiltonian simulations, QSP is used for approximating the propagator with a given absolute error, as $f(\mathcal{H}) \approx \exp(-\yi \mathcal{H} t) + \oO(\epsilon_\yqsp)$, and
encodes the polynomial into a quantum circuit as
\begin{equation}
    U_\yqsp = 
    \begin{pmatrix}
        f(\mathcal{H}) & . \\
        .    & . \\
    \end{pmatrix}.\label{eq:qsp-f-enc}
\end{equation}
Here, the unitary matrix $U_\yqsp$ depends on $(2N_\phi + 1)$ real parameters $\phi_j$, henceforth called \textit{phases}, and is represented by a set of rotations~\cite{Low17,Haah20} (Fig.~\ref{fig:qsp-gen}):
\begin{equation}
    U_\yqsp = R_{z,b}(\phi_{2N_\phi})\ \Phi_{N_\phi-1}(\phi_{2N_\phi-1}, \phi_{2N_\phi-2})\ ...\ \Phi_0(\phi_1, \phi_0).\label{eq:qsp}
\end{equation}
Unlike in graphical representations of quantum circuits, the rightmost operator $\Phi_0$ is applied first, and the leftmost operator $R_{z,b}$ is applied last.
The latter is the Pauli rotation gate:
\begin{equation}
R_z(\phi) = 
    \begin{pmatrix}
        \exp(-i\phi/2) & 0 \\
        0 & \exp(i\phi/2)
    \end{pmatrix},\label{eq:rz}
\end{equation}
which acts on the ancilla qubit $\ket{b}$.
The operator $\Phi_j(\phi_{2j+1}, \phi_{2j})$ is described in Fig.~\ref{fig:qsp-element}.
It consists of alternating rotation gates $R_z$ (Eq.~\ref{eq:rz}) and an operator $W$ (Eq.~\ref{eq:w}), which encodes the Hamiltonian (Sec.~\ref{ssec:qubitization}).
The copies of $W$ create higher powers of $\mathcal{H}$, which are used to construct the polynomial $f(\mathcal{H})$, while the rotations $R_z$ generate the appropriate coefficients in $f(\mathcal{H})$~\cite{Low17,Low19}. 

An initial state $\psi(0)$ is stored in the input register $\ket{s}$ (in our case, the combination of registers $\ket{d}$ and $\ket{j}$), and the ancillae $\ket{b}$, $\ket{q}$ and $\ket{a}$ are initialized in the zero state.
Then, QHS is implemented via application of the unitary $U_\yqsp$ to the system state vector.
Because the propagator is encoded as the upper-left block of $U_\yqsp$, the desired state $\exp(-\yi \mathcal{H} t)\psi(0)$ is yielded as the output in the register $\ket{s}$ when all the ancillae are measured in the zero state.
In Hamiltonian simulations by the QSP method, the probability of the post-selected state is close to unity, $1-\epsilon$, with an arbitrary small $\epsilon$, which is determined by the QSP approximation error.\cite{Martyn-22}

The phases are chosen depending on the simulation time and error tolerance $\epsilon_\yqsp$, and are calculated on a classical computer using known algorithms.
The corresponding details can be found in Refs.~\onlinecite{Haah20, Chao20, Dong21}.
The codes for the calculation of the phases are presented in Refs.~\onlinecite{Haah20code, Dong21code}.
In our work, we use the code described in Ref.~\onlinecite{Haah20} and presented in Ref.~\onlinecite{Haah20code}.

\subsection{Block-encoding}
The operator $W$ (Fig.~\ref{fig:w}) contains some oracle $U_H$, which is the mapping of the Hamiltonian $\mathcal{H}$ on a unitary matrix.
In this work, circuit representation of any function or mapping is called oracle.
Generally, Hamiltonian is a non-unitary matrix, it cannot be directly described by a quantum circuit, which can contain only unitary gates.
Therefore, to operate with $\mathcal{H}$ in the QSP circuit, one has to encode $\mathcal{H}$ within a unitary matrix.
The corresponding mapping $\mathcal{H}\rightarrow U_H$ is called block-encoding.
This transformation can be done by extending Hilbert space such that the matrix $\mathcal{H}$ becomes a sub-block of the unitary $U_H$:
\begin{equation}
    U_H = 
    \begin{pmatrix}
        \frac{\mathcal{H}}{\varsigma M} & . \\
        . & .
    \end{pmatrix}\label{eq:u}
\end{equation}
where the original matrix $\mathcal{H}$ must be normalized to its norm
\begin{equation}
    M\equiv||\mathcal{H}||_{\rm max} = \max_{i} \sum_j\sqrt{|\mathcal{H}_{ij}|^2},
\end{equation}
and to its sparsity $\varsigma$ (a maximum number of nonzero elements in every row and column).
The extension of the space is achieved by using the ancilla register $\ket{a}$.
The block-encoding as shown in Eq.~\ref{eq:u} indicates that the normalized Hamiltonian can be extracted from $U_H$ when the ancillae have zero input and output states:
\begin{equation}
    \bra k_s \bra 0_a U_H \ket 0_a \ket j_s = \frac{\mathcal{H}_{jk}}{\varsigma M}.\label{eq:be-main-theory}
\end{equation}
Here, the row index $j$ is encoded as an input in the register $\ket{s}$, and the column index $k$ is an output in the same register after the block-encoding operation.
This can be understood as a decomposition of the Hamiltonian into a collection of its nonzero elements, $\mathcal{H} = \sum\mathcal{H}_{jk}\ket{k}\bra{j}$, 
which is optimal for sparse $\mathcal{H}$.
Efficiency of the QSP technique is usually estimated in a number of queries to the oracle $U_H$.
The block-encoding of our wave Hamiltonian (Eq.~\ref{eq:H}) is discussed in Section~\ref{sec:oracle}.

\subsection{Qubitization}\label{ssec:qubitization}

The purpose of the QSP is to build the desired polynomial $f(\mathcal{H})$.
Because $U_H$ is linear in $\mathcal{H}$, multiple applications of $U_H$ are required, where the block-encoding matrix acts as:
\begin{equation}
    U_H\ket0_a\ket\lambda_s = \lambda\ket0_a\ket\lambda_s + \sqrt{1 - \lambda^2}\ket{\perp}_{a,s}.
\end{equation}
Here, $\lambda, \ket\lambda$ are the eigenvalue and eigenvector of $\mathcal{H}$, respectively;
the state $\ket{\perp}_{a,s}$ is created by the undefined part of $U_H$ marked with dots in Eq.~\ref{eq:u}.
To build higher powers of $\mathcal{H}$, we need to stay within the space spanned by $(\ket0_a\ket\lambda_s, U_H\ket0_a\ket\lambda_s)$. 
The problem is that in general, this space is not invariant under the action of $U_H$.
Every next application of $U_H$ adds additional perpendicular vectors that are different from the original $\ket{\perp}_{a,s}$.
To overcome this issue, one can decompose (qubitize)~\cite{Low19} the entire Hilbert space into two-dimensional orthogonal subspaces, $\bigoplus_\lambda\mathcal{P}_\lambda$, where each subspace $\mathcal{P}_\lambda$ corresponds to a particular $\ket{\lambda}$ of $\mathcal{H}$.
After that, one replaces $U_H$ by a new operator $W$ that performs rotation in each of these disjoint subspaces.
As a result, the operator $W$ block-encodes $\mathcal{H}$ and by acting in an invariant space $\mathcal{P}_\lambda = (\ket0_a\ket\lambda_s, W\ket0_a\ket\lambda_s)$ for each $\ket{\lambda}$ can produce necessary moments of $\mathcal{H}$ to construct the polynomial $f(\mathcal{H})$.

According to Lemma 10 from Ref.~\onlinecite{Low19}, $W$ can always be constructed using the following procedure.
First of all, one extends again Hilbert space with an ancilla $\ket{q}$ initialized in the superposition state $\ket+$ (Eq.~\ref{eq:plus-state}):
\begin{equation}
\ket{0}_a \rightarrow \ket{+}_q\ket{0}_a.\label{eq:a-to-qa}
\end{equation}
Then, one applies an $X$ gate (Eq.~\ref{eq:x-gate}) to the ancilla $\ket{q}$:
\begin{equation}
S = X_q\otimes I_{a,s}\label{eq:s-operator}\\
\end{equation}
and uses two copies of controlled $U_H$:
\begin{equation}
U_H^\prime = \ket0_q\bra0_q\otimes U_H + \ket1_q\bra1_q\otimes U_H^\dagger,\label{eq:u-prime}
\end{equation}
Combining the above operators, one constructs
\begin{equation}
W = \ylb U_R\otimes I_s\yrb S U_H^\prime,\label{eq:w}
\end{equation}
where the so-called reflector $U_R$:
\begin{equation}
    U_R = 2\ket{+}_q\ket{0}_a\bra0_a\bra+_q - I_{q,a}\label{eq:reflector}
\end{equation}
keeps unchanged the zero-ancilla state and inverts the sign of the perpendicular state.
Here, $I_{y}$ is the unit operator that acts on an ancilla $\ket{y}$. 
The corresponding circuit is shown in Fig.~\ref{fig:w}.
As it is proven in Lemma 8 and Lemma 10 from Ref.~\onlinecite{Low19}, the above construction guarantees that the matrix form of the operator $W$ is represented by the following direct sum:
\begin{equation}
    W = \bigoplus_\lambda 
    \begin{pmatrix}
    \lambda            & -\sqrt{1-\lambda^2} \\
    \sqrt{1-\lambda^2} & \lambda
    \end{pmatrix},
\end{equation}
which ensures the invariance of each subspace $\mathcal{P}_\lambda$ under the action of $W$.

%% file: BE.tex
\section{Block-encoding of the wave Hamiltonian}
\label{sec:oracle}


\begin{figure}[!t]
\centering
\includegraphics[]{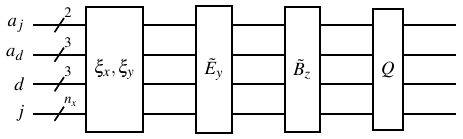}
\caption{\label{fig:of-general} General circuit of the oracle $O_F$. Its components are shown in Fig.~\ref{fig:of-components}.
According to Table~\ref{table:of}, the block $\tilde E_x$ does not modify ancillae $\ket{a_j}\ket{a_d}$.}
\end{figure}

\begin{figure}[!t]
\centering
\begin{quantikz}[row sep={0.5cm,between origins},column sep=0.1cm]
    \lstick{$a_{d,1}$}&\qw\gategroup[wires=6, steps=5,style={white, dashed, inner sep=0pt}, label style={yshift=-0.05 cm}]{$\xi_x, \xi_y$}              
            &\targ{} &\qw &\targ{}&\targ{} &\qw \\
    \lstick{$a_{d,0}$}&\gate{H}         &\octrl{-1}&\gate{H}  &\ctrl{-1} &\ctrl{-1} &\qw  \\
    \lstick{$d_2$}    &\octrl{-1}       &\octrl{-1}&\octrl{-1}&\octrl{-1}&\octrl{-1}&\qw  \\
    \lstick{$d_1$}    &\octrl{-1}       &\octrl{-1}&\octrl{-1}&\octrl{-1}&\octrl{-1}&\qw  \\
    \lstick{$d_0$}    &\qw              &\octrl{-1}&\ctrl{-1} &\ctrl{-1} &\ctrl{-1} &\qw  \\
    \lstick{$j$}      &\qw\qwbundle{n_x}&\qw       &\ygfl     &\ygfl     &\qw       &\qw
\end{quantikz}
\begin{quantikz}[row sep={0.5cm,between origins},column sep=0.1cm]
    \lstick{$a_{d,2}$}&\ytarg{2}\gategroup[wires=7, steps=3,style={white, dashed, inner sep=0pt}, label style={yshift=-0.05 cm}]{$Q$}
        &\qw    &\qw       &\qw  \\
    \lstick{$a_{d,1}$}&\qw              &\qw       &\qw       &\qw  \\
    \lstick{$a_{d,0}$}&\targ{}          &\targ{}   &\gate{H}  &\qw  \\
    \lstick{$d_2$}    &\ctrl{-1}        &\ctrl{-1} &\ctrl{-1} &\qw  \\
    \lstick{$d_1$}    &\octrl{-1}       &\octrl{-1}&\octrl{-1}&\qw  \\
    \lstick{$d_0$}    &\ctrl{-1}        &\ctrl{-1} &\ctrl{-1} &\qw  \\
    \lstick{$j$}      &\qw\qwbundle{n_x}&\ygq      &\ygq      &\qw
\end{quantikz}
\\
\begin{quantikz}[row sep={0.5cm,between origins},column sep=0.1cm]
    \lstick{$a_{j,1}$}&\qw\gategroup[wires=9, steps=9,style={white, dashed, inner sep=0pt}, label style={yshift=-0.05 cm}]{$\tilde E_y$}              
        &\qw       &\qw       &\qw       &\qw       &\qw       &\qw       &\qw       &\targ{}   &\qw  \\
    \lstick{$a_{j,0}$}&\qw              &\qw       &\qw       &\qw       &\qw       &\gate{H}  &\gate{H}  &\targ{}   &\octrl{-1}&\qw  \\
    \lstick{$a_{d,2}$}&\qw              &\targ{}   &\targ{}   &\qw       &\qw       &\ctrl{-1} &\ctrl{-1} &\ctrl{-1} &\ctrl{-1} &\qw  \\
    \lstick{$a_{d,1}$}&\qw              &\qw       &\qw       &\qw       &\ytarg{1} &\qw       &\qw       &\qw       &\qw       &\qw  \\
    \lstick{$a_{d,0}$}&\gate{H}         &\octrl{-2}&\octrl{-2}&\gate{H}  &\targ{}   &\octrl{-2}&\octrl{-2}&\octrl{-2}&\octrl{-2}&\qw  \\
    \lstick{$d_2$}    &\octrl{-1}       &\octrl{-1}&\octrl{-1}&\octrl{-1}&\octrl{-1}&\octrl{-1}&\octrl{-1}&\octrl{-1}&\octrl{-1}&\qw  \\
    \lstick{$d_1$}    &\ctrl{-1}        &\ctrl{-1} &\ctrl{-1} &\ctrl{-1} &\ctrl{-1} &\ctrl{-1} &\ctrl{-1} &\ctrl{-1} &\ctrl{-1} &\qw  \\
    \lstick{$d_0$}    &\ctrl{-1}        &\ctrl{-1} &\ctrl{-1} &\ctrl{-1} &\ctrl{-1} &\ctrl{-1} &\ctrl{-1} &\ctrl{-1} &\ctrl{-1} &\qw  \\
    \lstick{$j$}      &\qw\qwbundle{n_x}&\qw       &\ygfl     &\ygfl     &\ygfl     &\qw       &\ygsp     &\ygp      &\qw       &\qw
\end{quantikz}
\\
\begin{quantikz}[row sep={0.5cm,between origins},column sep=0.1cm]
    \lstick{$a_{j,1}$}&\qw\gategroup[wires=9, steps=9,style={white, dashed, inner sep=0pt}, label style={yshift=-0.05 cm}]{$\tilde B_z$}
        &\qw       &\qw       &\qw       &\qw       &\qw       &\qw       &\qw       &\targ{}   &\qw  \\
    \lstick{$a_{j,0}$}&\qw              &\qw       &\qw       &\qw       &\qw       &\gate{H}  &\gate{H}  &\targ{}   &\octrl{-1}&\qw  \\
    \lstick{$a_{d,2}$}&\qw              &\ytarg{1} &\qw       &\qw       &\targ{}   &\qw       &\qw       &\qw       &\qw       &\qw  \\
    \lstick{$a_{d,1}$}&\ytarg{1}        &\ytarg{1} &\targ{}   &\gate{H}  &\octrl{-1}&\ctrl{-2} &\ctrl{-2} &\ctrl{-2} &\ctrl{-2} &\qw  \\
    \lstick{$a_{d,0}$}&\targ{}          &\targ{}   &\qw       &\qw       &\qw       &\ctrl{-1} &\ctrl{-1} &\ctrl{-1} &\ctrl{-1} &\qw  \\
    \lstick{$d_2$}    &\ctrl{-1}        &\ctrl{-1} &\ctrl{-2} &\ctrl{-2} &\ctrl{-2} &\ctrl{-1} &\ctrl{-1} &\ctrl{-1} &\ctrl{-1} &\qw  \\
    \lstick{$d_1$}    &\octrl{-1}       &\octrl{-1}&\octrl{-1}&\octrl{-1}&\octrl{-1}&\octrl{-1}&\octrl{-1}&\octrl{-1}&\octrl{-1}&\qw  \\
    \lstick{$d_0$}    &\octrl{-1}       &\octrl{-1}&\octrl{-1}&\octrl{-1}&\octrl{-1}&\octrl{-1}&\octrl{-1}&\octrl{-1}&\octrl{-1}&\qw  \\
    \lstick{$j$}      &\qw\qwbundle{n_x}&\ygfl     &\ygq      &\ygq      &\ygq      &\qw       &\ygsp     &\ygp      &\qw       &\qw
\end{quantikz}
\\
\begin{quantikz}[align equals at=1, row sep={0.5cm,between origins}]
    \lstick{$j$} &\ygfirstlast\qwbundle{} &\qw
\end{quantikz}
    =
\begin{quantikz}[row sep={0.5cm,between origins},column sep=0.3cm]
    \lstick{$j_{1,n_x-1}$}
    &\ocontrol{}\qwbundle{}\gategroup[wires=2, steps=1,style={gray, dashed, inner xsep=-20pt, inner ysep=6pt,fill=gray!20}, background, label style={yshift=0.05 cm}]{$0$}
    &\control{}            \gategroup[wires=2, steps=1,style={gray, dashed, inner xsep=-20pt, inner ysep=6pt,fill=gray!20}, background, label style={yshift=0.05 cm}]{$-1$}
    &\qw  \\
    \lstick{$j_0$}       &\octrl{-1}            &\ctrl{-1} &\qw
\end{quantikz}
\\
\begin{quantikz}[align equals at=1, row sep={0.5cm,between origins},column sep=0.3cm]
    \lstick{$j$} &\ygsecondprev\qwbundle{} &\qw
    \end{quantikz}
    =
    \begin{quantikz}[align equals at=1, row sep={0.5cm,between origins},column sep=0.3cm]
    \lstick{$j$} &\ygsecond\qwbundle{}&\ygprev &\qw
    \end{quantikz}
    =
    \begin{quantikz}[row sep={0.5cm,between origins},column sep=0.3cm]
    \lstick{$j_{1,n_x-1}$}&\ocontrol{}\qwbundle{}\gategroup[wires=2, steps=1,style={gray, dashed, inner xsep=-20pt, inner ysep=6pt,fill=gray!20}, background, label style={yshift=0.05 cm}]{$1$}
        &\control{}                              \gategroup[wires=2, steps=1,style={gray, dashed, inner xsep=-20pt, inner ysep=6pt,fill=gray!20}, background, label style={yshift=0.05 cm}]{$-2$}
        &\qw  \\
    \lstick{$j_0$}       &\ctrl{-1}             &\octrl{-1} &\qw
\end{quantikz}
\caption{\label{fig:of-components} 
Sub-blocks of the $O_F$ quantum circuit shown in Fig.~\ref{fig:of-general}. 
The circuits are constructed following the Table~\ref{table:of}. 
The control block $j_Q$ is illustrated in Fig.~\ref{fig:block-jq}.
The control nodes in the register $\ket{d}$ encode the variable label, while the register $\ket{j}$ encodes the spatial coordinate.
}
\end{figure}
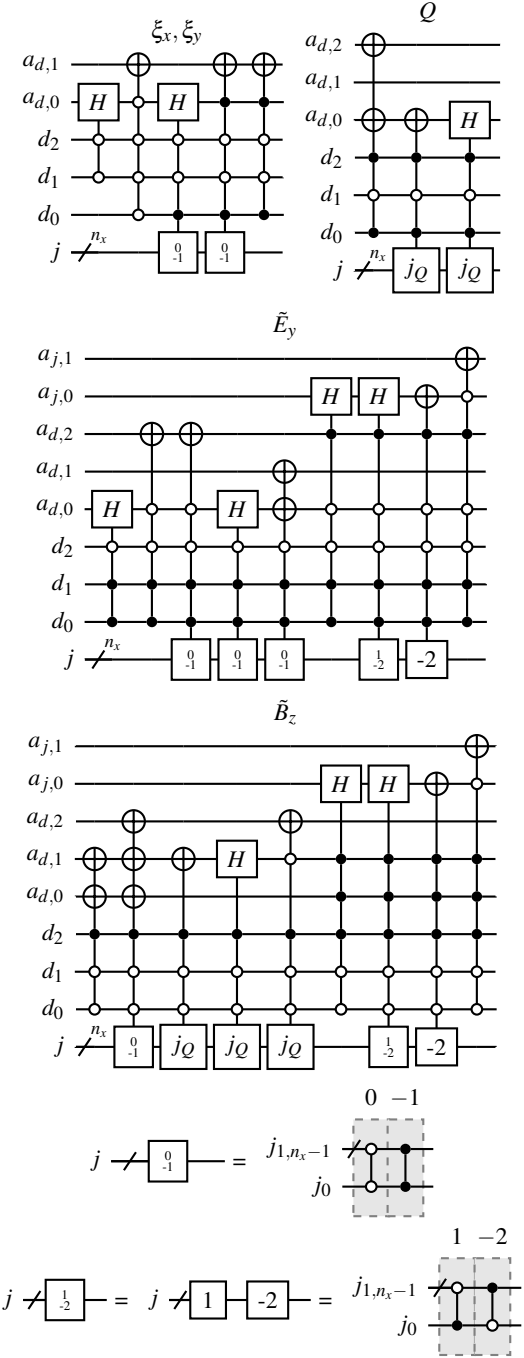

\begin{figure}[t]
\begin{quantikz}[align equals at=1, row sep={0.5cm,between origins}, column sep=0.3cm]
\lstick{$j$} &\qw\qwbundle{n_x} &\gate{j_Q} &\qw
\end{quantikz}
    =
\begin{quantikz}[align equals at=2.5, row sep={0.5cm,between origins},column sep={0.5cm,between origins}]
\lstick{$j_3$} &\control{}&\control{}&\qw \\
\lstick{$j_2$} &\octrl{-1}&\octrl{-1}&\qw \\
\lstick{$j_1$} &\octrl{-1}&\octrl{-1}&\qw \\
\lstick{$j_0$} &\octrl{-1}&\ctrl{-1} &\qw 
\end{quantikz}
\caption{\label{fig:block-jq} Circuit of the control block $j_Q$ in a system with $n_x = 4$, where the source $Q$ is placed at spatial points with indices $j_{q,1} = 8$ and $j_{q,2} = 9$.
}
\end{figure}
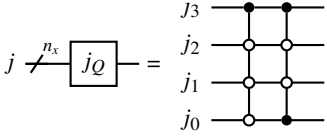

\begin{figure*}[t!]
\centering
\includegraphics[]{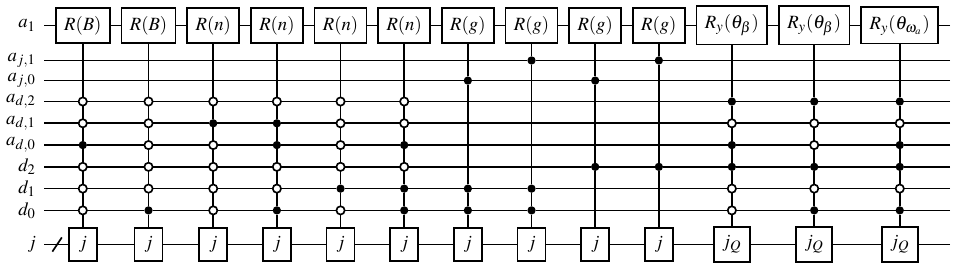}
\caption{\label{fig:oh} Quantum circuit of the oracle $O_{\sqrt{H},a_1}$. 
Every box $j$ indicates the conditional dependence on the register $\ket{j}$. 
Here, the conditional-rotation gate $R(B)$ is used to obtain the profile $\sqrt{\beta_H B_{0,j}}$; $R(n)$ encodes the profile $\sqrt{\beta_H n_j^{1/2}}$; $R(g)$ encodes a constant $\sqrt{\beta_H/(2h)}$ with different superpositions for bulk and boundary $x$-points. 
In our model, the conditional-rotation gates are represented by multi-qubit matrices.
The circuit is built according to Table~\ref{table:oh}.}
\end{figure*}

\begin{figure*}[t!]
\centering
\includegraphics[]{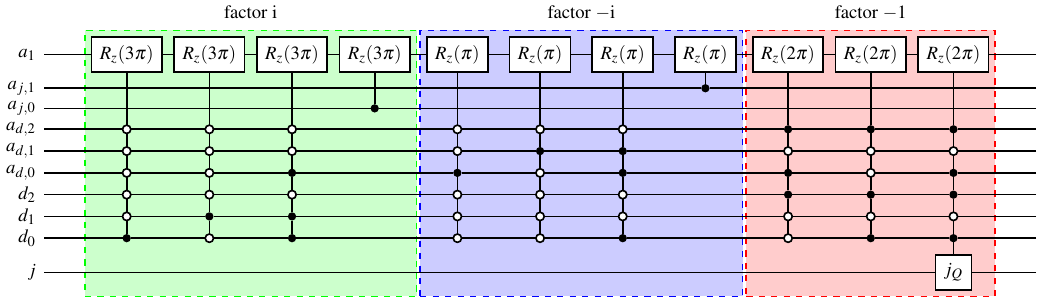}
\caption{\label{fig:os} Quantum circuit of the oracle $O_{S,a_1}$.
The circuit is constructed according to Table~\ref{table:os}.
Each colored box multiplies the amplitude stored in $\ket{a_1}$ by the factor specified in the figure ($\yi$, $-\yi$, $-1$).
The rotation gate $R_z$ acts as shown in Eq.~\ref{eq:rz}.}
\end{figure*}


\begin{figure}[!t]
\centering
\includegraphics[]{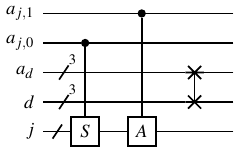}
\caption{\label{fig:oracle-OM} Quantum circuit of the oracle $O_{M}$. The subtractor $S$ and the adder $A$ are shown in Fig.~\ref{fig:add-sub}.
The oracle converts the relative positions stored in the register $\ket{a_j}$ into the corresponding absolute positions and stores them in the register $\ket{j}$.
Then, the circuit swaps (Fig.~\ref{fig:swap-phase-schemes}) the registers $\ket d$ and $\ket{a_d}$.}
\end{figure}

\begin{figure}[!t]
\centering
\subfloat{
    \begin{quantikz}[align equals at=1, row sep={0.5cm,between origins}]
    \lstick{$j$} &\gate{A}\qwbundle{4} &\qw
    \end{quantikz}
    =
    \begin{quantikz}[row sep={0.5cm,between origins}]
    \lstick{$j_3$} &\targ{}   &\qw       &\qw       &\qw      &\qw  \\
    \lstick{$j_2$} &\ctrl{-1} &\targ{}   &\qw       &\qw      &\qw  \\
    \lstick{$j_1$} &\ctrl{-1} &\ctrl{-1} &\targ{}   &\qw      &\qw  \\
    \lstick{$j_0$} &\ctrl{-1} &\ctrl{-1} &\ctrl{-1} &\gate{X} &\qw
    \end{quantikz}
}\\
\subfloat{
    \begin{quantikz}[align equals at=1, row sep={0.5cm,between origins}]
    \lstick{$j$} &\gate{S}\qwbundle{4} &\qw
    \end{quantikz}
    =
    \begin{quantikz}[row sep={0.5cm,between origins}]
    \lstick{$j_3$} &\qw      &\qw       &\qw       &\targ{}   &\qw  \\
    \lstick{$j_2$} &\qw      &\qw       &\targ{}   &\ctrl{-1} &\qw  \\
    \lstick{$j_1$} &\qw      &\targ{}   &\ctrl{-1} &\ctrl{-1} &\qw  \\
    \lstick{$j_0$} &\gate{X} &\ctrl{-1} &\ctrl{-1} &\ctrl{-1} &\qw
    \end{quantikz}
}
\caption{\label{fig:add-sub} Quantum circuits of the adder and of the subtractor by $1$ for the case with four qubits.}
\end{figure}
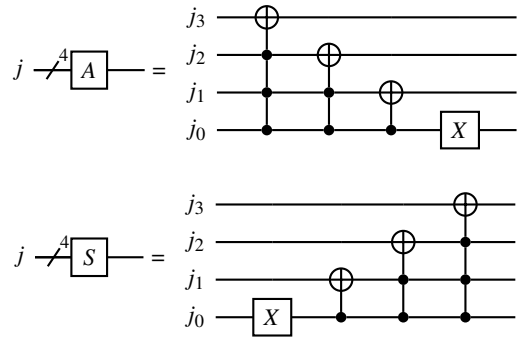

\subsection{General algorithm}
To encode the wave Hamiltonian (Eq.~\ref{eq:H}), we apply the state-preparation technique~\cite{Low19}.
The standard procedure is described in Appendix~\ref{app:state-preparation-tech}.
In this algorithm, one finds positions of all nonzero elements within $\mathcal{H}$ and then encodes values of these elements into the amplitudes of quantum states.
First of all, for each row of $\mathcal{H}$ one stores the column indices of nonzero matrix elements in an ancilla register.
Being an integer, each index is encoded as a quantum state represented by a bit-string of qubits.
After that, knowing the column and row indices, one finds the values of the corresponding nonzero elements.
The values are encoded in the amplitudes of the quantum states.
Since any element of the rescaled Hamiltonian $\mathcal{H}/(\varsigma M)$ (Eq.~\ref{eq:u}) is less than (or equal to) unity by the absolute value, it can be represented as $\cos(\theta/2)$ of a certain angle $\theta$.
The cosine can be generated by the rotation gate $R_y(\theta)$:
\begin{equation}
R_y(\theta) = 
    \begin{pmatrix}
    \cos(\theta/2) & -\sin(\theta/2)\\
    \sin(\theta/2) & \cos(\theta/2)
    \end{pmatrix}\label{eq:ry-matrix}
\end{equation}
acting on the zero state:
\begin{eqnarray}
R_y(\theta)\ket{0} = \cos(\theta/2)\ket{0} + \sin(\theta/2)\ket{1}.\label{eq:ry-v}
\end{eqnarray}

\subsection{Normalization}
To normalize the wave Hamiltonian (Eq.~\ref{eq:H}), we adopt:
\begin{eqnarray}
\mathcal{H}_\yqsp = \beta_H \mathcal{H},\quad t_\yqsp = t/\beta_H,\quad \beta_H = \frac{1}{d_H^2 M},
\end{eqnarray}
or more explicitly
\begin{eqnarray}
\beta_H = \frac{1}{d_H^2\sqrt{\omega_{\text{UH,max}}^2 + \frac{1}{2h^2} + \beta^2 + \omega_a^2}}.\label{eq:beta-h}
\end{eqnarray}
Here, $\mathcal{H}_\yqsp$ is the rescaled Hamiltonian, $t_\yqsp$ is the time interval to be simulated by the QSP circuit, $\omega_{\text{UH,max}} = \sqrt{B^2_{0,\text{max}} + n_\text{max}}$, and $n_\text{max}$ and $B_{0,\text{max}}$ are the maximum values of the background density and magnetic field, respectively.
In our case, $d^2_H = 4$, which is related to the sparsity $\varsigma$ in Eq.~\ref{eq:be-main-theory} as it is explained in Appendix~\ref{app:tables}.

Note that larger $t_\yqsp$ requires calculation of a higher number of QSP phases, which can be challenging.
Instead of doing that, we represent the QSP circuit with $n_t$ sequential copies of a shorter circuit constructed for the time interval $t_\yqsp/n_t$.
In this case, performing simulations on a digital emulator, one can also analyse intermediate quantum states. 

\subsection{Ancilla registers}\label{ssec:ancilla-registers}
To implement the state-preparation method (Appendix~\ref{app:state-preparation-tech}), we introduce several ancilla registers (Fig.~\ref{fig:of-general}) to store positions of matrix nonzero elements.
The register $\ket{a_d}$ is responsible for the location of sub-blocks such as $M_h$ or $M_\beta$ in Eq.~\ref{eq:H}.
Given a variable index $\ket{d}$ (Eq.~\ref{eq:qc-enc}), the register $\ket{a_d}$ stores the absolute column indices of all sub-blocks that contain nonzero elements.
The maximum number of sub-blocks in a row (including zero sub-blocks) coincides with the number of variables in our plasma system. Thus, the size of $\ket{a_d}$ is the same as that of the register $\ket{d}$.

Then, we describe the location of nonzero elements within each sub-block.
The ancilla register $\ket{a_j}$ is introduced for this purpose.
To decrease the total number of ancillae, this register stores not an absolute column index but the relative position of a nonzero element with respect to the sub-block diagonal:
\begin{eqnarray}
&&\ket{00}_{a_j} \rightarrow i_c = i_r,\\ 
&&\ket{01}_{a_j} \rightarrow i_c = i_r-1,\\ 
&&\ket{10}_{a_j} \rightarrow i_c = i_r+1, 
\end{eqnarray}
where $i_r$ is the row index, which is the index of a given point on $x$-grid of a variable. 
The row index is stored in the register $\ket{j}$ (Eq.~\ref{eq:qc-enc}).
The integer $i_c$ is the column index of the nonzero value within the sub-block $\ket{a_d}$.
For instance, $\ket{00}_{a_j}$ means that a nonzero element lies on the local diagonal of the sub-block.
One should note that the size of the register $\ket{a_j}$ does not depend on $N_x$, which is not the case in the standard technique (Appendix~\ref{app:state-preparation-tech}).
This feature allows us to increase $N_x$ without increasing the number of ancillae.
The size of $\ket{a_j}$ increases, however, with the discretization order
and it may also depend on the boundary conditions.

Finally, two more single-qubit ancilla registers $\ket{a_1}$ and $\ket{a_2}$ are introduced.
The rotation gates $R_y(\theta)$ act on these qubits to store the nonzero elements of the Hamiltonian (Eq.~\ref{eq:ry-v}).
A thinner space grid has a smaller difference between neighboring Hamiltonian elements, and as a result, requires that the rotation angle $\theta$ be calculated with a higher precision.

\subsection{Block-encoding operator}
Now, when we have introduced all necessary ancillae and know how they store the structure of $\mathcal{H}$, we can represent the block-encoding as a product of several operators:
\begin{equation}
    U_H = O_F^\dagger O_{\sqrt{H},a_2}^\dagger O_M O_{S,a_1}O_{\sqrt{H},a_1}O_F,\label{eq:u-be}
\end{equation}
where every operator is responsible for a particular part of the block-encoding procedure.
The oracle $O_F$ (Table~\ref{table:of}, Figs.~\ref{fig:of-general}-\ref{fig:of-components} and Fig.~\ref{fig:block-jq}) defines the location of nonzero elements for a given variable index stored in $\ket{d}$ and for a local row index saved in $\ket{j}$.
This oracle writes the column indices of sub-blocks with nonzero elements into the register $\ket{a_d}$ and local relative positions of these nonzero elements into the register $\ket{a_j}$.
To construct the quantum circuit of the oracle $O_F$, we consider it as a sequence of sub-circuits for different variables (Fig.~\ref{fig:of-general}).
Every sub-circuit corresponds to one block from Table~\ref{table:of}.
If one assumes that multi-controlled gates are physically realizable, then the depth of the $O_F$ circuit does not change with the system size $N_x$.
However, whether it will be possible to efficiently connect non-neighboring qubits in future quantum computers is yet to be seen.
Transpiling the multi-controlled gates into one-qubit and two-qubit gates using a non-optimized algorithm makes the circuit depth grow exponentially with the number of qubits.
However, as shown in Ref.~\onlinecite{Barenco95}, by using a sufficient number of ancillae one can decompose an arbitrary $n_q$-controlled unitary matrix into $\oO(n_q^2)$ elementary gates.
For instance, a $n_q$-controlled Pauli $X$ gate can be transpiled into $4(n_q - 2)$ $2$-controlled $X$ gates (so-called Toffoli gates) by adding $n_q - 2$ ancilla qubits.

The oracle $O_{\sqrt{H}}$ (Table~\ref{table:oh}, Fig.~\ref{fig:oh}) reads the row and column indices and provides the square root of the absolute value of the corresponding nonzero element.
It acts on the ancilla $\ket{a_1}$ or ancilla $\ket{a_2}$.
The oracle $O_{S,a_1}$ (Table~\ref{table:os}, Fig.~\ref{fig:os}) describes whether an element is imaginary or real, as well as whether it is positive or negative.

In the standard state-preparation algorithm, ancilla registers store the absolute column indices (Eqs.~\ref{eq:T-psi}-\ref{eq:T-chi}).
As a result, the index exchange between the ancilla and input registers can be implemented by a simple swap operator.
In our case, the register $\ket{a_j}$ works with relative indices.
Because of that, we need to implement the mapping between the absolute and relative indices during the index exchange.
This is provided by the oracle $O_M$ (Fig.~\ref{fig:oracle-OM}) by using a subtractor and an adder by $1$, which are depicted in Fig.~\ref{fig:add-sub}. 
After the application of the oracle $O_M$, the register $\ket{j}$ encodes the absolute column indices of nonzero elements within a sub-block, and the register $\ket{d}$ contains sub-block column indices. 
The depth of the circuit $O_M$ is proportional to $\log_2(N_x)$ because of the adder and subtractor.

To sum up, taking the row index as an input in $\ket{d}\ket{j}$ and the ancillae initialized in the zero state, the resulting oracle $U_H$ outputs column indices encoded in $\ket{d}\ket{j}$ as states with amplitudes equal to the corresponding Hamiltonian elements when all ancillae registers are returned in the zero state (Eq.~\ref{eq:be-main-theory}).

%% file: results.tex
\section{Simulation results}
\label{sec:results}


\begin{figure*}[!t]
\centering
\subfloat{\includegraphics[]{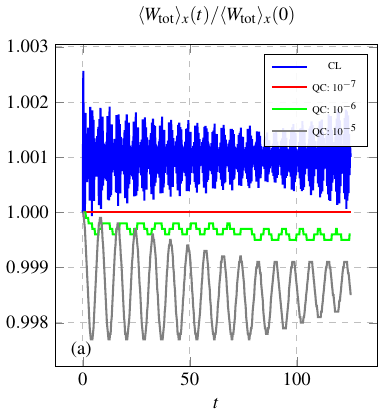}\label{fig:W-full-scan}}
\hspace{0.5cm}
\subfloat{\includegraphics[]{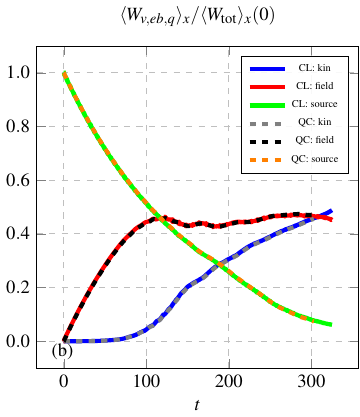}\label{fig:Wcomp-comp}}
\caption{
(a): space--integrated total system energy for different values of the QSP error $\epsilon_\yqsp$. 
(b): comparison of the corresponding energy components (Eqs.~\ref{eq:wv}-\ref{eq:wq}) in classical and quantum (with $\epsilon_\yqsp=10^{-6}$) simulations.
Here, CL stands for classical simulations and QC stands for emulated quantum simulations.}
\end{figure*}


\begin{figure*}[!t]
\centering
\subfloat{\includegraphics[]{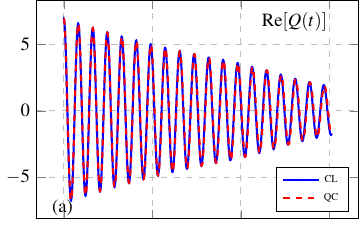}\label{fig:comp-Q-real-t}}
\hspace{0.5cm}
\subfloat{\includegraphics[]{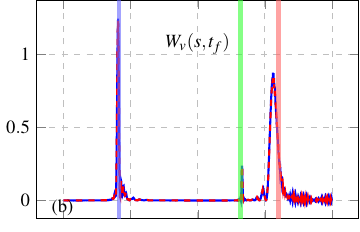}\label{fig:Wkin-comp-x}}\\
\subfloat{\includegraphics[]{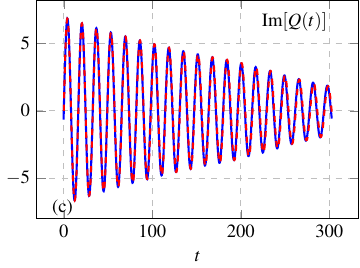}\label{fig:comp-Q-imag-t}}
\hspace{0.5cm}
\subfloat{\includegraphics[]{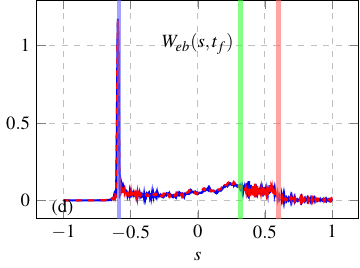}\label{fig:Wfield-comp-x}}
\caption{
Comparison of classical (solid blue curves) and emulated quantum (red dashed curves) simulations.
Real (a) and imaginary (c) components of the source $Q$.
(b): kinetic energy $W_\nu$ (Eq.~\ref{eq:wv}) at time $t_f$ as a function of $s$.
(d): field energy $W_{eb}$ (Eq.~\ref{eq:web}) at time $t_f$ as a function of $s$. 
The vertical lines show the HCR (blue), the LCR (green), the IUHR (red).}
\end{figure*}

\begin{figure*}[!t]
\centering
\subfloat{\includegraphics[]{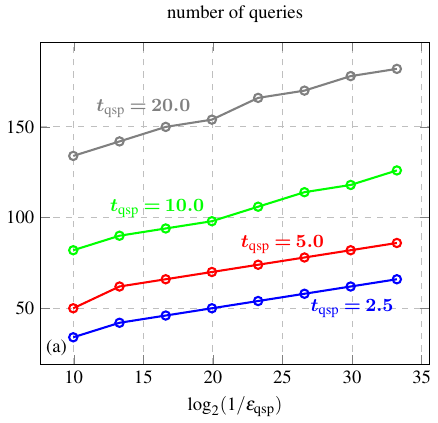}\label{fig:scan-na-eps}}
\hspace{0.5cm}
\subfloat{\includegraphics[]{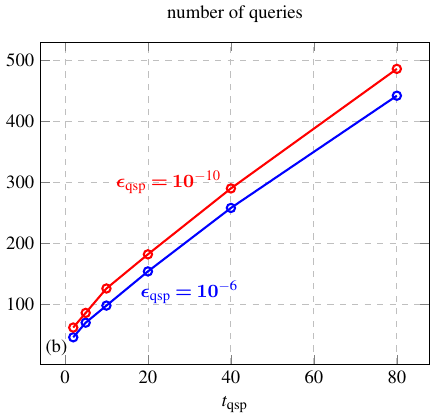}\label{fig:scan-na-t}}
\caption{Dependence of the query complexity on the QSP error $\epsilon_\yqsp \in [10^{-3}, 10^{-4},...,10^{-10}]$ (a) and on the simulation time $t_\yqsp$ (b).}
\end{figure*}

\subsection{Comparison of classical and quantum simulations}
We implement QHS using the circuit described above on a classical emulator of a quantum computer using QuEST toolkit~\cite{Jones19}.
The code and the corresponding input data for the simulations can be found in Ref.~\onlinecite{QSP-code}.
Unlike in an actual quantum simulation, this gives us access to the whole output space at all moments of time. 
The results presented below are directly extracted from $\psi$ (Eq.~\ref{eq:qc-enc}) without performing, or emulating, quantum measurements. 
Then, we compare our results with those of classical simulations, which have been obtained by directly solving Eqs.~\ref{sys:1d-model} using the central finite difference scheme for both space and time. 
As a reminder, the normalized background profiles are shown in Fig.~\ref{fig:nB-profiles}.
The size of the system is $r_0 = 20~\text{cm}$ with $N_x = 1024$ spatial points, which corresponds to $n_x = 10$. The simulation time is $t_f = 300.5 \omega_p^{-1}$, which is split into $n_t = 1200$ time steps with duration $\tau = 0.2504 \omega_p^{-1}$. The corresponding Courant number is $0.76$.

The background density and magnetic field profiles are calculated by the following equations (here $r = x\kappa_x$):
\begin{subequations}\label{sys:nB-profiles}
\begin{eqnarray}
&&n(r) = n_0 e^{-\frac{(r-r_n)^2}{2\Delta^2_n}} + n_{0,{\rm aux}} e^{-\frac{(r-r_{n,{\rm aux}})^2}{2\Delta^2_{n,{\rm aux}}}},\label{eq:n-profile}\\
&&B_0(r) = 
    \begin{cases}
      \frac{B_c R_0}{R_0 + r}, & r < r_{\rm aux},\\
      \frac{B_{c,{\rm aux}} R_{0,{\rm aux}}}{R_{0,{\rm aux}} + (r - r_{\rm aux})}, & r \geq r_{\rm aux}.\\
    \end{cases}\label{eq:b-profile}
\end{eqnarray}
\end{subequations}
The maximum background density is $n_0 = 2\times 10^{13}~\text{cm}^{-3}$ at $r_n = -0.99 r_0$, while the peak of the auxiliary density profile is $n_{0,{\rm aux}} = 10^{12}~\text{cm}^{-3}$ at $r_{n,{\rm aux}} = 0.90 r_0$. 
The widths of the density peaks are $\Delta_n = 0.2 r_0$ and $\Delta_{n,{\rm aux}} = 0.18 r_0$, correspondingly. 
The background magnetic field at the center ($r = 0$) is $B_c = 1~\text{kG}$, and the auxiliary magnetic peak at $s = 0.4$ ($r_{\rm aux} = 0.4r_0$) has $B_{c,{\rm aux}} = 7~\text{kG}$. 
The shape of the field is defined by two parameters: the major radius $R_0 = 167~\text{cm}$ and the auxiliary radius $R_{0,{\rm aux}} = 10~\text{cm}$. The profiles of the background magnetic field are combined by the cubic interpolation in a narrow domain near $r = r_{\rm aux}$. 
The source-field coupling coefficient is $\beta = 0.1$, and $Q$ oscillates with the frequency $\omega_a = 0.38 \omega_{p,0}$. The corresponding wavenumber is $k_x = 64.56r_0^{-1}$. With these parameters, the normalization of the Hamiltonian (Eq.~\ref{eq:beta-h}) becomes $\beta_H = 0.102$. Therefore, the QSP time step is $\tau_\yqsp = \tau/\beta_H = 2.455$ with the resulting time interval $t_\yqsp = n_t \tau_\yqsp = 2946$ to simulate. The QSP error is $\epsilon_\yqsp = 10^{-6}$. As seen from Fig.~\ref{fig:W-full-scan}, such $\epsilon_\yqsp$ corresponds to $\sim 10^{-4}$ error in the energy conservation. For this QSP error, the number of QSP angles is equal to $25$ for the time interval $\tau_\yqsp$.

According to Ref.~\onlinecite{Low19}, the query complexity (the number of copies of the oracle $U_H$) of the QSP circuit is $\oO(t_\yqsp + \log_2(1/\epsilon_\yqsp))$. The asymptotic dependence is confirmed by our direct computation (Figs.~\ref{fig:scan-na-eps} and \ref{fig:scan-na-t}). There, the total number of queries is calculated as twice the number of the QSP phases, because each phase corresponds to two calls to the block-encoding oracle $U_H$ (Figs.~\ref{fig:qsp-element} and~\ref{fig:w}). The number of the phases is found using the code from Ref.~\onlinecite{Haah20code}. In our particular case, where the whole QSP circuit is split into $n_t$ subcircuits, the query complexity scales as $\oO(n_t \tau_\yqsp + n_t\log_2(1/\epsilon_\yqsp))$.

We compare the time evolution of the energy components (Eqs.~\ref{eq:wv}-\ref{eq:wq}) integrated in space.
Figure~\ref{fig:Wcomp-comp} shows that both the classical and quantum simulations produce the same time histories of various energy components (Eqs.~\ref{eq:wv}-\ref{eq:wq}).
When the wave reaches the HCR (at $t\approx 100$, according to Fig.~\ref{fig:W-magn-xt}), the field energy converts partly into the kinetic plasma energy.
Figures~\ref{fig:Wkin-comp-x}-\ref{fig:Wfield-comp-x} show that the kinetic (field) energy has a similar spatial distribution in both the simulations, and the X wave accumulates in the UH resonance of the HCR.
The wave also passes the LCR practically without interaction and deposits its energy at the IUHR, as anticipated.

\subsection{Oracle scaling}
Assuming that multi-qubit controlled gates are realizable, the depths of the oracles $O_F$ (Fig.~\ref{fig:of-general}) and $O_S$ (Fig.~\ref{fig:os}) are independent of $n_x = \log_2 N_x$. 
However, these depths can change if the discretization order is increased or if the source $Q$ is distributed over multiple grid points.
The depth of the oracle $O_M$ increases linearly with $n_x$ due to the adder and subtractor (Fig.~\ref{fig:add-sub}).

The oracle $O_{\sqrt{H}}$ (Fig.~\ref{fig:oh}) depends on $n_x$ due to the conditional-rotation gates, as it is discussed in Appendix~\ref{app:tables}.
Similar gates are considered in Ref.~\onlinecite{Burg21} (see supplemental materials there), where they are called multiplexed unitaries.
These gates can be implemented via arithmetic functions that usually scale as \begin{math}\oO(\text{poly}(n_x))\end{math}.
For instance, as shown in Ref.~\onlinecite{Suau21}, the depth of a general adder with one of the summands predefined scales as $\oO(n_x)$, and the subtractor depth scales as $\oO(n_x) + 2n_x$.
In general, there is a trade between the number of ancillae used to store intermediate data and the depth of the resulting circuit.
In our case, a conditional-rotation gate must implement a smooth function that depends on the space variable $x$ encoded inside the register $\ket{j}$.
As explained in Ref.~\onlinecite{Haner18}, a polynomial of order $D$ can be evaluated by using the Horner scheme. 
A given polynomial $y_D$ with coefficients $a_i$ can be obtained by $D$ subsequent iterations:
\begin{eqnarray*}
&&y_1 = a_D x + a_{D-1},\nonumber\\
&&y_2 = y_1 x + a_{D-2},\nonumber\\
&&...\nonumber\\
&&y_D = y_{D-1}x + a_0.
\end{eqnarray*}
The total number of gates necessary to implement the whole polynomial scales as $\oO(Dn_x^2)$.

\subsection{Quantum measurements}\label{ssec:meas}

\begin{figure}[t!]
\centering
\includegraphics[]{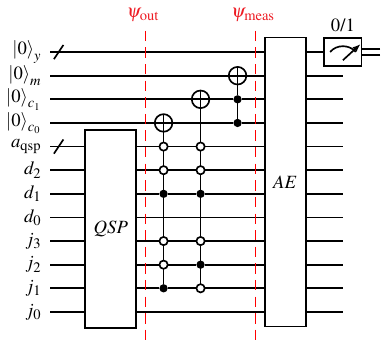}
\caption{\label{fig:meas} 
Schematic circuit of the measurement of the electric energy density $W_e$ summed over the spatial points with indices $j=[2,...5]$, for $n_x = 4$.
The register $\ket{a_\yqsp}$ represents all ancillae necessary for the QSP technique.
The electric energy components are addressed using the control nodes on $\ket{d} = \ket{2}$ (for $\tilde{E}_x$) and $\ket{d} = \ket{3}$ (for $\tilde{E}_y$) (Eqs.~\ref{eq:psi} and~\ref{sys:label-d}).
The resulting state is $\psi_\text{meas} = \sum_{k=2}^5 (\tilde{E}_{x}(x_k)\ket{2}_d + \tilde{E}_{y}(x_k)\ket{3}_d)\ket{k}_j\ket{1}_m + ...\ket{0}_m$ (the other registers are omitted), where the desired information is stored in the probability amplitude of the state $\ket{1}_m$.
Information about the state of the register $m$ is then written to the register $y$ by the AE (Appendix~\ref{app:meas}).
}
\end{figure}

In an actual quantum simulation, the output vector cannot be accessed directly.
One can measure only the expectation value of a given operator on the output state. 
How to do this for potentially practical RF simulations is a problem separate from QSP that we discuss here, so it is left to future work. 
However, here is how at least one of the quantities of interest can be measured, namely, the wave energy within a given spatial volume.

The field components are encoded into the quantum state $\psi$ as shown in Eq.~\ref{eq:qc-enc}.
To compute the electric energy, one needs to sum the squares of the electric components in a desired spatial interval.
As an example, a case with $n_x = 4$ is considered in Fig.~\ref{fig:meas}.
The QSP algorithm outputs $\psi_\yout$ that encodes plasma variables if the QSP ancillae $\ket{a_\yqsp}$ are all in the zero state.
In this example, we sum up $W_e = \tilde E_x^2 + \tilde E_y^2$ over the spatial points with indices $j=[2,...5]$.
The first controlled Pauli $X$ gate entangles the superposition of the amplitudes of both $\tilde{E}_x(j\in[2,3])$ and $\tilde{E}_y(j\in[2,3])$ with the state $\ket{1}_{c_0}$.
The second $X$ gate entangles $\tilde{E}_x(j\in[4,5])$ and $\tilde{E}_y(j\in[4,5])$ with the state $\ket{1}_{c_1}$.
The last controlled $X$ gate finds the conjunction (logical AND) of the above states, and as a result, stores $\sum_{j=2}^5 W_e(x_j)$ as the probability of the state $\ket{1}_m$.

If the qubit $m$ has the state $\ket{1}$ with amplitude $a$, then it takes at least $\oO(|a|^{-2})$ repetitions of the whole QSP operator before the direct measurement returns $\ket{m}=\ket{1}$. 
However, there is a quantum \textit{Amplitude--Amplification} (AA) algorithm~\cite{Grover97, Grover98, Brassard02} that requires only $\oO(|a|^{-1})$ iterations of the operator to measure the desired state with probability at least $\max(1-|a|^2,|a|^2)$.
This method is the basis of so-called \textit{Amplitude--Estimation} (AE) techniques\cite{Brassard02,Suzuki20,Uno21,Grinko21} that allow to find the state probability with a predefined absolute error $\delta$.
The AA is based on the Grover-like~\cite{Grover97, Grover98} rotation $R_{\rm AA}$ in the quantum space spanned by a state of interest $\ket{\rm G}$ ($\sim\ket{1}_m$ in our case) and a ``garbage'' state $\ket{\rm B}$ ($\sim\ket{0}_m$) in such a way that the amplitude of the rotated state $\ket{\rm G}$ becomes a sinusoidal function of the number of applications $n_{AA}$ of $R_{\rm AA}$: $\sin((2n_{AA}+1)\theta)$, where $\sin^2\theta = |a|^2$ and $\theta\in[0,\pi/2]$.
In our case, every rotation includes the whole QSP operator and its inverse.

As shown in Ref.~\onlinecite{Brassard02}, the operator $R_{\rm AA}$ has eigenvalues $\exp(\pm\yi2\theta)$.
Therefore, one can calculate the probability $|a|^2$ from the estimation of the angle $\theta$ by constructing a superposition of states rotated with several $n_{AA}$, and by applying a subsequent \textit{Quantum Fourier Transform} (QFT).
That is the essence of the conventional AE algorithm described in Ref.~\onlinecite{Brassard02}.
This method estimates the probability $|a|^2$ with an absolute error $\delta$ by applying $M=\oO(1/\delta)$ queries (in our case, calls to the QSP) and by using $\log_2(M)$ ancilla qubits, while classically one would need $\oO(1/\delta^2)$ queries due to the central limit theorem.
In our case, the error $\delta$ corresponds to the absolute error of the measured space-integrated energy.
More details are given in Appendix~\ref{app:meas}.

There are also state-of-the-art AE techniques~\cite{Suzuki20, Grinko21, Uno21} with a similar asymptotic scaling but smaller number of ancillae and controlled gates.
For instance, the algorithm proposed in Ref.~\onlinecite{Suzuki20} also uses a set of AA operators with a various number of rotations $R_{\rm AA}$.
However, instead of the QFT, it performs statistical post-processing of measurement results by implementing the maximum likehood estimation of $|a|^2$.

The numerical implementation of the quantum measurements for classical RF systems is left to future work.

%% file: conclusions.tex
\section{Discussion and conclusions}
\label{sec:conclusions}

We have proposed how to apply the Quantum Signal Processing (QSP) technique to simulating cold-plasma waves and explicitly developed a quantum algorithm for modeling one-dimensional X-wave propagation in cold electron plasma.
We have demonstrated how to construct an oracle to encode the wave Hamiltonian into a quantum circuit. The number of the ancillae in this oracle does not depend on the spatial resolution, so one can use a resolution higher than in the case with the standard state-preparation method. Since the oracle complexity scales as $\oO(\text{poly}(\log_2{N_x}))$, QSP simulations can provide a near-exponential speedup in comparison to classical simulations, which scale as $\oO(N_x)$. This approach can be particularly helpful in simulations with high spatial resolution, which is advantageous, for example, for modeling the wave dynamics near resonances.

Our quantum simulations have been performed on a digital emulator of the quantum circuit and have shown agreement with the corresponding classical modeling. For emulation, we used the QuEST numerical toolkit\cite{Jones19} that operates with a whole $2^{n_q}$ state vector in a circuit with $n_q$ qubits. For our one-dimensional QSP simulations, the emulator have shown efficient parallelization on GPUs. However, one might need a more advanced emulator for higher-dimensional simulations. One of the promising approaches in this regard is the model proposed in Ref.~\onlinecite{Jaques21}. It uses the fact that a quantum state is mostly sparse, i.e., many elements of the state vector are zero. If only its nonzero elements were stored (e.g. in a form of a hash table), one could significantly reduce memory usage and the simulation runtime. Moreover, one could model oracles with conditional rotations implemented via actual arithmetic functions with a big number of ancillae, since the ancillae act only locally and are zeroed otherwise thus being removed from the memory. Yet, this model needs to be extended with a GPU parallelization and the corresponding implementation of dynamic hash tables\cite{Li21}.

Based on our results, we can also assess the overall utility of the QSP technique in application to linear problems. Being a universal algorithm with a clear hierarchical structure, the QSP can be easily coded as a set of subsequent subroutines where only the block-encoding module needs to be modified for different plasma problems. The QSP provides an optimal dependence of the query complexity (the number of calls to the oracle) on the simulation time and on the error tolerance (as was pointed out in Ref.~\onlinecite{Low19}). Also, the QSP requires only two ancillae in addition to the qubits used by the oracle. This reduces the circuit width.
That said, the QSP circuits and our oracle in particular have many multi-controlled operators, where one gate is controlled by several nodes. 
Such a configuration is not directly realizable on existing quantum computers. 
The decomposition of these operators strongly depends on what gates are available on a chosen quantum computer, and on how the circuit is mapped on a specific quantum processor. Both of these aspects will be hardware-dependent, so the limitations of the QSP will become clearer when practical error-corrected quantum computers with sufficiently many qubits become available.
In general, an arbitrary single-target $n_q$-controlled gate can be decomposed into $\oO(n_q^2)$ elementary gates by adding $\oO(n_q)$ ancillae.\cite{Barenco95}
Another possibility is the hardware implementation of these gates.
For example, a possible realization of $n$-controlled iSWAP gate is proposed in Ref.~\onlinecite{Rasmussen20}.
Also, application of the so-called diamond gates, which are native in superconducting circuits, to reach higher connectivity in quantum circuits is discussed in Refs.~\onlinecite{Loft20, Bahnsen22}.

Taking into account the block-encoding scaling with $N_x$, the dependence of the QSP circuit on the length of the simulated time interval $t$, and the number of the QSP queries needed for measurements, the final scaling of our circuit depth is
\begin{equation}
    \oO\ylb\frac{\text{poly}(\log_2 N_x)}{\delta}\ylb t_\yqsp + \log_2(1/\delta)\yrb\yrb,\label{eq:final-scaling}
\end{equation}
where we take the absolute error $\delta$, which appears from measurements, equal to the QSP approximation error $\epsilon_\yqsp$. 
The numerical coefficient in the scaling is determined by the specific implementation of the conditional rotations and of the measurement algorithm. Studying these subjects is left to future work.

Also note that the plasma model assumed in our paper is limited. One obvious limitation is that the waves are considered linear; but this is also true for most RF codes. A bigger limitation is that thermal effects and dissipation are neglected, including collisional and collisionless damping and also the possible wave-energy leaking through the boundaries of the simulation domain. (Remember that reflective boundary conditions are assumed in our algorithm.) Reinstating these effects makes the Hamiltonian in Eq.~\ref{eq:schrodinger} non-Hermitian\cite{Dodin20}, and then Hamiltonian simulations cannot be done using the QSP. A possible solution to this is to consider stationary waves ($\partial_t = - \yi\omega$) and solve the corresponding boundary-value problem, as commonly done in classical RF modeling. Instead of Hamiltonian simulations, solving a boundary-value problem requires only inverting a non-Hermitian matrix.\cite{Dodin20} A possible way to do it is by using the known algorithm based on the Quantum Singular Value Transformation (QSVT)\cite{Gilyen19, Martyn21, Dong21}. 
Because such simulations would be very different from those considered in this paper, we leave them to future work as well.

%% file: appendix.tex
\appendix

\section{Basic notation and terminology of quantum computing}
\label{app:qc-basics}

The elementary memory cell of a quantum computer is a qubit, whose state can be characterized by a two-dimensional vector.
In the computational basis, the basis vectors are
\begin{eqnarray}
&\ket{0} = 
\begin{pmatrix}
    1 \\
    0
\end{pmatrix},\quad  
&\ket{1} = 
\begin{pmatrix}
    0 \\
    1
\end{pmatrix}.\label{eq:qubit-classical-vector}
\end{eqnarray}
A qubit can also be in a superposition state
\begin{equation}
    \alpha\ket0 + \beta\ket1 = 
    \begin{pmatrix}
    \alpha \\
    \beta
\end{pmatrix},\label{eq:sup-one-qubit}
\end{equation}
where $\alpha$ and $\beta$ are the complex amplitudes of the states $\ket{0}$ and $\ket{1}$, respectively, such that $|\alpha|^2 + |\beta|^2=1$. 
A quantum computer typically operates with $n > 1$ qubits, so its quantum state is characterized by a tensor product of the qubit states
\begin{equation}
    \ket{k_{n-1}}\otimes\ket{k_{n-2}}\otimes...\ket{k_1}\otimes\ket{k_0},
\end{equation}
where $k_i = 0, 1$ in the computational basis.
For convenience, one can organize qubits in quantum registers.
A combination of $n_r$ qubits in a register $\ket{r}$ can be written either as a bit--string $\ket{k_{n_r-1}k_{n_r-2}...k_{1}k_{0}}_r$, or as an integer $\ket{k}_r$, where 
\begin{equation}
    k = \sum_{i=0}^{n_r-1} k_i 2^i.
\end{equation}
Several qubits can also form superposition states:
\begin{equation}
    \ket{\psi}=\sum_k \alpha_k \ket{k},
\end{equation} 
where $|\alpha_k|^2$ is the probability amplitude of the state $\ket{k}$, and $\sum_k|\alpha_k|^2 = 1$.

\begin{figure}
\centering
\includegraphics[]{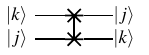}\\
\vspace{0.5cm}
\includegraphics[]{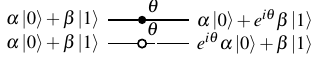}
\caption{\label{fig:swap-phase-schemes} 
Top: SWAP gate of two qubits $\ket k$ and $\ket j$.
Bottom: phase gates with an angle $\theta$. 
If the dot is hollow, then the phase gate acts on the zero state.
}
\end{figure}

\begin{figure}
\centering
\includegraphics[]{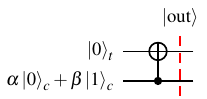}
\caption{\label{fig:control-basics} CNOT gate, or controlled $X$ gate.
The $1$-controlled gate takes action only if the control qubit is in the state $\ket{1}$: $\ket{\text{out}} = \alpha\ket{0}_t\ket{0}_c + \beta\ket{1}_t\ket{1}_c$.
A $0$-controlled (in this case, the dot is hollow) gate takes action only if the control node is in the state $\ket{0}$.
Here, the black dot is called 1-control node. If the dot is hollow, it is called 0-control node.}
\end{figure}

To modify a quantum state of a circuit, one applies gates, which are unitary operators acting on $n_g \ge 1$ qubits and can be represented by matrices of size $2^{n_g} \times 2^{n_g}$.
The gates that are used in this work are mainly the $X$ Pauli, Hadamard $H$, Phase $P$, and SWAP operators.
The $X$ gate inverses the qubit value:
\begin{eqnarray}
&&X = 
\begin{pmatrix}
    0 & 1 \\
    1 & 0
\end{pmatrix},\label{eq:paulix}\\
&&X(\alpha\ket{0} + \beta\ket{1}) = \beta\ket{0} + \alpha\ket{1}\label{eq:x-gate}.
\end{eqnarray}
The Hadamard gate creates a superposition of states:
\begin{eqnarray}
&&H = \frac{1}{\sqrt{2}}
\begin{pmatrix}
    1 & 1 \\
    1 & -1
\end{pmatrix},\label{eq:hadamard}\\
&&H\ket{0} = \ket+ = \frac{1}{\sqrt{2}}(\ket 0 + \ket 1),\label{eq:plus-state}\\ 
&&H\ket{1} = \ket- = \frac{1}{\sqrt{2}}(\ket 0 - \ket 1).
\end{eqnarray}
The Phase gate modifies the phase of the state $\ket{1}$, and keeps the zero state $\ket{0}$ unchanged (Fig.~\ref{fig:swap-phase-schemes}):
\begin{eqnarray}
&&P(\theta) = 
\begin{pmatrix}
    1 & 0 \\
    0 & e^{i\theta}
\end{pmatrix},\\
&&P(\theta)(\alpha\ket{0} + \beta\ket{1}) = \alpha\ket{0} + e^{\yi\theta}\beta\ket{1}.
\end{eqnarray}
The SWAP gate exchanges the qubit values (Fig.~\ref{fig:swap-phase-schemes}):
\begin{eqnarray}
&&\text{SWAP} =
\begin{pmatrix}
    1 & 0 & 0 & 0 \\
    0 & 0 & 1 & 0 \\
    0 & 1 & 0 & 0 \\
    0 & 0 & 0 & 1
\end{pmatrix},\\
&&\text{SWAP}\ket{kj} = \ket{jk}.
\end{eqnarray}
A gate can be controlled by one or several other qubits (control nodes) as shown in Fig.~\ref{fig:control-basics}.
For instance, a $1$-controlled $H$ gate can be written as 
\begin{equation}
    CH = \ket0\bra0_c\otimes I_t + \ket1\bra1_c\otimes H_t,
\end{equation}
where the Hadamard gate acts on the target qubit $t$ only if the control qubit $c$ is in the state $\ket{1}$.


\begin{figure}[t!]
\centering
\includegraphics[]{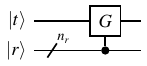}
\caption{\label{fig:control-register} 
Circuit containing a gate $G$ controlled by the register $\ket{r}$.
The register has $n_r$ qubits. 
The gate $G$ modifies the target qubit $\ket{t}$ only if all $n_r$ qubits of the register $\ket{r}$ are in the  state $\ket{1}$.}
\end{figure}

Also note that a gate can be controlled by a whole register. 
In this case, the gate is executed only if each qubit in the register is in the $\ket0$ or $\ket1$ state, as illustrated in Fig.~\ref{fig:control-register}.


\begin{table*}[t!]
\caption{\label{table:of} Action of the oracle $O_F$. An input variable is encoded into the register $\ket d$, and its space dependence is taken from the register $\ket{j}$.
The operator $O_F$ returns the registers $\ket{a_j}\ket{a_d}$ in the indicated output states, while the states of the registers $\ket{d}\ket{j}$ remain unchanged.}
\begin{ruledtabular}
\begin{tabular}{lll}
Variable     
& Input\footnote{$\ket{0;N_x-1}_j$ denotes $\ket 0_j$ or $\ket{N_x-1}_j$ states; $\ket{[1,N_x-2]}_j$ denotes all states from $\ket 1_j$ up to $\ket{N_x-2}_j$;
$\ket{j_Q}_j$ corresponds to the spatial position of the source $Q$.}                                  
& Output                                                                                    \\ \hline
$\xi_x$      & $\ket{000}_d\ket{j}_j$                 & $\ket{00}_{a_j}\ket{001}_{a_d} + \ket{00}_{a_j}\ket{010}_{a_d}$                               \\ \hline
\multirow{2}{*}{$\xi_y$}      & $\ket{001}_d\ket{0;N_x-1}_j$           & $\ket{00}_{a_j}\ket{000}_{a_d}$                                                             \\
             & $\ket{001}_d\ket{[1,N_x-2]}_j$         & $\ket{00}_{a_j}\ket{000}_{a_d} + \ket{00}_{a_j}\ket{011}_{a_d}$                               \\ \hline
$\tilde E_x$ & $\ket{010}_d\ket{j}_j$                 & $\ket{00}_{a_j}\ket{000}_{a_d}$                                                             \\ \hline
\multirow{4}{*}{$\tilde E_y$} & $\ket{011}_d\ket{0;N_x-1}_j$           & $\ket{00}_{a_j}\ket{011}_{a_d}$                                                             \\
             & $\ket{011}_d\ket{1}_j$                 & $\ket{00}_{a_j}\ket{001}_{a_d} + \ket{10}_{a_j}\ket{100}_{a_d}$                               \\
             & $\ket{011}_d\ket{[2,N_x-3]}_j$         & $\ket{00}_{a_j}\ket{001}_{a_d} + \ket{01}_{a_j}\ket{100}_{a_d} + \ket{10}_{a_j}\ket{100}_{a_d}$ \\
             & $\ket{011}_d\ket{N_x-2}_j$             & $\ket{00}_{a_j}\ket{001}_{a_d} + \ket{01}_{a_j}\ket{100}_{a_d}$                               \\ \hline
\multirow{5}{*}{$\tilde B_z$} & $\ket{100}_d\ket{0;N_x-1}_j$           & $\ket{00}_{a_j}\ket{100}_{a_d}$                                                             \\ 
             & $\ket{100}_d\ket{1}_j$                 & $\ket{10}_{a_j}\ket{011}_{a_d}$                                                             \\
             & $\ket{100}_d\ket{[2,N_x-3]\neq j_Q}_j$ & $\ket{01}_{a_j}\ket{011}_{a_d} + \ket{10}_{a_j}\ket{011}_{a_d}$                               \\
             & $\ket{100}_d\ket{j_Q}_j$               & $\ket{01}_{a_j}\ket{011}_{a_d} + \ket{10}_{a_j}\ket{011}_{a_d} + \ket{00}_{a_j}\ket{101}_{a_d}$ \\
             & $\ket{100}_d\ket{N_x-2}_j$             & $\ket{01}_{a_j}\ket{011}_{a_d}$                                                             \\ \hline
\multirow{2}{*}{$Q$}          & $\ket{101}_d\ket{j\neq j_Q}_j$         & $\ket{00}_{a_j}\ket{101}_{a_d}$                                                             \\ 
             & $\ket{101}_d\ket{j_Q}_j$               & $\ket{00}_{a_j}\ket{100}_{a_d} + \ket{00}_{a_j}\ket{101}_{a_d}$                              
\end{tabular}
\end{ruledtabular}
\end{table*}

\begin{table}
\caption{\label{table:oh} Action of the oracle $O_{\sqrt{H}}$. 
The operator is controlled by the indicated input registers.
For various bit arrays encoded in these registers, it returns the indicated output amplitudes on the ancilla $a_1$ (or $a_2$) as explained in Eq.~\ref{eq:ry-v}.}
\begin{ruledtabular}
\begin{tabular}{ll}
Input                       & Output amplitude                            \\ \hline
$\ket{001}_{a_d}\ket{000}_d$ & \multirow{2}{*}{$\sqrt{\beta_H B_0}$}       \\
$\ket{000}_{a_d}\ket{001}_d$ &                                             \\ \hline
$\ket{010}_{a_d}\ket{000}_d$ & \multirow{4}{*}{$\sqrt{\beta_H n_0^{1/2}}$} \\
$\ket{011}_{a_d}\ket{001}_d$ &                                             \\
$\ket{000}_{a_d}\ket{010}_d$ &                                             \\
$\ket{001}_{a_d}\ket{011}_d$ &                                             \\ \hline
$\ket{01}_{a_j}$             & \multirow{2}{*}{$\sqrt{\beta_H/(2h)}$}      \\
$\ket{10}_{a_j}$             &                                             \\ \hline
$\ket{101}_{a_d}\ket{100}_d$ & \multirow{2}{*}{$\sqrt{\beta_H\beta}$}      \\
$\ket{100}_{a_d}\ket{101}_d$ &                                             \\ \hline
$\ket{101}_{a_d}\ket{101}_d$ & $\sqrt{\beta_H\omega_a}$                    \\ 
\end{tabular}
\end{ruledtabular}
\end{table}

\begin{table}
\caption{\label{table:os} Action of the oracle $O_{S,a_1}$. 
For a given indicated state, the operator outputs the corresponding coefficient that is multiplied by the value returned by the oracle $O_{\sqrt{H},a_1}$.}
\begin{ruledtabular}
\begin{tabular}{ll}
Input                       & Output multiplier       \\ \hline
$\ket{000}_{a_d}\ket{001}_d$ & \multirow{4}{*}{$\yi$}  \\
$\ket{000}_{a_d}\ket{010}_d$ &                         \\
$\ket{001}_{a_d}\ket{011}_d$ &                         \\
$\ket{01}_{a_j}$             &                         \\ \hline
$\ket{001}_{a_d}\ket{000}_d$ & \multirow{4}{*}{$-\yi$} \\
$\ket{010}_{a_d}\ket{000}_d$ &                         \\
$\ket{011}_{a_d}\ket{001}_d$ &                         \\
$\ket{10}_{a_j}$             &                         \\ \hline
$\ket{101}_{a_d}\ket{100}_d$ & \multirow{3}{*}{$-1$}   \\
$\ket{100}_{a_d}\ket{101}_d$ &                         \\
$\ket{101}_{a_d}\ket{101}_d$ &                        
\end{tabular}
\end{ruledtabular}
\end{table}

\section{Block-encoding by state-preparation method}
\label{app:state-preparation-tech}

The block-encoding $U_H$ of a Hermitian operator $\mathcal{H}$ can be constructed by applying the state-preparation algorithm\cite{Low19} as a product of two unitary matrices:
\begin{equation}
    U_H = T^\dagger_2 T_1,\label{eq:u-t1t2}
\end{equation}
where
\begin{eqnarray}
&&T_1 = \sum_j \ket{\psi_j}\bra0_a\bra j_s,\\
&&T_2 = \sum_k \ket{\chi_k}\bra0_a\bra k_s.
\end{eqnarray}
Each $T_i$ involves the sum of states $\ket\psi$ and $\ket\chi$ defined as 
\begin{eqnarray}
&&\ket{\psi_j} = \sum_{p\in F_j} \frac{\ket p_{a_3}}{\sqrt{\varsigma}}\ket{\mathcal{H}_{jp}}_{a_1}\ket0_{a_2}\ket j_s,\label{eq:T-psi}\\
&&\ket{\chi_k} = \sum_{p\in F_k} \frac{\ket k_{a_3}}{\sqrt{\varsigma}}\ket{\mathcal{H}_{kp}}_{a_2}\ket0_{a_1}\ket p_s\label{eq:T-chi},
\end{eqnarray}
where
\begin{eqnarray}
&&\ket{\mathcal{H}_{jk}}_a = \sqrt{\frac{|\mathcal{H}_{jk}|}{M}}\ket0_a + \sqrt{1 - \frac{|\mathcal{H}_{jk}|}{M}}\ket1_a.\label{eq:h-ket}
\end{eqnarray}
Using the oracle $O_F$ (Eq.~\ref{eq:u-be}), one finds a set of column indices $F_j$ of all nonzero elements on the row $j$. 
The operator $T_1$ reads the row index $j$ and saves the corresponding column indices to the ancilla register $\ket{a_3}$.
After that, it rotates the ancilla qubit $\ket{a_1}$, so the necessary matrix element (its square root) becomes the amplitude of the zero state $\ket{0}_{a_1}$ (Eq.~\ref{eq:h-ket}).
This is implemented by the oracle $O_{\sqrt{H}}$ in Eq.~\ref{eq:u-be}.
The operator $T_2$ rotates the ancilla qubit $\ket{a_2}$ in a similar way by taking a row index from the ancilla register $\ket{a_3}$ and column indices from the register $\ket{s}$. 
Since the encoding of the row and column indices in the registers $a_3$ and $s$ in $\ket{\chi_k}$ is swapped in comparison with the state $\ket{\psi_j}$, we introduce the oracle $O_M$, which performs the corresponding index swapping in Eq.~\ref{eq:u-be}.

\section{Tabular description of the oracle circuits}\label{app:tables}

In Tables~\ref{table:of}-\ref{table:os}, one can find the descriptions of the various oracles whose circuits are shown in Sec.~\ref{sec:oracle}.
The action of each oracle is defined by a set of output states, which should be returned by the oracle for the indicated input states.
The notations of the main and ancilla registers ($d$, $a_d$, etc.) coincide with those introduced in Secs.~\ref{ssec:main-registers} and~\ref{ssec:ancilla-registers}. 
Together, the tables describe the Hamiltonian~\ref{eq:H}.

One can consider Table~\ref{table:of} as a set of instructions.
Each instruction says which column indices encoded in the ancilla registers $a_j$ and $a_d$ should be returned by the oracle $O_F$, when the oracle is initialized with a row index encoded in the main registers $d$ and $j$. 
In other words, this table indicates where nonzero matrix elements sit inside the Hamiltonian. According to Eq.~\ref{eq:H}, several rows of the Hamiltonian have only zero elements. In this case, to simplify the construction of the oracle $O_F$, we output the column index equal to the row index, as one can see, for instance, for the input $\ket{101}_d\ket{j\neq j_Q}_j$ (in the block $Q$) in Table~\ref{table:of}.
The presence of these diagonal elements does not affect the action of the oracles $O_{\sqrt{H}}$ and $O_S$.

Table~\ref{table:oh} presents the square roots of the absolute values of the matrix elements given the element row and column indices.
The sign and the additional unitary factor $\yi$ contained in these elements are presented in Table~\ref{table:os}.

To create the $O_{\sqrt{H}}$ quantum circuit, we use gates that perform rotations $R_y(\theta_j)$ conditioned on the register $j$. 
Every combination of qubits in $\ket{j}$ corresponds to a particular angle $\theta_j$ expressing the space dependence of a given field on $x$.
For instance, to obtain the profile of the background magnetic field, one can use angles $\theta_j = 2\arccos{\sqrt{\beta_H B_{0,j}}}$.
For convenience, we denote the corresponding conditional gate as $R(B)$ in our quantum circuit (Fig.~\ref{fig:oh}).

The conditional-rotation gates must be expressed via arithmetic functions with a set of additional qubits to store the angles $\theta_j$. In our work, every conditional gate is coded as a multi-qubit gate with inner sub-blocks as in Eq.~\ref{eq:ry-matrix} on the main diagonal. Every sub-block corresponds to one $\theta_j$ for a particular $j$.
The circuit depth of $O_{\sqrt{H}}$ may strongly depend on how efficient the implementation of the conditional gates is for given profiles and how they depend on the system size $N_x$.
However, the general tendency is that the depth of quantum arithmetic circuits scales as \begin{math}\oO(\text{poly}(n_x))\end{math} (Refs.~\onlinecite{Haner18, Suau21}).

According to Eq.~\ref{eq:be-main-theory}, the normalization of the Hamiltonian depends on the matrix sparsity, which we define as the number of nonzero elements in a matrix row maximized over the row index. This dependence is reflected in the fact that $d_H^2$ appears in the normalization coefficient $\beta_H$ in Eq.~\ref{eq:beta-h}. In our case, $d^2_H = 4$, which is close to the sparsity $\varsigma = 3$ of our wave Hamiltonian (Eq.~\ref{eq:H}), as it should be according to Eq.~\ref{eq:u}.
If $\varsigma > 1$, then one must encode the positions of several nonzero elements knowing only a single row index.
This means that a superposition of several quantum states has to be created from a single input state. For instance, the input state $\ket{011}_d\ket{1}_j$  must create a superposition of two states, $\ket{00}_{a_j}\ket{001}_{a_d} + \ket{10}_{a_j}\ket{100}_{a_d}$.
This can be done by applying a Hadamard gate (Eq.~\ref{eq:hadamard}), and one needs at least $n_H = 2$ Hadamard gates to produce the superposition of $\varsigma = 3$ states. However, the amplitude of each state in the resulting superposition has an additional multiplier $1/2^{n_H/2}$.
Then, according to Eq.~\ref{eq:ry-v}, we encode the square root of the absolute value of the matrix element $v$ as
\begin{equation}
    \sqrt{\frac{|v|}{d_H^2 M}} = \frac{1}{2^{n_H/2}} \cos(\theta/2),
\end{equation}
where $\cos(\theta/2) = \sqrt{|v|/M} \leq 1$, and therefore, $d_H = 2^{n_H/2}$.

\section{Amplitude estimation}\label{app:meas}


\begin{figure}[!b]
\centering
\includegraphics[]{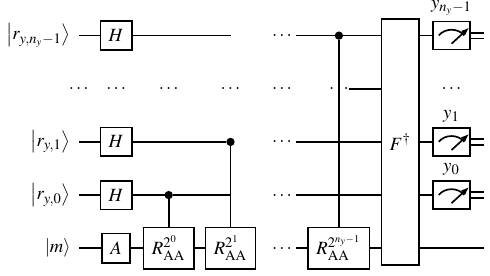}
\caption{\label{fig:ae} 
Circuit for the amplitude estimation.
Here, $F^\dagger$ is the inverse quantum Fourier transform.
The circuit for the Fourier transform, $F$, can be found in Ref.~\onlinecite{Nielsen10}.
Here, as in the rest of the paper, it is assumed that the top qubit is the most significant one.
}
\end{figure}

As it is shown in Fig.~\ref{fig:meas}, we can store the energy integrated over a given spatial volume as the probability $p_{\rm G}$ of the state $\ket{1}$ on the ancilla qubit $m$:
\begin{subequations}
\begin{eqnarray}
    &&\psi_{\rm meas} = a_{\rm B}\ket{0} + a\ket{1},\\
    &&p_{\rm G} \equiv |a|^2 = \langle W\rangle_x(t_f),\\
    &&p_{\rm B} \equiv |a_{\rm B}|^2 = 1 - p_{\rm G}.\label{eq:pB}
\end{eqnarray}
\end{subequations}
To save the energy in the ancilla $m$, we perform the post-selection by choosing only those states where all QSP ancillae qubits are in the zero state.
If we calculate the total system energy, $p_{\rm G}$, the resulting probability of the post-selected state will be close to unity, $p_{\rm G} \approx 1$, with an absolute error defined by the QSP approximation error $\epsilon_{\yqsp}$.
However, if we want to find an energy component in a small spatial volume, then the corresponding probability can be much less than unity, $p_{\rm G} \ll 1$.
Once the state $\psi_{\rm meas}$ is formed, one can use the Amplitude Estimation (AE) technique to measure $p_{\rm G}$. 

There is a wide variety of modern AE techniques as it has been discussed in Sec.~\ref{ssec:meas}.
Here, however, we use the standard AE first introduced in Ref.~\onlinecite{Brassard02} to show how the $1/\delta$ factor appears in the final scaling of our circuit, Eq.~\ref{eq:final-scaling}.
For any AE method, we need an operator $A$ that creates the state whose probability $p_{\rm G}$ we need to measure.
In Hamiltonian simulations, $A$ is equal to the product of the QSP operator and of the energy integration over a given spatial volume.

\begin{figure}[!t]
\centering
\includegraphics[]{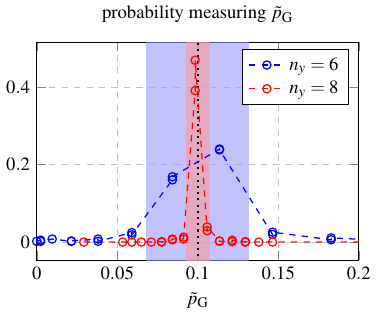}
\caption{\label{fig:AE-meas} 
Probability distribution of measurement outcomes $\tilde{p}_{\rm G}$.
Amplitude estimation of known $p_{\rm G} = 0.1$ is performed with $n_y = 6$ (blue markers) and $n_y = 8$ (red markers).
The shaded areas mark the corresponding theoretical intervals presented in Eq.~\ref{eq:th-abs-error}.
}
\end{figure}

The value $p_{\rm G}$ can be changed by using the Amplitude-Amplification (AA) operator $R_{\rm AA}$:
\begin{subequations}
\begin{eqnarray}
&&R_{\rm AA} = A S_0 A^\dagger S_{\rm G},\\
&&S_0 = 1 - 2\ket{0}_m\bra{0}_m,\\
&&S_{\rm G} = 1 - 2\ket{1}_m\bra{1}_m,
\end{eqnarray}
\end{subequations}
where the operator $S_0$ inverses the sign of the initial state, which is $\ket{0}_m$; the operator $S_{\rm G}$ inverses the sign of the state to measure, which is $\ket{1}_m$ in our case.
To find $p_{\rm G}$, we need to estimate the eigenphase $2\theta_{\rm B}$ of the operator $R_{\rm AA}$.
This can be done by using the phase estimation circuit as it is shown in Fig.~\ref{fig:ae}.
The circuit requires $\sum_{k=0}^{n_y-1} 2^k = M - 1$ queries to $R_{\rm AA}$, where $M = 2^{n_y}$.
This results in $\oO(M)$ requests to the oracle $A$.
More precisely, one needs $M$ queries to $A$ and $M - 1$ queries to $A^\dagger$ to perform the AE.
The eigenphase is then calculated by using the following integer:
\begin{equation}
    y = \sum_{k=0}^{n_y-1} y_k 2^k, 
\end{equation}
where each $y_k = 0$ or $1$ is measured on the corresponding qubit $r_{y,k}$.
The resulting phase and the corresponding probability are estimated as 
\begin{subequations}
\begin{eqnarray}
&&\theta_{\rm B} = \frac{\pi y}{2^{n_y}},\\
&&\tilde{p}_{\rm B} = \sin^2(\theta_{\rm B}).
\end{eqnarray}
\end{subequations}
After that, one can estimate $p_{\rm G}$ by using Eq.~\ref{eq:pB}.
With a probability at least $8/\pi^2\approx 0.81$, the absolute error of the estimation is\cite{Brassard02}
\begin{equation}
    \delta \equiv |p_{\rm G} - \tilde{p}_{\rm G}| \leq 2\pi\frac{\sqrt{p_{\rm G}(1 - p_{\rm G})}}{M} + \frac{\pi^2}{M^2}.\label{eq:th-abs-error}
\end{equation}
This means that $\delta = \oO(1/M)$, which results in $\oO(1/\delta)$ queries to the oracle $A$ in the AE circuit.
As mentioned above, in the general case of QSP simulations, the oracle $A$ includes the whole QSP circuit.
Therefore, to calculate $p_{\rm G}$, we need $\oO(1/\delta)$ requests to the QSP operator.
This explains the appearance of the prefactor $\oO(1/\delta)$ in the final scaling~\ref{eq:final-scaling}.
As an example, numerical estimation of $p_{\rm G} = 0.1$ is given in Fig.~\ref{fig:AE-meas} for two different $n_y$.

%% file: main.bbl
\begin{thebibliography}{67}%
\makeatletter
\providecommand \@ifxundefined [1]{%
 \@ifx{#1\undefined}
}%
\providecommand \@ifnum [1]{%
 \ifnum #1\expandafter \@firstoftwo
 \else \expandafter \@secondoftwo
 \fi
}%
\providecommand \@ifx [1]{%
 \ifx #1\expandafter \@firstoftwo
 \else \expandafter \@secondoftwo
 \fi
}%
\providecommand \natexlab [1]{#1}%
\providecommand \enquote  [1]{``#1''}%
\providecommand \bibnamefont  [1]{#1}%
\providecommand \bibfnamefont [1]{#1}%
\providecommand \citenamefont [1]{#1}%
\providecommand \href@noop [0]{\@secondoftwo}%
\providecommand \href [0]{\begingroup \@sanitize@url \@href}%
\providecommand \@href[1]{\@@startlink{#1}\@@href}%
\providecommand \@@href[1]{\endgroup#1\@@endlink}%
\providecommand \@sanitize@url [0]{\catcode `\\12\catcode `\$12\catcode
  `\&12\catcode `\#12\catcode `\^12\catcode `\_12\catcode `\%12\relax}%
\providecommand \@@startlink[1]{}%
\providecommand \@@endlink[0]{}%
\providecommand \url  [0]{\begingroup\@sanitize@url \@url }%
\providecommand \@url [1]{\endgroup\@href {#1}{\urlprefix }}%
\providecommand \urlprefix  [0]{URL }%
\providecommand \Eprint [0]{\href }%
\providecommand \doibase [0]{http://dx.doi.org/}%
\providecommand \selectlanguage [0]{\@gobble}%
\providecommand \bibinfo  [0]{\@secondoftwo}%
\providecommand \bibfield  [0]{\@secondoftwo}%
\providecommand \translation [1]{[#1]}%
\providecommand \BibitemOpen [0]{}%
\providecommand \bibitemStop [0]{}%
\providecommand \bibitemNoStop [0]{.\EOS\space}%
\providecommand \EOS [0]{\spacefactor3000\relax}%
\providecommand \BibitemShut  [1]{\csname bibitem#1\endcsname}%
\let\auto@bib@innerbib\@empty
\bibitem [{\citenamefont {Suau}, \citenamefont {Staffelbach},\ and\
  \citenamefont {Calandra}(2021)}]{Suau21}%
  \BibitemOpen
  \bibfield  {author} {\bibinfo {author} {\bibfnamefont {A.}~\bibnamefont
  {Suau}}, \bibinfo {author} {\bibfnamefont {G.}~\bibnamefont {Staffelbach}}, \
  and\ \bibinfo {author} {\bibfnamefont {H.}~\bibnamefont {Calandra}},\
  }\bibfield  {title} {\enquote {\bibinfo {title} {Practical quantum
  computing},}\ }\href {\doibase 10.1145/3430030} {\bibfield  {journal}
  {\bibinfo  {journal} {ACM Transactions on Quantum Computing}\ }\textbf
  {\bibinfo {volume} {2}},\ \bibinfo {pages} {1–35} (\bibinfo {year}
  {2021})}\BibitemShut {NoStop}%
\bibitem [{\citenamefont {Cao}\ \emph {et~al.}(2013)\citenamefont {Cao},
  \citenamefont {Papageorgiou}, \citenamefont {Petras}, \citenamefont {Traub},\
  and\ \citenamefont {Kais}}]{Cao13}%
  \BibitemOpen
  \bibfield  {author} {\bibinfo {author} {\bibfnamefont {Y.}~\bibnamefont
  {Cao}}, \bibinfo {author} {\bibfnamefont {A.}~\bibnamefont {Papageorgiou}},
  \bibinfo {author} {\bibfnamefont {I.}~\bibnamefont {Petras}}, \bibinfo
  {author} {\bibfnamefont {J.}~\bibnamefont {Traub}}, \ and\ \bibinfo {author}
  {\bibfnamefont {S.}~\bibnamefont {Kais}},\ }\bibfield  {title} {\enquote
  {\bibinfo {title} {Quantum algorithm and circuit design solving the {Poisson}
  equation},}\ }\href {\doibase 10.1088/1367-2630/15/1/013021} {\bibfield
  {journal} {\bibinfo  {journal} {New Journal of Physics}\ }\textbf {\bibinfo
  {volume} {15}},\ \bibinfo {pages} {013021} (\bibinfo {year}
  {2013})}\BibitemShut {NoStop}%
\bibitem [{\citenamefont {Wang}\ \emph {et~al.}(2020)\citenamefont {Wang},
  \citenamefont {Wang}, \citenamefont {Li}, \citenamefont {Fan}, \citenamefont
  {Wei},\ and\ \citenamefont {Gu}}]{Wang20}%
  \BibitemOpen
  \bibfield  {author} {\bibinfo {author} {\bibfnamefont {S.}~\bibnamefont
  {Wang}}, \bibinfo {author} {\bibfnamefont {Z.}~\bibnamefont {Wang}}, \bibinfo
  {author} {\bibfnamefont {W.}~\bibnamefont {Li}}, \bibinfo {author}
  {\bibfnamefont {L.}~\bibnamefont {Fan}}, \bibinfo {author} {\bibfnamefont
  {Z.}~\bibnamefont {Wei}}, \ and\ \bibinfo {author} {\bibfnamefont
  {Y.}~\bibnamefont {Gu}},\ }\bibfield  {title} {\enquote {\bibinfo {title}
  {Quantum fast {Poisson} solver: the algorithm and complete and modular
  circuit design},}\ }\href {\doibase 10.1007/s11128-020-02669-7} {\bibfield
  {journal} {\bibinfo  {journal} {Quantum Information Processing}\ }\textbf
  {\bibinfo {volume} {19}},\ \bibinfo {pages} {170} (\bibinfo {year}
  {2020})}\BibitemShut {NoStop}%
\bibitem [{\citenamefont {Sinha}\ and\ \citenamefont {Russer}(2010)}]{Sinha10}%
  \BibitemOpen
  \bibfield  {author} {\bibinfo {author} {\bibfnamefont {S.}~\bibnamefont
  {Sinha}}\ and\ \bibinfo {author} {\bibfnamefont {P.}~\bibnamefont {Russer}},\
  }\bibfield  {title} {\enquote {\bibinfo {title} {Quantum computing algorithm
  for electromagnetic field simulation},}\ }\href {\doibase
  10.1007/s11128-009-0133-x} {\bibfield  {journal} {\bibinfo  {journal}
  {Quantum Information Processing}\ }\textbf {\bibinfo {volume} {9}},\ \bibinfo
  {pages} {385--404} (\bibinfo {year} {2010})}\BibitemShut {NoStop}%
\bibitem [{\citenamefont {Scherer}\ \emph {et~al.}(2017)\citenamefont
  {Scherer}, \citenamefont {Valiron}, \citenamefont {Mau}, \citenamefont
  {Alexander}, \citenamefont {van~den Berg},\ and\ \citenamefont
  {Chapuran}}]{Scherer17}%
  \BibitemOpen
  \bibfield  {author} {\bibinfo {author} {\bibfnamefont {A.}~\bibnamefont
  {Scherer}}, \bibinfo {author} {\bibfnamefont {B.}~\bibnamefont {Valiron}},
  \bibinfo {author} {\bibfnamefont {S.-C.}\ \bibnamefont {Mau}}, \bibinfo
  {author} {\bibfnamefont {S.}~\bibnamefont {Alexander}}, \bibinfo {author}
  {\bibfnamefont {E.}~\bibnamefont {van~den Berg}}, \ and\ \bibinfo {author}
  {\bibfnamefont {T.~E.}\ \bibnamefont {Chapuran}},\ }\bibfield  {title}
  {\enquote {\bibinfo {title} {Concrete resource analysis of the quantum
  linear-system algorithm used to compute the electromagnetic scattering cross
  section of a {2D} target},}\ }\href {\doibase 10.1007/s11128-016-1495-5}
  {\bibfield  {journal} {\bibinfo  {journal} {Quantum Information Processing}\
  }\textbf {\bibinfo {volume} {16}},\ \bibinfo {pages} {60} (\bibinfo {year}
  {2017})}\BibitemShut {NoStop}%
\bibitem [{\citenamefont {Fillion-Gourdeau}\ and\ \citenamefont
  {Lorin}(2019)}]{Gourdeau19}%
  \BibitemOpen
  \bibfield  {author} {\bibinfo {author} {\bibfnamefont {F.}~\bibnamefont
  {Fillion-Gourdeau}}\ and\ \bibinfo {author} {\bibfnamefont {E.}~\bibnamefont
  {Lorin}},\ }\bibfield  {title} {\enquote {\bibinfo {title} {Simple digital
  quantum algorithm for symmetric first-order linear hyperbolic systems},}\
  }\href {\doibase 10.1007/s11075-018-0639-3} {\bibfield  {journal} {\bibinfo
  {journal} {Numerical Algorithms}\ }\textbf {\bibinfo {volume} {82}},\
  \bibinfo {pages} {1009--1045} (\bibinfo {year} {2019})}\BibitemShut {NoStop}%
\bibitem [{\citenamefont {Gaitan}(2020)}]{Gaitan20}%
  \BibitemOpen
  \bibfield  {author} {\bibinfo {author} {\bibfnamefont {F.}~\bibnamefont
  {Gaitan}},\ }\bibfield  {title} {\enquote {\bibinfo {title} {Finding flows of
  a {Navier--Stokes} fluid through quantum computing},}\ }\href {\doibase
  10.1038/s41534-020-00291-0} {\bibfield  {journal} {\bibinfo  {journal} {npj
  Quantum Information}\ }\textbf {\bibinfo {volume} {6}},\ \bibinfo {pages}
  {61} (\bibinfo {year} {2020})}\BibitemShut {NoStop}%
\bibitem [{\citenamefont {Gaitan}(2021)}]{Gaitan21}%
  \BibitemOpen
  \bibfield  {author} {\bibinfo {author} {\bibfnamefont {F.}~\bibnamefont
  {Gaitan}},\ }\bibfield  {title} {\enquote {\bibinfo {title} {Finding
  solutions of the {Navier-Stokes} equations through quantum computing—recent
  progress, a generalization, and next steps forward},}\ }\href {\doibase
  https://doi.org/10.1002/qute.202100055} {\bibfield  {journal} {\bibinfo
  {journal} {Advanced Quantum Technologies}\ }\textbf {\bibinfo {volume} {4}},\
  \bibinfo {pages} {2100055} (\bibinfo {year} {2021})},\ \Eprint
  {http://arxiv.org/abs/https://onlinelibrary.wiley.com/doi/pdf/10.1002/qute.202100055}
  {https://onlinelibrary.wiley.com/doi/pdf/10.1002/qute.202100055} \BibitemShut
  {NoStop}%
\bibitem [{\citenamefont {Todorova}\ and\ \citenamefont
  {Steijl}(2020)}]{Todorova20}%
  \BibitemOpen
  \bibfield  {author} {\bibinfo {author} {\bibfnamefont {B.~N.}\ \bibnamefont
  {Todorova}}\ and\ \bibinfo {author} {\bibfnamefont {R.}~\bibnamefont
  {Steijl}},\ }\bibfield  {title} {\enquote {\bibinfo {title} {Quantum
  algorithm for the collisionless {Boltzmann} equation},}\ }\href {\doibase
  10.1016/j.jcp.2020.109347} {\bibfield  {journal} {\bibinfo  {journal}
  {Journal of Computational Physics}\ }\textbf {\bibinfo {volume} {409}},\
  \bibinfo {pages} {109347} (\bibinfo {year} {2020})}\BibitemShut {NoStop}%
\bibitem [{\citenamefont {Budinski}(2021)}]{Budinski21}%
  \BibitemOpen
  \bibfield  {author} {\bibinfo {author} {\bibfnamefont {L.}~\bibnamefont
  {Budinski}},\ }\bibfield  {title} {\enquote {\bibinfo {title} {Quantum
  algorithm for the advection--diffusion equation simulated with the lattice
  {Boltzmann} method},}\ }\href {\doibase 10.1007/s11128-021-02996-3}
  {\bibfield  {journal} {\bibinfo  {journal} {Quantum Information Processing}\
  }\textbf {\bibinfo {volume} {20}},\ \bibinfo {pages} {57} (\bibinfo {year}
  {2021})}\BibitemShut {NoStop}%
\bibitem [{\citenamefont {Dodin}\ and\ \citenamefont
  {Startsev}(2021)}]{Dodin20}%
  \BibitemOpen
  \bibfield  {author} {\bibinfo {author} {\bibfnamefont {I.~Y.}\ \bibnamefont
  {Dodin}}\ and\ \bibinfo {author} {\bibfnamefont {E.~A.}\ \bibnamefont
  {Startsev}},\ }\bibfield  {title} {\enquote {\bibinfo {title} {On
  applications of quantum computing to plasma simulations},}\ }\href {\doibase
  10.1063/5.0056974} {\bibfield  {journal} {\bibinfo  {journal} {Physics of
  Plasmas}\ }\textbf {\bibinfo {volume} {28}},\ \bibinfo {pages} {092101}
  (\bibinfo {year} {2021})},\ \Eprint
  {http://arxiv.org/abs/https://doi.org/10.1063/5.0056974}
  {https://doi.org/10.1063/5.0056974} \BibitemShut {NoStop}%
\bibitem [{\citenamefont {Stix}(1992)}]{Stix92}%
  \BibitemOpen
  \bibfield  {author} {\bibinfo {author} {\bibfnamefont {T.~H.}\ \bibnamefont
  {Stix}},\ }\href@noop {} {\emph {\bibinfo {title} {Waves in Plasmas}}}\
  (\bibinfo  {publisher} {AIP Press},\ \bibinfo {year} {1992})\BibitemShut
  {NoStop}%
\bibitem [{\citenamefont {Pinsker}(2001)}]{Pinsker01}%
  \BibitemOpen
  \bibfield  {author} {\bibinfo {author} {\bibfnamefont {R.~I.}\ \bibnamefont
  {Pinsker}},\ }\bibfield  {title} {\enquote {\bibinfo {title} {Introduction to
  wave heating and current drive in magnetized plasmas},}\ }\href {\doibase
  10.1063/1.1343512} {\bibfield  {journal} {\bibinfo  {journal} {Physics of
  Plasmas}\ }\textbf {\bibinfo {volume} {8}},\ \bibinfo {pages} {1219--1228}
  (\bibinfo {year} {2001})},\ \Eprint
  {http://arxiv.org/abs/https://aip.scitation.org/doi/pdf/10.1063/1.1343512}
  {https://aip.scitation.org/doi/pdf/10.1063/1.1343512} \BibitemShut {NoStop}%
\bibitem [{\citenamefont {Prater}(2004)}]{Prater04}%
  \BibitemOpen
  \bibfield  {author} {\bibinfo {author} {\bibfnamefont {R.}~\bibnamefont
  {Prater}},\ }\bibfield  {title} {\enquote {\bibinfo {title} {Heating and
  current drive by electron cyclotron waves},}\ }\href {\doibase
  10.1063/1.1690762} {\bibfield  {journal} {\bibinfo  {journal} {Physics of
  Plasmas}\ }\textbf {\bibinfo {volume} {11}},\ \bibinfo {pages} {2349--2376}
  (\bibinfo {year} {2004})},\ \Eprint
  {http://arxiv.org/abs/https://doi.org/10.1063/1.1690762}
  {https://doi.org/10.1063/1.1690762} \BibitemShut {NoStop}%
\bibitem [{\citenamefont {Fisch}(1987)}]{Fisch87}%
  \BibitemOpen
  \bibfield  {author} {\bibinfo {author} {\bibfnamefont {N.~J.}\ \bibnamefont
  {Fisch}},\ }\bibfield  {title} {\enquote {\bibinfo {title} {Theory of current
  drive in plasmas},}\ }\href {\doibase 10.1103/RevModPhys.59.175} {\bibfield
  {journal} {\bibinfo  {journal} {Reviews of Modern Physics}\ }\textbf
  {\bibinfo {volume} {59}},\ \bibinfo {pages} {175--234} (\bibinfo {year}
  {1987})}\BibitemShut {NoStop}%
\bibitem [{\citenamefont {Fisch}(1978)}]{Fisch78}%
  \BibitemOpen
  \bibfield  {author} {\bibinfo {author} {\bibfnamefont {N.~J.}\ \bibnamefont
  {Fisch}},\ }\bibfield  {title} {\enquote {\bibinfo {title} {Confining a
  tokamak plasma with {RF}-driven currents},}\ }\href {\doibase
  10.1103/PhysRevLett.41.873} {\bibfield  {journal} {\bibinfo  {journal}
  {Physical Review Letters}\ }\textbf {\bibinfo {volume} {41}},\ \bibinfo
  {pages} {873--876} (\bibinfo {year} {1978})}\BibitemShut {NoStop}%
\bibitem [{\citenamefont {Fisch}\ and\ \citenamefont {Boozer}(1980)}]{Fisch80}%
  \BibitemOpen
  \bibfield  {author} {\bibinfo {author} {\bibfnamefont {N.~J.}\ \bibnamefont
  {Fisch}}\ and\ \bibinfo {author} {\bibfnamefont {A.~H.}\ \bibnamefont
  {Boozer}},\ }\bibfield  {title} {\enquote {\bibinfo {title} {Creating an
  asymmetric plasma resistivity with waves},}\ }\href {\doibase
  10.1103/PhysRevLett.45.720} {\bibfield  {journal} {\bibinfo  {journal}
  {Physical Review Letters}\ }\textbf {\bibinfo {volume} {45}},\ \bibinfo
  {pages} {720--722} (\bibinfo {year} {1980})}\BibitemShut {NoStop}%
\bibitem [{\citenamefont {Reiman}(1983)}]{Reiman83}%
  \BibitemOpen
  \bibfield  {author} {\bibinfo {author} {\bibfnamefont {A.~H.}\ \bibnamefont
  {Reiman}},\ }\bibfield  {title} {\enquote {\bibinfo {title} {Suppression of
  magnetic islands by {RF} driven currents},}\ }\href {\doibase
  10.1063/1.864258} {\bibfield  {journal} {\bibinfo  {journal} {The Physics of
  Fluids}\ }\textbf {\bibinfo {volume} {26}},\ \bibinfo {pages} {1338--1340}
  (\bibinfo {year} {1983})},\ \Eprint
  {http://arxiv.org/abs/https://aip.scitation.org/doi/pdf/10.1063/1.864258}
  {https://aip.scitation.org/doi/pdf/10.1063/1.864258} \BibitemShut {NoStop}%
\bibitem [{\citenamefont {Reiman}\ and\ \citenamefont
  {Fisch}(2018)}]{Reiman18}%
  \BibitemOpen
  \bibfield  {author} {\bibinfo {author} {\bibfnamefont {A.~H.}\ \bibnamefont
  {Reiman}}\ and\ \bibinfo {author} {\bibfnamefont {N.~J.}\ \bibnamefont
  {Fisch}},\ }\bibfield  {title} {\enquote {\bibinfo {title} {Suppression of
  tearing modes by radio frequency current condensation},}\ }\href {\doibase
  10.1103/PhysRevLett.121.225001} {\bibfield  {journal} {\bibinfo  {journal}
  {Physical Review Letters}\ }\textbf {\bibinfo {volume} {121}},\ \bibinfo
  {pages} {225001} (\bibinfo {year} {2018})}\BibitemShut {NoStop}%
\bibitem [{\citenamefont {Ding}\ \emph {et~al.}(2018)\citenamefont {Ding},
  \citenamefont {Bonoli}, \citenamefont {Tuccillo}, \citenamefont {Goniche},
  \citenamefont {Kirov}, \citenamefont {Li}, \citenamefont {Li}, \citenamefont
  {Cesario}, \citenamefont {Peysson}, \citenamefont {Ekedahl}, \citenamefont
  {Amicucci}, \citenamefont {Baek}, \citenamefont {Faust}, \citenamefont
  {Parker}, \citenamefont {Shiraiwa}, \citenamefont {Wallace}, \citenamefont
  {Cardinali}, \citenamefont {Castaldo}, \citenamefont {Ceccuzzi},
  \citenamefont {Mailloux}, \citenamefont {Napoli}, \citenamefont {Liu},\ and\
  \citenamefont {Wan}}]{Ding18}%
  \BibitemOpen
  \bibfield  {author} {\bibinfo {author} {\bibfnamefont {B.~J.}\ \bibnamefont
  {Ding}}, \bibinfo {author} {\bibfnamefont {P.~T.}\ \bibnamefont {Bonoli}},
  \bibinfo {author} {\bibfnamefont {A.}~\bibnamefont {Tuccillo}}, \bibinfo
  {author} {\bibfnamefont {M.}~\bibnamefont {Goniche}}, \bibinfo {author}
  {\bibfnamefont {K.}~\bibnamefont {Kirov}}, \bibinfo {author} {\bibfnamefont
  {M.}~\bibnamefont {Li}}, \bibinfo {author} {\bibfnamefont {Y.}~\bibnamefont
  {Li}}, \bibinfo {author} {\bibfnamefont {R.}~\bibnamefont {Cesario}},
  \bibinfo {author} {\bibfnamefont {Y.}~\bibnamefont {Peysson}}, \bibinfo
  {author} {\bibfnamefont {A.}~\bibnamefont {Ekedahl}}, \bibinfo {author}
  {\bibfnamefont {L.}~\bibnamefont {Amicucci}}, \bibinfo {author}
  {\bibfnamefont {S.}~\bibnamefont {Baek}}, \bibinfo {author} {\bibfnamefont
  {I.}~\bibnamefont {Faust}}, \bibinfo {author} {\bibfnamefont
  {R.}~\bibnamefont {Parker}}, \bibinfo {author} {\bibfnamefont
  {S.}~\bibnamefont {Shiraiwa}}, \bibinfo {author} {\bibfnamefont {G.~M.}\
  \bibnamefont {Wallace}}, \bibinfo {author} {\bibfnamefont {A.}~\bibnamefont
  {Cardinali}}, \bibinfo {author} {\bibfnamefont {C.}~\bibnamefont {Castaldo}},
  \bibinfo {author} {\bibfnamefont {S.}~\bibnamefont {Ceccuzzi}}, \bibinfo
  {author} {\bibfnamefont {J.}~\bibnamefont {Mailloux}}, \bibinfo {author}
  {\bibfnamefont {F.}~\bibnamefont {Napoli}}, \bibinfo {author} {\bibfnamefont
  {F.}~\bibnamefont {Liu}}, \ and\ \bibinfo {author} {\bibfnamefont
  {B.}~\bibnamefont {Wan}},\ }\bibfield  {title} {\enquote {\bibinfo {title}
  {Review of recent experimental and modeling advances in the understanding of
  lower hybrid current drive in {ITER}-relevant regimes},}\ }\href {\doibase
  10.1088/1741-4326/aad0aa} {\bibfield  {journal} {\bibinfo  {journal} {Nuclear
  Fusion}\ }\textbf {\bibinfo {volume} {58}},\ \bibinfo {pages} {095003}
  (\bibinfo {year} {2018})}\BibitemShut {NoStop}%
\bibitem [{\citenamefont {Tsujimura}\ \emph {et~al.}(2020)\citenamefont
  {Tsujimura}, \citenamefont {Yanai}, \citenamefont {Mizuno}, \citenamefont
  {Tanaka}, \citenamefont {Yoshimura}, \citenamefont {Tokuzawa}, \citenamefont
  {Nishiura}, \citenamefont {Sakamoto}, \citenamefont {Motojima}, \citenamefont
  {Kubo}, \citenamefont {Shimozuma}, \citenamefont {Igami}, \citenamefont
  {Takahashi}, \citenamefont {Yoshinuma},\ and\ \citenamefont
  {Ohshima}}]{Tsujimura20}%
  \BibitemOpen
  \bibfield  {author} {\bibinfo {author} {\bibfnamefont {T.~I.}\ \bibnamefont
  {Tsujimura}}, \bibinfo {author} {\bibfnamefont {R.}~\bibnamefont {Yanai}},
  \bibinfo {author} {\bibfnamefont {Y.}~\bibnamefont {Mizuno}}, \bibinfo
  {author} {\bibfnamefont {K.}~\bibnamefont {Tanaka}}, \bibinfo {author}
  {\bibfnamefont {Y.}~\bibnamefont {Yoshimura}}, \bibinfo {author}
  {\bibfnamefont {T.}~\bibnamefont {Tokuzawa}}, \bibinfo {author}
  {\bibfnamefont {M.}~\bibnamefont {Nishiura}}, \bibinfo {author}
  {\bibfnamefont {R.}~\bibnamefont {Sakamoto}}, \bibinfo {author}
  {\bibfnamefont {G.}~\bibnamefont {Motojima}}, \bibinfo {author}
  {\bibfnamefont {S.}~\bibnamefont {Kubo}}, \bibinfo {author} {\bibfnamefont
  {T.}~\bibnamefont {Shimozuma}}, \bibinfo {author} {\bibfnamefont
  {H.}~\bibnamefont {Igami}}, \bibinfo {author} {\bibfnamefont
  {H.}~\bibnamefont {Takahashi}}, \bibinfo {author} {\bibfnamefont
  {M.}~\bibnamefont {Yoshinuma}}, \ and\ \bibinfo {author} {\bibfnamefont
  {S.}~\bibnamefont {Ohshima}},\ }\bibfield  {title} {\enquote {\bibinfo
  {title} {Improved performance of electron cyclotron resonance heating by
  perpendicular injection in the {Large Helical Device}},}\ }\href {\doibase
  10.1088/1741-4326/abc977} {\bibfield  {journal} {\bibinfo  {journal} {Nuclear
  Fusion}\ }\textbf {\bibinfo {volume} {61}},\ \bibinfo {pages} {026012}
  (\bibinfo {year} {2020})}\BibitemShut {NoStop}%
\bibitem [{\citenamefont {Bonoli}\ and\ \citenamefont
  {Porkolab}(1987)}]{Bonoli87}%
  \BibitemOpen
  \bibfield  {author} {\bibinfo {author} {\bibfnamefont {P.~T.}\ \bibnamefont
  {Bonoli}}\ and\ \bibinfo {author} {\bibfnamefont {M.}~\bibnamefont
  {Porkolab}},\ }\bibfield  {title} {\enquote {\bibinfo {title} {Radiofrequency
  current generation by lower hybrid slow waves in the presence of fusion
  generated alpha particles in the reactor regime},}\ }\href {\doibase
  10.1088/0029-5515/27/8/013} {\bibfield  {journal} {\bibinfo  {journal}
  {Nuclear Fusion}\ }\textbf {\bibinfo {volume} {27}},\ \bibinfo {pages}
  {1341--1346} (\bibinfo {year} {1987})}\BibitemShut {NoStop}%
\bibitem [{\citenamefont {Cesario}\ \emph {et~al.}(2010)\citenamefont
  {Cesario}, \citenamefont {Amicucci}, \citenamefont {Cardinali}, \citenamefont
  {Castaldo}, \citenamefont {Marinucci}, \citenamefont {Panaccione},
  \citenamefont {Santini}, \citenamefont {Tudisco}, \citenamefont {Apicella},
  \citenamefont {Calabr{\`o}}, \citenamefont {Cianfarani}, \citenamefont
  {Frigione}, \citenamefont {Galli}, \citenamefont {Mazzitelli}, \citenamefont
  {Mazzotta}, \citenamefont {Pericoli}, \citenamefont {Schettini},
  \citenamefont {Tuccillo}, \citenamefont {Angelini}, \citenamefont
  {Apruzzese}, \citenamefont {Barbato}, \citenamefont {Belli}, \citenamefont
  {Bin}, \citenamefont {Boncagni}, \citenamefont {Botrugno}, \citenamefont
  {Briguglio}, \citenamefont {Bruschi}, \citenamefont {Ceccuzzi}, \citenamefont
  {Centioli}, \citenamefont {Cirant}, \citenamefont {Crisanti}, \citenamefont
  {D'Arcangelo}, \citenamefont {De~Angelis}, \citenamefont {Di~Matteo},
  \citenamefont {Di~Troia}, \citenamefont {Esposito}, \citenamefont {Fogaccia},
  \citenamefont {Fusco}, \citenamefont {Gabellieri}, \citenamefont
  {Garavaglia}, \citenamefont {Giovannozzi}, \citenamefont {Granucci},
  \citenamefont {Grossetti}, \citenamefont {Grosso}, \citenamefont {Iannone},
  \citenamefont {Maddaluno}, \citenamefont {Marocco}, \citenamefont {Micozzi},
  \citenamefont {Milovanov}, \citenamefont {Mirizzi}, \citenamefont {Monari},
  \citenamefont {Moro}, \citenamefont {Novak}, \citenamefont {Orsitto},
  \citenamefont {Panella}, \citenamefont {Pucella}, \citenamefont {Ravera},
  \citenamefont {Sternini}, \citenamefont {Vitale}, \citenamefont {Vlad},
  \citenamefont {Zanza}, \citenamefont {Zerbini}, \citenamefont {Zonca},\ and\
  \citenamefont {{the FTU Team}}}]{Cesario10}%
  \BibitemOpen
  \bibfield  {author} {\bibinfo {author} {\bibfnamefont {R.}~\bibnamefont
  {Cesario}}, \bibinfo {author} {\bibfnamefont {L.}~\bibnamefont {Amicucci}},
  \bibinfo {author} {\bibfnamefont {A.}~\bibnamefont {Cardinali}}, \bibinfo
  {author} {\bibfnamefont {C.}~\bibnamefont {Castaldo}}, \bibinfo {author}
  {\bibfnamefont {M.}~\bibnamefont {Marinucci}}, \bibinfo {author}
  {\bibfnamefont {L.}~\bibnamefont {Panaccione}}, \bibinfo {author}
  {\bibfnamefont {F.}~\bibnamefont {Santini}}, \bibinfo {author} {\bibfnamefont
  {O.}~\bibnamefont {Tudisco}}, \bibinfo {author} {\bibfnamefont {M.~L.}\
  \bibnamefont {Apicella}}, \bibinfo {author} {\bibfnamefont {G.}~\bibnamefont
  {Calabr{\`o}}}, \bibinfo {author} {\bibfnamefont {C.}~\bibnamefont
  {Cianfarani}}, \bibinfo {author} {\bibfnamefont {D.}~\bibnamefont
  {Frigione}}, \bibinfo {author} {\bibfnamefont {A.}~\bibnamefont {Galli}},
  \bibinfo {author} {\bibfnamefont {G.}~\bibnamefont {Mazzitelli}}, \bibinfo
  {author} {\bibfnamefont {C.}~\bibnamefont {Mazzotta}}, \bibinfo {author}
  {\bibfnamefont {V.}~\bibnamefont {Pericoli}}, \bibinfo {author}
  {\bibfnamefont {G.}~\bibnamefont {Schettini}}, \bibinfo {author}
  {\bibfnamefont {A.~A.}\ \bibnamefont {Tuccillo}}, \bibinfo {author}
  {\bibfnamefont {B.}~\bibnamefont {Angelini}}, \bibinfo {author}
  {\bibfnamefont {G.}~\bibnamefont {Apruzzese}}, \bibinfo {author}
  {\bibfnamefont {E.}~\bibnamefont {Barbato}}, \bibinfo {author} {\bibfnamefont
  {G.}~\bibnamefont {Belli}}, \bibinfo {author} {\bibfnamefont
  {W.}~\bibnamefont {Bin}}, \bibinfo {author} {\bibfnamefont {L.}~\bibnamefont
  {Boncagni}}, \bibinfo {author} {\bibfnamefont {A.}~\bibnamefont {Botrugno}},
  \bibinfo {author} {\bibfnamefont {S.}~\bibnamefont {Briguglio}}, \bibinfo
  {author} {\bibfnamefont {A.}~\bibnamefont {Bruschi}}, \bibinfo {author}
  {\bibfnamefont {S.}~\bibnamefont {Ceccuzzi}}, \bibinfo {author}
  {\bibfnamefont {C.}~\bibnamefont {Centioli}}, \bibinfo {author}
  {\bibfnamefont {S.}~\bibnamefont {Cirant}}, \bibinfo {author} {\bibfnamefont
  {F.}~\bibnamefont {Crisanti}}, \bibinfo {author} {\bibfnamefont
  {O.}~\bibnamefont {D'Arcangelo}}, \bibinfo {author} {\bibfnamefont
  {R.}~\bibnamefont {De~Angelis}}, \bibinfo {author} {\bibfnamefont
  {L.}~\bibnamefont {Di~Matteo}}, \bibinfo {author} {\bibfnamefont
  {C.}~\bibnamefont {Di~Troia}}, \bibinfo {author} {\bibfnamefont
  {B.}~\bibnamefont {Esposito}}, \bibinfo {author} {\bibfnamefont
  {G.}~\bibnamefont {Fogaccia}}, \bibinfo {author} {\bibfnamefont
  {V.}~\bibnamefont {Fusco}}, \bibinfo {author} {\bibfnamefont
  {L.}~\bibnamefont {Gabellieri}}, \bibinfo {author} {\bibfnamefont
  {A.}~\bibnamefont {Garavaglia}}, \bibinfo {author} {\bibfnamefont
  {E.}~\bibnamefont {Giovannozzi}}, \bibinfo {author} {\bibfnamefont
  {G.}~\bibnamefont {Granucci}}, \bibinfo {author} {\bibfnamefont
  {G.}~\bibnamefont {Grossetti}}, \bibinfo {author} {\bibfnamefont
  {G.}~\bibnamefont {Grosso}}, \bibinfo {author} {\bibfnamefont
  {F.}~\bibnamefont {Iannone}}, \bibinfo {author} {\bibfnamefont
  {G.}~\bibnamefont {Maddaluno}}, \bibinfo {author} {\bibfnamefont
  {D.}~\bibnamefont {Marocco}}, \bibinfo {author} {\bibfnamefont
  {P.}~\bibnamefont {Micozzi}}, \bibinfo {author} {\bibfnamefont
  {A.}~\bibnamefont {Milovanov}}, \bibinfo {author} {\bibfnamefont
  {F.}~\bibnamefont {Mirizzi}}, \bibinfo {author} {\bibfnamefont
  {G.}~\bibnamefont {Monari}}, \bibinfo {author} {\bibfnamefont
  {A.}~\bibnamefont {Moro}}, \bibinfo {author} {\bibfnamefont {S.}~\bibnamefont
  {Novak}}, \bibinfo {author} {\bibfnamefont {F.~P.}\ \bibnamefont {Orsitto}},
  \bibinfo {author} {\bibfnamefont {M.}~\bibnamefont {Panella}}, \bibinfo
  {author} {\bibfnamefont {G.}~\bibnamefont {Pucella}}, \bibinfo {author}
  {\bibfnamefont {G.}~\bibnamefont {Ravera}}, \bibinfo {author} {\bibfnamefont
  {E.}~\bibnamefont {Sternini}}, \bibinfo {author} {\bibfnamefont
  {E.}~\bibnamefont {Vitale}}, \bibinfo {author} {\bibfnamefont
  {G.}~\bibnamefont {Vlad}}, \bibinfo {author} {\bibfnamefont {V.}~\bibnamefont
  {Zanza}}, \bibinfo {author} {\bibfnamefont {M.}~\bibnamefont {Zerbini}},
  \bibinfo {author} {\bibfnamefont {F.}~\bibnamefont {Zonca}}, \ and\ \bibinfo
  {author} {\bibnamefont {{the FTU Team}}},\ }\bibfield  {title} {\enquote
  {\bibinfo {title} {Current drive at plasma densities required for
  thermonuclear reactors},}\ }\href {\doibase 10.1038/ncomms1052} {\bibfield
  {journal} {\bibinfo  {journal} {Nature Communications}\ }\textbf {\bibinfo
  {volume} {1}},\ \bibinfo {pages} {55} (\bibinfo {year} {2010})}\BibitemShut
  {NoStop}%
\bibitem [{\citenamefont {Wan}\ \emph {et~al.}(2019)\citenamefont {Wan},
  \citenamefont {Liang}, \citenamefont {Gong}, \citenamefont {Xiang},
  \citenamefont {Xu}, \citenamefont {Sun}, \citenamefont {Wang}, \citenamefont
  {Qian}, \citenamefont {Liu}, \citenamefont {Zeng}, \citenamefont {Zhang},
  \citenamefont {Zhang}, \citenamefont {Ding}, \citenamefont {Zang},
  \citenamefont {Lyu}, \citenamefont {Garofalo}, \citenamefont {Ekedahl},
  \citenamefont {Li}, \citenamefont {Ding}, \citenamefont {Ding}, \citenamefont
  {Du}, \citenamefont {Kong}, \citenamefont {Yu}, \citenamefont {Yang},
  \citenamefont {Luo}, \citenamefont {Huang}, \citenamefont {Zhang},
  \citenamefont {Zhang}, \citenamefont {Li}, \citenamefont {Xia},\ and\
  \citenamefont {{the EAST team and Collaborators}}}]{Wan19}%
  \BibitemOpen
  \bibfield  {author} {\bibinfo {author} {\bibfnamefont {B.~N.}\ \bibnamefont
  {Wan}}, \bibinfo {author} {\bibfnamefont {Y.}~\bibnamefont {Liang}}, \bibinfo
  {author} {\bibfnamefont {X.~Z.}\ \bibnamefont {Gong}}, \bibinfo {author}
  {\bibfnamefont {N.}~\bibnamefont {Xiang}}, \bibinfo {author} {\bibfnamefont
  {G.~S.}\ \bibnamefont {Xu}}, \bibinfo {author} {\bibfnamefont
  {Y.}~\bibnamefont {Sun}}, \bibinfo {author} {\bibfnamefont {L.}~\bibnamefont
  {Wang}}, \bibinfo {author} {\bibfnamefont {J.~P.}\ \bibnamefont {Qian}},
  \bibinfo {author} {\bibfnamefont {H.~Q.}\ \bibnamefont {Liu}}, \bibinfo
  {author} {\bibfnamefont {L.}~\bibnamefont {Zeng}}, \bibinfo {author}
  {\bibfnamefont {L.}~\bibnamefont {Zhang}}, \bibinfo {author} {\bibfnamefont
  {X.~J.}\ \bibnamefont {Zhang}}, \bibinfo {author} {\bibfnamefont {B.~J.}\
  \bibnamefont {Ding}}, \bibinfo {author} {\bibfnamefont {Q.}~\bibnamefont
  {Zang}}, \bibinfo {author} {\bibfnamefont {B.}~\bibnamefont {Lyu}}, \bibinfo
  {author} {\bibfnamefont {A.~M.}\ \bibnamefont {Garofalo}}, \bibinfo {author}
  {\bibfnamefont {A.}~\bibnamefont {Ekedahl}}, \bibinfo {author} {\bibfnamefont
  {M.~H.}\ \bibnamefont {Li}}, \bibinfo {author} {\bibfnamefont
  {F.}~\bibnamefont {Ding}}, \bibinfo {author} {\bibfnamefont {S.~Y.}\
  \bibnamefont {Ding}}, \bibinfo {author} {\bibfnamefont {H.~F.}\ \bibnamefont
  {Du}}, \bibinfo {author} {\bibfnamefont {D.~F.}\ \bibnamefont {Kong}},
  \bibinfo {author} {\bibfnamefont {Y.}~\bibnamefont {Yu}}, \bibinfo {author}
  {\bibfnamefont {Y.}~\bibnamefont {Yang}}, \bibinfo {author} {\bibfnamefont
  {Z.~P.}\ \bibnamefont {Luo}}, \bibinfo {author} {\bibfnamefont
  {J.}~\bibnamefont {Huang}}, \bibinfo {author} {\bibfnamefont
  {T.}~\bibnamefont {Zhang}}, \bibinfo {author} {\bibfnamefont
  {Y.}~\bibnamefont {Zhang}}, \bibinfo {author} {\bibfnamefont {G.~Q.}\
  \bibnamefont {Li}}, \bibinfo {author} {\bibfnamefont {T.~Y.}\ \bibnamefont
  {Xia}}, \ and\ \bibinfo {author} {\bibnamefont {{the EAST team and
  Collaborators}}},\ }\bibfield  {title} {\enquote {\bibinfo {title} {Recent
  advances in {EAST} physics experiments in support of steady-state operation
  for {ITER} and {CFETR}},}\ }\href {\doibase 10.1088/1741-4326/ab0396}
  {\bibfield  {journal} {\bibinfo  {journal} {Nuclear Fusion}\ }\textbf
  {\bibinfo {volume} {59}},\ \bibinfo {pages} {112003} (\bibinfo {year}
  {2019})}\BibitemShut {NoStop}%
\bibitem [{\citenamefont {Wallace}\ \emph {et~al.}(2021)\citenamefont
  {Wallace}, \citenamefont {Ding}, \citenamefont {Li}, \citenamefont {Chen},
  \citenamefont {Baek}, \citenamefont {Bonoli}, \citenamefont {Shiraiwa},
  \citenamefont {Liu},\ and\ \citenamefont {Wu}}]{Wallace21}%
  \BibitemOpen
  \bibfield  {author} {\bibinfo {author} {\bibfnamefont {G.~M.}\ \bibnamefont
  {Wallace}}, \bibinfo {author} {\bibfnamefont {B.~J.}\ \bibnamefont {Ding}},
  \bibinfo {author} {\bibfnamefont {M.~H.}\ \bibnamefont {Li}}, \bibinfo
  {author} {\bibfnamefont {J.}~\bibnamefont {Chen}}, \bibinfo {author}
  {\bibfnamefont {S.~G.}\ \bibnamefont {Baek}}, \bibinfo {author}
  {\bibfnamefont {P.~T.}\ \bibnamefont {Bonoli}}, \bibinfo {author}
  {\bibfnamefont {S.}~\bibnamefont {Shiraiwa}}, \bibinfo {author}
  {\bibfnamefont {L.}~\bibnamefont {Liu}}, \ and\ \bibinfo {author}
  {\bibfnamefont {C.}~\bibnamefont {Wu}},\ }\bibfield  {title} {\enquote
  {\bibinfo {title} {Scoping study of lower hybrid current drive for
  {CFETR}},}\ }\href {\doibase 10.1088/1741-4326/ac1ae1} {\bibfield  {journal}
  {\bibinfo  {journal} {Nuclear Fusion}\ }\textbf {\bibinfo {volume} {61}},\
  \bibinfo {pages} {106009} (\bibinfo {year} {2021})}\BibitemShut {NoStop}%
\bibitem [{\citenamefont {Fasoli}\ \emph {et~al.}(2016)\citenamefont {Fasoli},
  \citenamefont {Brunner}, \citenamefont {Cooper}, \citenamefont {Graves},
  \citenamefont {Ricci}, \citenamefont {Sauter},\ and\ \citenamefont
  {Villard}}]{Fasoli16}%
  \BibitemOpen
  \bibfield  {author} {\bibinfo {author} {\bibfnamefont {A.}~\bibnamefont
  {Fasoli}}, \bibinfo {author} {\bibfnamefont {S.}~\bibnamefont {Brunner}},
  \bibinfo {author} {\bibfnamefont {W.~A.}\ \bibnamefont {Cooper}}, \bibinfo
  {author} {\bibfnamefont {J.~P.}\ \bibnamefont {Graves}}, \bibinfo {author}
  {\bibfnamefont {P.}~\bibnamefont {Ricci}}, \bibinfo {author} {\bibfnamefont
  {O.}~\bibnamefont {Sauter}}, \ and\ \bibinfo {author} {\bibfnamefont
  {L.}~\bibnamefont {Villard}},\ }\bibfield  {title} {\enquote {\bibinfo
  {title} {Computational challenges in magnetic-confinement fusion physics},}\
  }\href {\doibase 10.1038/nphys3744} {\bibfield  {journal} {\bibinfo
  {journal} {Nature Physics}\ }\textbf {\bibinfo {volume} {12}},\ \bibinfo
  {pages} {411--423} (\bibinfo {year} {2016})}\BibitemShut {NoStop}%
\bibitem [{\citenamefont {Svidzinski}\ \emph {et~al.}(2018)\citenamefont
  {Svidzinski}, \citenamefont {Kim}, \citenamefont {Zhao}, \citenamefont
  {Galkin},\ and\ \citenamefont {Spencer}}]{Svidzinski18}%
  \BibitemOpen
  \bibfield  {author} {\bibinfo {author} {\bibfnamefont {V.~A.}\ \bibnamefont
  {Svidzinski}}, \bibinfo {author} {\bibfnamefont {J.~S.}\ \bibnamefont {Kim}},
  \bibinfo {author} {\bibfnamefont {L.}~\bibnamefont {Zhao}}, \bibinfo {author}
  {\bibfnamefont {S.~A.}\ \bibnamefont {Galkin}}, \ and\ \bibinfo {author}
  {\bibfnamefont {J.~A.}\ \bibnamefont {Spencer}},\ }\bibfield  {title}
  {\enquote {\bibinfo {title} {Hybrid iterative approach for simulation of
  radio-frequency fields in plasma},}\ }\href {\doibase 10.1063/1.5037110}
  {\bibfield  {journal} {\bibinfo  {journal} {Physics of Plasmas}\ }\textbf
  {\bibinfo {volume} {25}},\ \bibinfo {pages} {082509} (\bibinfo {year}
  {2018})},\ \Eprint {http://arxiv.org/abs/https://doi.org/10.1063/1.5037110}
  {https://doi.org/10.1063/1.5037110} \BibitemShut {NoStop}%
\bibitem [{\citenamefont {Tracy}\ \emph {et~al.}(2014)\citenamefont {Tracy},
  \citenamefont {Brizard}, \citenamefont {Richardson},\ and\ \citenamefont
  {Kaufman}}]{book:tracy}%
  \BibitemOpen
  \bibfield  {author} {\bibinfo {author} {\bibfnamefont {E.~R.}\ \bibnamefont
  {Tracy}}, \bibinfo {author} {\bibfnamefont {A.~J.}\ \bibnamefont {Brizard}},
  \bibinfo {author} {\bibfnamefont {A.~S.}\ \bibnamefont {Richardson}}, \ and\
  \bibinfo {author} {\bibfnamefont {A.~N.}\ \bibnamefont {Kaufman}},\
  }\href@noop {} {\emph {\bibinfo {title} {Ray Tracing and Beyond: Phase Space
  Methods in Plasma Wave Theory}}}\ (\bibinfo  {publisher} {Cambridge
  University Press},\ \bibinfo {address} {New York},\ \bibinfo {year}
  {2014})\BibitemShut {NoStop}%
\bibitem [{Notea()}]{Notea}%
  \BibitemOpen
  \bibinfo {note} {For example, see the recent series of Refs.~\protect
  \rev@citealpnum {my:quasiop1, my:quasiop2, my:quasiop3, my:quasiop4,
  Yanagihara21} and the references cited therein.}\BibitemShut {Stop}%
\bibitem [{\citenamefont {Nielsen}\ and\ \citenamefont
  {Chuang}(2010)}]{Nielsen10}%
  \BibitemOpen
  \bibfield  {author} {\bibinfo {author} {\bibfnamefont {M.~A.}\ \bibnamefont
  {Nielsen}}\ and\ \bibinfo {author} {\bibfnamefont {I.~L.}\ \bibnamefont
  {Chuang}},\ }\href@noop {} {\emph {\bibinfo {title} {Quantum Computation and
  Quantum Information}}}\ (\bibinfo  {publisher} {Cambridge University Press;
  10th Anniversary edition},\ \bibinfo {year} {2010})\BibitemShut {NoStop}%
\bibitem [{\citenamefont {Rieffel}\ and\ \citenamefont
  {Polak}(2011)}]{Rieffel11}%
  \BibitemOpen
  \bibfield  {author} {\bibinfo {author} {\bibfnamefont {E.~G.}\ \bibnamefont
  {Rieffel}}\ and\ \bibinfo {author} {\bibfnamefont {W.~H.}\ \bibnamefont
  {Polak}},\ }\href@noop {} {\emph {\bibinfo {title} {Quantum Computing: A
  Gentle Introduction}}},\ \bibinfo {edition} {1st}\ ed.\ (\bibinfo
  {publisher} {MIT Press},\ \bibinfo {year} {2011})\BibitemShut {NoStop}%
\bibitem [{Cir(2021)}]{CircuitDepth}%
  \BibitemOpen
  \href@noop {} {\enquote {\bibinfo {title} {Circuit depth explanation},}\
  }\bibinfo {howpublished}
  {\url{https://qiskit.org/textbook/ch-labs/Lab01_QuantumCircuits.html\#step-3-interpret-the-result}}
  (\bibinfo {year} {2021}),\ \bibinfo {note} {accessed: 10-2021}\BibitemShut
  {NoStop}%
\bibitem [{\citenamefont {Low}\ and\ \citenamefont {Chuang}(2017)}]{Low17}%
  \BibitemOpen
  \bibfield  {author} {\bibinfo {author} {\bibfnamefont {G.~H.}\ \bibnamefont
  {Low}}\ and\ \bibinfo {author} {\bibfnamefont {I.~L.}\ \bibnamefont
  {Chuang}},\ }\bibfield  {title} {\enquote {\bibinfo {title} {Optimal
  {Hamiltonian} simulation by quantum signal processing},}\ }\href {\doibase
  10.1103/PhysRevLett.118.010501} {\bibfield  {journal} {\bibinfo  {journal}
  {Physical Review Letters}\ }\textbf {\bibinfo {volume} {118}},\ \bibinfo
  {pages} {010501} (\bibinfo {year} {2017})}\BibitemShut {NoStop}%
\bibitem [{\citenamefont {Low}\ and\ \citenamefont {Chuang}(2019)}]{Low19}%
  \BibitemOpen
  \bibfield  {author} {\bibinfo {author} {\bibfnamefont {G.~H.}\ \bibnamefont
  {Low}}\ and\ \bibinfo {author} {\bibfnamefont {I.~L.}\ \bibnamefont
  {Chuang}},\ }\bibfield  {title} {\enquote {\bibinfo {title} {Hamiltonian
  simulation by qubitization},}\ }\href {\doibase 10.22331/q-2019-07-12-163}
  {\bibfield  {journal} {\bibinfo  {journal} {{Quantum}}\ }\textbf {\bibinfo
  {volume} {3}},\ \bibinfo {pages} {163} (\bibinfo {year} {2019})}\BibitemShut
  {NoStop}%
\bibitem [{\citenamefont {Gily\'{e}n}\ \emph {et~al.}(2019)\citenamefont
  {Gily\'{e}n}, \citenamefont {Su}, \citenamefont {Low},\ and\ \citenamefont
  {Wiebe}}]{Gilyen19}%
  \BibitemOpen
  \bibfield  {author} {\bibinfo {author} {\bibfnamefont {A.}~\bibnamefont
  {Gily\'{e}n}}, \bibinfo {author} {\bibfnamefont {Y.}~\bibnamefont {Su}},
  \bibinfo {author} {\bibfnamefont {G.~H.}\ \bibnamefont {Low}}, \ and\
  \bibinfo {author} {\bibfnamefont {N.}~\bibnamefont {Wiebe}},\ }\bibfield
  {title} {\enquote {\bibinfo {title} {Quantum singular value transformation
  and beyond: Exponential improvements for quantum matrix arithmetics},}\ }in\
  \href {\doibase 10.1145/3313276.3316366} {\emph {\bibinfo {booktitle}
  {Proceedings of the 51st Annual ACM SIGACT Symposium on Theory of
  Computing}}},\ \bibinfo {series and number} {STOC 2019}\ (\bibinfo
  {publisher} {Association for Computing Machinery},\ \bibinfo {address} {New
  York, NY, USA},\ \bibinfo {year} {2019})\ p.\ \bibinfo {pages}
  {193–204}\BibitemShut {NoStop}%
\bibitem [{\citenamefont {Haah}(2019)}]{Haah20}%
  \BibitemOpen
  \bibfield  {author} {\bibinfo {author} {\bibfnamefont {J.}~\bibnamefont
  {Haah}},\ }\bibfield  {title} {\enquote {\bibinfo {title} {Product
  decomposition of periodic functions in quantum signal processing},}\ }\href
  {\doibase 10.22331/q-2019-10-07-190} {\bibfield  {journal} {\bibinfo
  {journal} {Quantum}\ }\textbf {\bibinfo {volume} {3}},\ \bibinfo {pages}
  {190} (\bibinfo {year} {2019})}\BibitemShut {NoStop}%
\bibitem [{\citenamefont {Chao}\ \emph {et~al.}(2020)\citenamefont {Chao},
  \citenamefont {Ding}, \citenamefont {Gilyen}, \citenamefont {Huang},\ and\
  \citenamefont {Szegedy}}]{Chao20}%
  \BibitemOpen
  \bibfield  {author} {\bibinfo {author} {\bibfnamefont {R.}~\bibnamefont
  {Chao}}, \bibinfo {author} {\bibfnamefont {D.}~\bibnamefont {Ding}}, \bibinfo
  {author} {\bibfnamefont {A.}~\bibnamefont {Gilyen}}, \bibinfo {author}
  {\bibfnamefont {C.}~\bibnamefont {Huang}}, \ and\ \bibinfo {author}
  {\bibfnamefont {M.}~\bibnamefont {Szegedy}},\ }\href@noop {} {\enquote
  {\bibinfo {title} {Finding angles for quantum signal processing with machine
  precision},}\ } (\bibinfo {year} {2020}),\ \Eprint
  {http://arxiv.org/abs/2003.02831} {arXiv:2003.02831 [quant-ph]} \BibitemShut
  {NoStop}%
\bibitem [{\citenamefont {Dong}\ \emph {et~al.}(2021)\citenamefont {Dong},
  \citenamefont {Meng}, \citenamefont {Whaley},\ and\ \citenamefont
  {Lin}}]{Dong21}%
  \BibitemOpen
  \bibfield  {author} {\bibinfo {author} {\bibfnamefont {Y.}~\bibnamefont
  {Dong}}, \bibinfo {author} {\bibfnamefont {X.}~\bibnamefont {Meng}}, \bibinfo
  {author} {\bibfnamefont {K.~B.}\ \bibnamefont {Whaley}}, \ and\ \bibinfo
  {author} {\bibfnamefont {L.}~\bibnamefont {Lin}},\ }\bibfield  {title}
  {\enquote {\bibinfo {title} {Efficient phase-factor evaluation in quantum
  signal processing},}\ }\href {\doibase 10.1103/physreva.103.042419}
  {\bibfield  {journal} {\bibinfo  {journal} {Physical Review A}\ }\textbf
  {\bibinfo {volume} {103}},\ \bibinfo {pages} {042419} (\bibinfo {year}
  {2021})}\BibitemShut {NoStop}%
\bibitem [{\citenamefont {von Burg}\ \emph {et~al.}(2021)\citenamefont {von
  Burg}, \citenamefont {Low}, \citenamefont {H\"aner}, \citenamefont {Steiger},
  \citenamefont {Reiher}, \citenamefont {Roetteler},\ and\ \citenamefont
  {Troyer}}]{Burg21}%
  \BibitemOpen
  \bibfield  {author} {\bibinfo {author} {\bibfnamefont {V.}~\bibnamefont {von
  Burg}}, \bibinfo {author} {\bibfnamefont {G.~H.}\ \bibnamefont {Low}},
  \bibinfo {author} {\bibfnamefont {T.}~\bibnamefont {H\"aner}}, \bibinfo
  {author} {\bibfnamefont {D.~S.}\ \bibnamefont {Steiger}}, \bibinfo {author}
  {\bibfnamefont {M.}~\bibnamefont {Reiher}}, \bibinfo {author} {\bibfnamefont
  {M.}~\bibnamefont {Roetteler}}, \ and\ \bibinfo {author} {\bibfnamefont
  {M.}~\bibnamefont {Troyer}},\ }\bibfield  {title} {\enquote {\bibinfo {title}
  {Quantum computing enhanced computational catalysis},}\ }\href {\doibase
  10.1103/PhysRevResearch.3.033055} {\bibfield  {journal} {\bibinfo  {journal}
  {Physical Review Research}\ }\textbf {\bibinfo {volume} {3}},\ \bibinfo
  {pages} {033055} (\bibinfo {year} {2021})}\BibitemShut {NoStop}%
\bibitem [{\citenamefont {Engel}, \citenamefont {Smith},\ and\ \citenamefont
  {Parker}(2019)}]{Engel19}%
  \BibitemOpen
  \bibfield  {author} {\bibinfo {author} {\bibfnamefont {A.}~\bibnamefont
  {Engel}}, \bibinfo {author} {\bibfnamefont {G.}~\bibnamefont {Smith}}, \ and\
  \bibinfo {author} {\bibfnamefont {S.~E.}\ \bibnamefont {Parker}},\ }\bibfield
   {title} {\enquote {\bibinfo {title} {Quantum algorithm for the {Vlasov}
  equation},}\ }\href {\doibase 10.1103/PhysRevA.100.062315} {\bibfield
  {journal} {\bibinfo  {journal} {Physical Review A}\ }\textbf {\bibinfo
  {volume} {100}},\ \bibinfo {pages} {062315} (\bibinfo {year}
  {2019})}\BibitemShut {NoStop}%
\bibitem [{\citenamefont {Jones}\ \emph {et~al.}(2019)\citenamefont {Jones},
  \citenamefont {Brown}, \citenamefont {Bush},\ and\ \citenamefont
  {Benjamin}}]{Jones19}%
  \BibitemOpen
  \bibfield  {author} {\bibinfo {author} {\bibfnamefont {T.}~\bibnamefont
  {Jones}}, \bibinfo {author} {\bibfnamefont {A.}~\bibnamefont {Brown}},
  \bibinfo {author} {\bibfnamefont {I.}~\bibnamefont {Bush}}, \ and\ \bibinfo
  {author} {\bibfnamefont {S.~C.}\ \bibnamefont {Benjamin}},\ }\bibfield
  {title} {\enquote {\bibinfo {title} {{QuEST} and high performance simulation
  of quantum computers},}\ }\href {\doibase 10.1038/s41598-019-47174-9}
  {\bibfield  {journal} {\bibinfo  {journal} {Scientific Reports}\ }\textbf
  {\bibinfo {volume} {9}},\ \bibinfo {pages} {10736} (\bibinfo {year}
  {2019})}\BibitemShut {NoStop}%
\bibitem [{\citenamefont {Swanson}(2003)}]{book:swanson}%
  \BibitemOpen
  \bibfield  {author} {\bibinfo {author} {\bibfnamefont {D.~G.}\ \bibnamefont
  {Swanson}},\ }\href@noop {} {\emph {\bibinfo {title} {Plasma Waves}}}\
  (\bibinfo  {publisher} {IOP},\ \bibinfo {address} {Philadelphia},\ \bibinfo
  {year} {2003})\ \bibinfo {note} {2nd~ed.}\BibitemShut {Stop}%
\bibitem [{Noteb()}]{Noteb}%
  \BibitemOpen
  \bibinfo {note} {For an introduction on the geometrical-optics approximation,
  see, for example, Refs.~\protect \rev@citealpnum {book:tracy, Stix92} or
  Secs.~7.1-7.3 in I.~Y. Dodin, arXiv:2201.08562.}\BibitemShut {Stop}%
\bibitem [{\citenamefont {Low}, \citenamefont {Yoder},\ and\ \citenamefont
  {Chuang}(2016)}]{Low16}%
  \BibitemOpen
  \bibfield  {author} {\bibinfo {author} {\bibfnamefont {G.~H.}\ \bibnamefont
  {Low}}, \bibinfo {author} {\bibfnamefont {T.~J.}\ \bibnamefont {Yoder}}, \
  and\ \bibinfo {author} {\bibfnamefont {I.~L.}\ \bibnamefont {Chuang}},\
  }\bibfield  {title} {\enquote {\bibinfo {title} {Methodology of resonant
  equiangular composite quantum gates},}\ }\href {\doibase
  10.1103/PhysRevX.6.041067} {\bibfield  {journal} {\bibinfo  {journal}
  {Physical Review X}\ }\textbf {\bibinfo {volume} {6}},\ \bibinfo {pages}
  {041067} (\bibinfo {year} {2016})}\BibitemShut {NoStop}%
\bibitem [{\citenamefont {Martyn}\ \emph
  {et~al.}(2021{\natexlab{a}})\citenamefont {Martyn}, \citenamefont {Liu},
  \citenamefont {Chin},\ and\ \citenamefont {Chuang}}]{Martyn-22}%
  \BibitemOpen
  \bibfield  {author} {\bibinfo {author} {\bibfnamefont {J.~M.}\ \bibnamefont
  {Martyn}}, \bibinfo {author} {\bibfnamefont {Y.}~\bibnamefont {Liu}},
  \bibinfo {author} {\bibfnamefont {Z.~E.}\ \bibnamefont {Chin}}, \ and\
  \bibinfo {author} {\bibfnamefont {I.~L.}\ \bibnamefont {Chuang}},\ }\href
  {\doibase 10.48550/ARXIV.2110.11327} {\enquote {\bibinfo {title} {Efficient
  fully-coherent hamiltonian simulation},}\ } (\bibinfo {year}
  {2021}{\natexlab{a}})\BibitemShut {NoStop}%
\bibitem [{Haa(2020)}]{Haah20code}%
  \BibitemOpen
  \href@noop {} {\enquote {\bibinfo {title} {Computation of angles for quantum
  signal processing in {F}\#},}\ }\bibinfo {howpublished}
  {\url{https://github.com/microsoft/Quantum-NC/tree/main/src/simulation/qsp}}
  (\bibinfo {year} {2020}),\ \bibinfo {note} {accessed: 04-2021}\BibitemShut
  {NoStop}%
\bibitem [{Don(2021)}]{Dong21code}%
  \BibitemOpen
  \href@noop {} {\enquote {\bibinfo {title} {Efficient phase-factor evaluation
  in quantum signal processing: code},}\ }\bibinfo {howpublished}
  {\url{https://github.com/qsppack/QSPPACK}} (\bibinfo {year} {2021}),\
  \bibinfo {note} {accessed: 09-2021}\BibitemShut {NoStop}%
\bibitem [{\citenamefont {Barenco}\ \emph {et~al.}(1995)\citenamefont
  {Barenco}, \citenamefont {Bennett}, \citenamefont {Cleve}, \citenamefont
  {DiVincenzo}, \citenamefont {Margolus}, \citenamefont {Shor}, \citenamefont
  {Sleator}, \citenamefont {Smolin},\ and\ \citenamefont
  {Weinfurter}}]{Barenco95}%
  \BibitemOpen
  \bibfield  {author} {\bibinfo {author} {\bibfnamefont {A.}~\bibnamefont
  {Barenco}}, \bibinfo {author} {\bibfnamefont {C.~H.}\ \bibnamefont
  {Bennett}}, \bibinfo {author} {\bibfnamefont {R.}~\bibnamefont {Cleve}},
  \bibinfo {author} {\bibfnamefont {D.~P.}\ \bibnamefont {DiVincenzo}},
  \bibinfo {author} {\bibfnamefont {N.}~\bibnamefont {Margolus}}, \bibinfo
  {author} {\bibfnamefont {P.}~\bibnamefont {Shor}}, \bibinfo {author}
  {\bibfnamefont {T.}~\bibnamefont {Sleator}}, \bibinfo {author} {\bibfnamefont
  {J.~A.}\ \bibnamefont {Smolin}}, \ and\ \bibinfo {author} {\bibfnamefont
  {H.}~\bibnamefont {Weinfurter}},\ }\bibfield  {title} {\enquote {\bibinfo
  {title} {Elementary gates for quantum computation},}\ }\href {\doibase
  10.1103/PhysRevA.52.3457} {\bibfield  {journal} {\bibinfo  {journal} {Phys.
  Rev. A}\ }\textbf {\bibinfo {volume} {52}},\ \bibinfo {pages} {3457--3467}
  (\bibinfo {year} {1995})}\BibitemShut {NoStop}%
\bibitem [{QSP(2022)}]{QSP-code}%
  \BibitemOpen
  \href@noop {} {\enquote {\bibinfo {title} {Numerical \relax{QSP} framework
  implemented in c++},}\ }\bibinfo {howpublished}
  {\url{https://github.com/ivanNovikau/QSVT_framework}} (\bibinfo {year}
  {2022}),\ \bibinfo {note} {accessed: 03-2022}\BibitemShut {NoStop}%
\bibitem [{\citenamefont {H\"aner}, \citenamefont {Roetteler},\ and\
  \citenamefont {Svore}(2018)}]{Haner18}%
  \BibitemOpen
  \bibfield  {author} {\bibinfo {author} {\bibfnamefont {T.}~\bibnamefont
  {H\"aner}}, \bibinfo {author} {\bibfnamefont {M.}~\bibnamefont {Roetteler}},
  \ and\ \bibinfo {author} {\bibfnamefont {K.~M.}\ \bibnamefont {Svore}},\
  }\href@noop {} {\enquote {\bibinfo {title} {Optimizing quantum circuits for
  arithmetic},}\ } (\bibinfo {year} {2018}),\ \Eprint
  {http://arxiv.org/abs/1805.12445} {arXiv:1805.12445 [quant-ph]} \BibitemShut
  {NoStop}%
\bibitem [{\citenamefont {Grover}(1997)}]{Grover97}%
  \BibitemOpen
  \bibfield  {author} {\bibinfo {author} {\bibfnamefont {L.~K.}\ \bibnamefont
  {Grover}},\ }\bibfield  {title} {\enquote {\bibinfo {title} {Quantum
  mechanics helps in searching for a needle in a haystack},}\ }\href {\doibase
  10.1103/PhysRevLett.79.325} {\bibfield  {journal} {\bibinfo  {journal}
  {Physical Review Letters}\ }\textbf {\bibinfo {volume} {79}},\ \bibinfo
  {pages} {325--328} (\bibinfo {year} {1997})}\BibitemShut {NoStop}%
\bibitem [{\citenamefont {Grover}(1998)}]{Grover98}%
  \BibitemOpen
  \bibfield  {author} {\bibinfo {author} {\bibfnamefont {L.~K.}\ \bibnamefont
  {Grover}},\ }\bibfield  {title} {\enquote {\bibinfo {title} {Quantum
  computers can search rapidly by using almost any transformation},}\ }\href
  {\doibase 10.1103/PhysRevLett.80.4329} {\bibfield  {journal} {\bibinfo
  {journal} {Physical Review Letters}\ }\textbf {\bibinfo {volume} {80}},\
  \bibinfo {pages} {4329--4332} (\bibinfo {year} {1998})}\BibitemShut {NoStop}%
\bibitem [{\citenamefont {Brassard}\ \emph {et~al.}(2002)\citenamefont
  {Brassard}, \citenamefont {H{\o}yer}, \citenamefont {Mosca},\ and\
  \citenamefont {Tapp}}]{Brassard02}%
  \BibitemOpen
  \bibfield  {author} {\bibinfo {author} {\bibfnamefont {G.}~\bibnamefont
  {Brassard}}, \bibinfo {author} {\bibfnamefont {P.}~\bibnamefont {H{\o}yer}},
  \bibinfo {author} {\bibfnamefont {M.}~\bibnamefont {Mosca}}, \ and\ \bibinfo
  {author} {\bibfnamefont {A.}~\bibnamefont {Tapp}},\ }\bibfield  {title}
  {\enquote {\bibinfo {title} {Quantum amplitude amplification and
  estimation},}\ }\href {\doibase 10.1090/conm/305/05215} {\bibfield  {journal}
  {\bibinfo  {journal} {Quantum Computation and Information}\ }\textbf
  {\bibinfo {volume} {305}},\ \bibinfo {pages} {53–74} (\bibinfo {year}
  {2002})}\BibitemShut {NoStop}%
\bibitem [{\citenamefont {Suzuki}\ \emph {et~al.}(2020)\citenamefont {Suzuki},
  \citenamefont {Uno}, \citenamefont {Raymond}, \citenamefont {Tanaka},
  \citenamefont {Onodera},\ and\ \citenamefont {Yamamoto}}]{Suzuki20}%
  \BibitemOpen
  \bibfield  {author} {\bibinfo {author} {\bibfnamefont {Y.}~\bibnamefont
  {Suzuki}}, \bibinfo {author} {\bibfnamefont {S.}~\bibnamefont {Uno}},
  \bibinfo {author} {\bibfnamefont {R.}~\bibnamefont {Raymond}}, \bibinfo
  {author} {\bibfnamefont {T.}~\bibnamefont {Tanaka}}, \bibinfo {author}
  {\bibfnamefont {T.}~\bibnamefont {Onodera}}, \ and\ \bibinfo {author}
  {\bibfnamefont {N.}~\bibnamefont {Yamamoto}},\ }\bibfield  {title} {\enquote
  {\bibinfo {title} {Amplitude estimation without phase estimation},}\ }\href
  {\doibase 10.1007/s11128-019-2565-2} {\bibfield  {journal} {\bibinfo
  {journal} {Quantum Information Processing}\ }\textbf {\bibinfo {volume} {19}}
  (\bibinfo {year} {2020}),\ 10.1007/s11128-019-2565-2}\BibitemShut {NoStop}%
\bibitem [{\citenamefont {Uno}\ \emph {et~al.}(2021)\citenamefont {Uno},
  \citenamefont {Suzuki}, \citenamefont {Hisanaga}, \citenamefont {Raymond},
  \citenamefont {Tanaka}, \citenamefont {Onodera},\ and\ \citenamefont
  {Yamamoto}}]{Uno21}%
  \BibitemOpen
  \bibfield  {author} {\bibinfo {author} {\bibfnamefont {S.}~\bibnamefont
  {Uno}}, \bibinfo {author} {\bibfnamefont {Y.}~\bibnamefont {Suzuki}},
  \bibinfo {author} {\bibfnamefont {K.}~\bibnamefont {Hisanaga}}, \bibinfo
  {author} {\bibfnamefont {R.}~\bibnamefont {Raymond}}, \bibinfo {author}
  {\bibfnamefont {T.}~\bibnamefont {Tanaka}}, \bibinfo {author} {\bibfnamefont
  {T.}~\bibnamefont {Onodera}}, \ and\ \bibinfo {author} {\bibfnamefont
  {N.}~\bibnamefont {Yamamoto}},\ }\bibfield  {title} {\enquote {\bibinfo
  {title} {Modified {Grover} operator for quantum amplitude estimation},}\
  }\href {\doibase 10.1088/1367-2630/ac19da} {\bibfield  {journal} {\bibinfo
  {journal} {New Journal of Physics}\ }\textbf {\bibinfo {volume} {23}},\
  \bibinfo {pages} {083031} (\bibinfo {year} {2021})}\BibitemShut {NoStop}%
\bibitem [{\citenamefont {Grinko}\ \emph {et~al.}(2021)\citenamefont {Grinko},
  \citenamefont {Gacon}, \citenamefont {Zoufal},\ and\ \citenamefont
  {Woerner}}]{Grinko21}%
  \BibitemOpen
  \bibfield  {author} {\bibinfo {author} {\bibfnamefont {D.}~\bibnamefont
  {Grinko}}, \bibinfo {author} {\bibfnamefont {J.}~\bibnamefont {Gacon}},
  \bibinfo {author} {\bibfnamefont {C.}~\bibnamefont {Zoufal}}, \ and\ \bibinfo
  {author} {\bibfnamefont {S.}~\bibnamefont {Woerner}},\ }\bibfield  {title}
  {\enquote {\bibinfo {title} {Iterative quantum amplitude estimation},}\
  }\href {\doibase 10.1038/s41534-021-00379-1} {\bibfield  {journal} {\bibinfo
  {journal} {npj Quantum Information}\ }\textbf {\bibinfo {volume} {7}}
  (\bibinfo {year} {2021}),\ 10.1038/s41534-021-00379-1}\BibitemShut {NoStop}%
\bibitem [{\citenamefont {Jaques}\ and\ \citenamefont
  {H\"aner}(2021)}]{Jaques21}%
  \BibitemOpen
  \bibfield  {author} {\bibinfo {author} {\bibfnamefont {S.}~\bibnamefont
  {Jaques}}\ and\ \bibinfo {author} {\bibfnamefont {T.}~\bibnamefont
  {H\"aner}},\ }\href@noop {} {\enquote {\bibinfo {title} {Leveraging state
  sparsity for more efficient quantum simulations},}\ } (\bibinfo {year}
  {2021}),\ \Eprint {http://arxiv.org/abs/2105.01533} {arXiv:2105.01533
  [quant-ph]} \BibitemShut {NoStop}%
\bibitem [{\citenamefont {Li}\ \emph {et~al.}(2021)\citenamefont {Li},
  \citenamefont {Zhu}, \citenamefont {Lyu}, \citenamefont {Huang},\ and\
  \citenamefont {Sun}}]{Li21}%
  \BibitemOpen
  \bibfield  {author} {\bibinfo {author} {\bibfnamefont {Y.}~\bibnamefont
  {Li}}, \bibinfo {author} {\bibfnamefont {Q.}~\bibnamefont {Zhu}}, \bibinfo
  {author} {\bibfnamefont {Z.}~\bibnamefont {Lyu}}, \bibinfo {author}
  {\bibfnamefont {Z.}~\bibnamefont {Huang}}, \ and\ \bibinfo {author}
  {\bibfnamefont {J.}~\bibnamefont {Sun}},\ }\bibfield  {title} {\enquote
  {\bibinfo {title} {{DyCuckoo}: Dynamic hash tables on {GPUs}},}\ }in\ \href
  {\doibase 10.1109/ICDE51399.2021.00070} {\emph {\bibinfo {booktitle} {2021
  IEEE 37th International Conference on Data Engineering (ICDE)}}}\ (\bibinfo
  {publisher} {IEEE Computer Society},\ \bibinfo {address} {Los Alamitos, CA,
  USA},\ \bibinfo {year} {2021})\ pp.\ \bibinfo {pages} {744--755}\BibitemShut
  {NoStop}%
\bibitem [{\citenamefont {Rasmussen}\ and\ \citenamefont
  {Zinner}(2020)}]{Rasmussen20}%
  \BibitemOpen
  \bibfield  {author} {\bibinfo {author} {\bibfnamefont {S.~E.}\ \bibnamefont
  {Rasmussen}}\ and\ \bibinfo {author} {\bibfnamefont {N.~T.}\ \bibnamefont
  {Zinner}},\ }\bibfield  {title} {\enquote {\bibinfo {title} {Simple
  implementation of high fidelity controlled-$i$swap gates and quantum circuit
  exponentiation of non-hermitian gates},}\ }\href {\doibase
  10.1103/PhysRevResearch.2.033097} {\bibfield  {journal} {\bibinfo  {journal}
  {Phys. Rev. Research}\ }\textbf {\bibinfo {volume} {2}},\ \bibinfo {pages}
  {033097} (\bibinfo {year} {2020})}\BibitemShut {NoStop}%
\bibitem [{\citenamefont {Loft}\ \emph {et~al.}(2020)\citenamefont {Loft},
  \citenamefont {Kjaergaard}, \citenamefont {Kristensen}, \citenamefont
  {Andersen}, \citenamefont {Larsen}, \citenamefont {Gustavsson}, \citenamefont
  {Oliver},\ and\ \citenamefont {Zinner}}]{Loft20}%
  \BibitemOpen
  \bibfield  {author} {\bibinfo {author} {\bibfnamefont {N.~J.~S.}\
  \bibnamefont {Loft}}, \bibinfo {author} {\bibfnamefont {M.}~\bibnamefont
  {Kjaergaard}}, \bibinfo {author} {\bibfnamefont {L.~B.}\ \bibnamefont
  {Kristensen}}, \bibinfo {author} {\bibfnamefont {C.~K.}\ \bibnamefont
  {Andersen}}, \bibinfo {author} {\bibfnamefont {T.~W.}\ \bibnamefont
  {Larsen}}, \bibinfo {author} {\bibfnamefont {S.}~\bibnamefont {Gustavsson}},
  \bibinfo {author} {\bibfnamefont {W.~D.}\ \bibnamefont {Oliver}}, \ and\
  \bibinfo {author} {\bibfnamefont {N.~T.}\ \bibnamefont {Zinner}},\ }\bibfield
   {title} {\enquote {\bibinfo {title} {Quantum interference device for
  controlled two-qubit operations},}\ }\href {\doibase
  10.1038/s41534-020-0275-3} {\bibfield  {journal} {\bibinfo  {journal} {npj
  Quantum Information}\ }\textbf {\bibinfo {volume} {6}} (\bibinfo {year}
  {2020}),\ 10.1038/s41534-020-0275-3}\BibitemShut {NoStop}%
\bibitem [{\citenamefont {Bahnsen}\ \emph {et~al.}(2022)\citenamefont
  {Bahnsen}, \citenamefont {Rasmussen}, \citenamefont {Loft},\ and\
  \citenamefont {Zinner}}]{Bahnsen22}%
  \BibitemOpen
  \bibfield  {author} {\bibinfo {author} {\bibfnamefont {E.}~\bibnamefont
  {Bahnsen}}, \bibinfo {author} {\bibfnamefont {S.}~\bibnamefont {Rasmussen}},
  \bibinfo {author} {\bibfnamefont {N.}~\bibnamefont {Loft}}, \ and\ \bibinfo
  {author} {\bibfnamefont {N.}~\bibnamefont {Zinner}},\ }\bibfield  {title}
  {\enquote {\bibinfo {title} {Application of the diamond gate in quantum
  fourier transformations and quantum machine learning},}\ }\href {\doibase
  10.1103/PhysRevApplied.17.024053} {\bibfield  {journal} {\bibinfo  {journal}
  {Phys. Rev. Applied}\ }\textbf {\bibinfo {volume} {17}},\ \bibinfo {pages}
  {024053} (\bibinfo {year} {2022})}\BibitemShut {NoStop}%
\bibitem [{\citenamefont {Martyn}\ \emph
  {et~al.}(2021{\natexlab{b}})\citenamefont {Martyn}, \citenamefont {Rossi},
  \citenamefont {Tan},\ and\ \citenamefont {Chuang}}]{Martyn21}%
  \BibitemOpen
  \bibfield  {author} {\bibinfo {author} {\bibfnamefont {J.~M.}\ \bibnamefont
  {Martyn}}, \bibinfo {author} {\bibfnamefont {Z.~M.}\ \bibnamefont {Rossi}},
  \bibinfo {author} {\bibfnamefont {A.~K.}\ \bibnamefont {Tan}}, \ and\
  \bibinfo {author} {\bibfnamefont {I.~L.}\ \bibnamefont {Chuang}},\ }\bibfield
   {title} {\enquote {\bibinfo {title} {Grand unification of quantum
  algorithms},}\ }\href {\doibase 10.1103/PRXQuantum.2.040203} {\bibfield
  {journal} {\bibinfo  {journal} {PRX Quantum}\ }\textbf {\bibinfo {volume}
  {2}},\ \bibinfo {pages} {040203} (\bibinfo {year}
  {2021}{\natexlab{b}})}\BibitemShut {NoStop}%
\bibitem [{\citenamefont {Dodin}\ \emph {et~al.}(2019)\citenamefont {Dodin},
  \citenamefont {Ruiz}, \citenamefont {Yanagihara}, \citenamefont {Zhou},\ and\
  \citenamefont {Kubo}}]{my:quasiop1}%
  \BibitemOpen
  \bibfield  {author} {\bibinfo {author} {\bibfnamefont {I.~Y.}\ \bibnamefont
  {Dodin}}, \bibinfo {author} {\bibfnamefont {D.~E.}\ \bibnamefont {Ruiz}},
  \bibinfo {author} {\bibfnamefont {K.}~\bibnamefont {Yanagihara}}, \bibinfo
  {author} {\bibfnamefont {Y.}~\bibnamefont {Zhou}}, \ and\ \bibinfo {author}
  {\bibfnamefont {S.}~\bibnamefont {Kubo}},\ }\bibfield  {title} {\enquote
  {\bibinfo {title} {Quasioptical modeling of wave beams with and without mode
  conversion. {I}. {Basic} theory},}\ }\href {\doibase 10.1063/1.5095076}
  {\bibfield  {journal} {\bibinfo  {journal} {Phys. Plasmas}\ }\textbf
  {\bibinfo {volume} {26}},\ \bibinfo {pages} {072110} (\bibinfo {year}
  {2019})}\BibitemShut {NoStop}%
\bibitem [{\citenamefont {Yanagihara}, \citenamefont {Dodin},\ and\
  \citenamefont {Kubo}(2019{\natexlab{a}})}]{my:quasiop2}%
  \BibitemOpen
  \bibfield  {author} {\bibinfo {author} {\bibfnamefont {K.}~\bibnamefont
  {Yanagihara}}, \bibinfo {author} {\bibfnamefont {I.~Y.}\ \bibnamefont
  {Dodin}}, \ and\ \bibinfo {author} {\bibfnamefont {S.}~\bibnamefont {Kubo}},\
  }\bibfield  {title} {\enquote {\bibinfo {title} {Quasioptical modeling of
  wave beams with and without mode conversion. {II}. {Numerical} simulations of
  single-mode beams},}\ }\href {\doibase 10.1063/1.5095173} {\bibfield
  {journal} {\bibinfo  {journal} {Phys. Plasmas}\ }\textbf {\bibinfo {volume}
  {26}},\ \bibinfo {pages} {072111} (\bibinfo {year}
  {2019}{\natexlab{a}})}\BibitemShut {NoStop}%
\bibitem [{\citenamefont {Yanagihara}, \citenamefont {Dodin},\ and\
  \citenamefont {Kubo}(2019{\natexlab{b}})}]{my:quasiop3}%
  \BibitemOpen
  \bibfield  {author} {\bibinfo {author} {\bibfnamefont {K.}~\bibnamefont
  {Yanagihara}}, \bibinfo {author} {\bibfnamefont {I.~Y.}\ \bibnamefont
  {Dodin}}, \ and\ \bibinfo {author} {\bibfnamefont {S.}~\bibnamefont {Kubo}},\
  }\bibfield  {title} {\enquote {\bibinfo {title} {Quasioptical modeling of
  wave beams with and without mode conversion. {III}. {Numerical} simulations
  of mode-converting beams},}\ }\href {\doibase 10.1063/1.5095174} {\bibfield
  {journal} {\bibinfo  {journal} {Phys. Plasmas}\ }\textbf {\bibinfo {volume}
  {26}},\ \bibinfo {pages} {072112} (\bibinfo {year}
  {2019}{\natexlab{b}})}\BibitemShut {NoStop}%
\bibitem [{\citenamefont {Yanagihara}, \citenamefont {Dodin},\ and\
  \citenamefont {Kubo}(2021)}]{my:quasiop4}%
  \BibitemOpen
  \bibfield  {author} {\bibinfo {author} {\bibfnamefont {K.}~\bibnamefont
  {Yanagihara}}, \bibinfo {author} {\bibfnamefont {I.~Y.}\ \bibnamefont
  {Dodin}}, \ and\ \bibinfo {author} {\bibfnamefont {S.}~\bibnamefont {Kubo}},\
  }\bibfield  {title} {\enquote {\bibinfo {title} {Quasioptical modeling of
  wave beams with and without mode conversion. {IV}. {Numerical} simulations of
  waves in dissipative media},}\ }\href {\doibase 10.1063/5.0057345} {\bibfield
   {journal} {\bibinfo  {journal} {Phys. Plasmas}\ }\textbf {\bibinfo {volume}
  {28}},\ \bibinfo {pages} {122102} (\bibinfo {year} {2021})}\BibitemShut
  {NoStop}%
\bibitem [{\citenamefont {Yanagihara}\ \emph {et~al.}(2021)\citenamefont
  {Yanagihara}, \citenamefont {Kubo}, \citenamefont {Dodin},\ and\
  \citenamefont {the {LHD} Experiment~Group}}]{Yanagihara21}%
  \BibitemOpen
  \bibfield  {author} {\bibinfo {author} {\bibfnamefont {K.}~\bibnamefont
  {Yanagihara}}, \bibinfo {author} {\bibfnamefont {S.}~\bibnamefont {Kubo}},
  \bibinfo {author} {\bibfnamefont {I.~Y.}\ \bibnamefont {Dodin}}, \ and\
  \bibinfo {author} {\bibnamefont {the {LHD} Experiment~Group}},\ }\bibfield
  {title} {\enquote {\bibinfo {title} {Quasioptical propagation and absorption
  of electron cyclotron waves: simulations and experiment},}\ }\href {\doibase
  10.1088/1741-4326/ac1d86} {\bibfield  {journal} {\bibinfo  {journal} {Nuclear
  Fusion}\ }\textbf {\bibinfo {volume} {61}},\ \bibinfo {pages} {106012}
  (\bibinfo {year} {2021})}\BibitemShut {NoStop}%
\end{thebibliography}%
